\documentclass[11pt]{article}

\usepackage[array,jhep,needs,young]{apturner}

\title{Generic matter representations in 6D supergravity theories}
\author{Washington Taylor}
\author{and Andrew P. Turner}
\affiliation{Massachusetts Institute of Technology, \\
77 Massachusetts Avenue, Cambridge, MA}
\emailAdd{wati {\rmfamily at} mit.edu}
\emailAdd{apturner {\rmfamily at} mit.edu}

\preprint{\today, MIT-CTP-5062}
{}
\abstract{
In six-dimensional supergravity, there is a natural sense in which matter
lying in certain representations of the gauge group is ``generic,'' in that
other ``exotic'' matter representations require more fine tuning. From
considerations of the dimensionality of the moduli space and anomaly
cancellation conditions, we find that the generic sets of matter
representations are well-defined for 6D supergravity theories with gauge
groups containing arbitrary numbers of nonabelian factors and $\U(1)$ factors.
These generic matter representations also match with those that arise in the
most generic F-theory constructions, both in 6D and in 4D, with non-generic
matter representations requiring more exotic singularity types. The analysis
of generic versus exotic matter illuminates long-standing puzzles regarding
F-theory models with multiple $\U(1)$ factors and provides a useful framework
for analyzing the 6D ``swampland'' of apparently consistent low-energy
theories that cannot be realized through known string constructions. We note
also that the matter content of the standard model is generic by the criteria
used here only if the global structure is $\SU(3)_\text{c} \times
\SU(2)_\text{L} \times \U(1)_Y / \Z_6$.
}

\arxivnumber{}

\begin{document}
\maketitle
\flushbottom

%%%%%%%%%%%%%%%%%%%%%%%%%%%%%%%%%%%%%%%%%%%%%%%%%%%%%%%%%%%%%%%%%%%%%%%%%%%%%%
%%%%%%%%%%%%%%%%%%%%%%%%%%%%%%%%%%%%%%%%%%%%%%%%%%%%%%%%%%%%%%%%%%%%%%%%%%%%%%
%%%%%%%%%%%%%%%%%%%%%%%%%%%%%%%%%%%%%%%%%%%%%%%%%%%%%%%%%%%%%%%%%%%%%%%%%%%%%%
\section{Introduction}\label{sec:intro}

In principle, it seems possible to define low-energy field theories in various
dimensions with a given gauge group $G$ and matter that can live in
essentially any representation of the group $G$. For example, one could
imagine a quantum field theory with a $\U(1)$ gauge group and elementary
particles associated with excitations of matter fields that transform with
fairly arbitrary combinations of charges $q_i$, or a QFT with an $\SU(2)$
gauge group and elementary particles transforming in a representation of
$\SU(2)$ with arbitrarily large dimension. From this point of view, it may
seem surprising that the standard model of particle physics that we observe in
nature has only very simple types of charges under the electromagnetic, weak,
and strong forces.

From the point of view of string theory, however, the most natural
constructions tend to lead to fairly simple representations, such as the
fundamental, adjoint, and two-index antisymmetric and symmetric
representations of $\SU(N)$.  There are also constraints from quantum
consistency conditions such as anomalies that limit the set of possible matter
representations, particularly in higher-dimensional theories with more
symmetry.

In this paper we consider the question of whether some matter representations
are more generic than others in a concrete context where the notion of
``generic'' can be given a quantitative meaning. In particular, we consider
matter spectra in six-dimensional $\cN = (1, 0)$ supergravity theories.  Six
is the largest dimension in which a supersymmetric theory with a gauge group
$G$ can have matter in any representation other than the adjoint.  Six
dimensional supergravity also has very strong conditions for the cancellation
of gauge, gravitational, and mixed gauge--gravitational anomalies
\cite{GreenSchwarzWest6DAnom,SagnottiGS}; these anomaly cancellation
conditions, for example, restrict the set of possible matter representations
of nonabelian gauge groups to a finite set (at least when the number of tensor
multiplets does not exceed 8)
\cite{KumarTaylorBound,KumarMorrisonTaylorGlobalAspects}, although they do not
place a bound on $\U(1)$ charges \cite{TaylorTurnerU1}. Perhaps most
importantly, however, the space of six-dimensional supergravity theories
consists of a set of interconnected\footnote{It is not proven that all
consistent 6D supergravity theories lie on a single connected moduli space.
This is true of all conventional F-theory models, since all elliptic
Calabi--Yau threefolds are connected through various topological transitions,
as discussed in, e.g., \cite{KumarMorrisonTaylorGlobalAspects}. There may,
however, be other branches associated with ``frozen phase'' F-theory models
that are not connected in the same way with the set of conventional F-theory
models \cite{BhardwajEtAlFrozen}.\label{foot:singleMod}} branches, where each
branch constitutes a moduli space of fixed dimensionality.  By comparing the
dimensionality of a branch of moduli space having, for example, a $\U(1)$
gauge group and elementary charges $q = 1, 2$ with a branch having a $\U(1)$
gauge group and elementary charges $q = 1, 2, 3$ and the same anomaly
coefficients, we can say that the former set of charges is more generic
because the dimensionality of the first branch of the moduli space is larger,
so that the set of theories with charges $q = 3$ requires more fine tuning of
the theory.

A precise definition of generic matter representations requires consideration
of how the matter content depends on the anomaly structure of the theory.
After a brief review of 6D anomaly conditions in \cref{sec:anomalies}, we
consider the notion of generic matter in 6D from the points of view of moduli
space dimension and anomalies in \cref{sec:generic}, and in
\cref{sec:generic-f-theory} we describe generic matter in F-theory and discuss
how generic matter illuminates questions about the 6D string landscape and
swampland.  By analyzing the dimensionality of various branches of the 6D
supergravity moduli space with different charge content, we find that in many
situations generic matter representations can be defined as those on the
branch of moduli space of greatest dimension, when the other discrete
parameters of the theory are taken to be fixed. In the simplest cases, with
at most one nonabelian gauge factor and one abelian $\U(1)$ factor, this
notion matches well with the anomaly cancellation conditions, in the sense
that the number of distinct generic matter representations matches the number
of anomaly cancellation conditions, so that the generic matter content in a
given theory is determined as a unique solution of the anomaly cancellation
equations restricted to the generic matter fields; these equations are fixed
in terms of the discrete parameters characterizing the number of tensor
multiplets, the gauge group, and the anomaly coefficients. Furthermore, we
find that the types of generic matter singled out by moduli space dimensions
and anomaly cancellation conditions match precisely with the types of matter
constructed in the most generic F-theory models with a given gauge group.  As
examples, in the case of a $\U(1)$ gauge group, all of these considerations
point to $q = 1, 2$ as the generic matter charges, and in the case of a single
$\SU(2)$ gauge group, generic matter is in the fundamental and adjoint
representations.  As the gauge group becomes more complicated, however,
generic matter becomes less uniquely defined; in particular, we find that when
there are multiple $\U(1)$ factors, the number of anomaly-inequivalent charge
combinations associated with generic matter becomes larger than the number of
anomaly cancellation conditions, so that in the ``generic'' models of maximal
moduli space dimension, only a subset of all the generic charge combinations
are realized.

While most of the considerations in this paper are based on 6D supergravity,
the fact that the generic matter representations identified in 6D match with
those constructed by generic F-theory models suggests that the same or similar
notions of generic matter may be relevant in four dimensions.  In
\cref{sec:further}, we discuss how the framework of this paper illuminates
various questions relevant to both 6D and 4D supergravity, including the
possible relevance of this kind of analysis to the standard model.  In
\cref{sec:conclusions}, we make some concluding remarks and describe some
further directions for related investigations.

We close this introduction with a brief discussion of the term ``generic'' and
the sense in which the matter representations we define as generic here are
natural from a physics perspective.  The word ``generic'' is a somewhat
dangerous term, which is used in many different ways by physicists.  We use
this term here in a very specific way.  In particular, we define generic
matter representations to be those that arise in 6D supergravity theories with
a fixed gauge group and number of tensor multiplets on the moduli space branch
of highest dimensionality given relatively small anomaly coefficients. We feel
this term is appropriate and useful in this context because it matches
structure associated with anomaly cancellation conditions and the geometry of
F-theory. The idea that the generic matter fields are those on the
largest-dimensional component of moduli space essentially captures the idea
that other matter representations involve some fine tuning, in the same sense
that an arbitrary point $x_i$ in a 37-dimensional space is more generic than
one that satisfies a system of 16 algebraic equations $f_j(x) = 0$.  An
important aspect of this definition is, however, that we have fixed the gauge
group of the theory, and the generic matter fields are defined only with
respect to this certain choice of gauge group.  More broadly speaking, some
gauge groups themselves may be non-generic from the point of view of string
compactifications.  For example, in 6D supergravity theories coming from
F-theory, if we fix the number of tensor multiplets and the string charge
lattice---associated with a given choice of elliptic Calabi--Yau threefold for
the F-theory geometry---the only generic gauge groups are those associated
with non-Higgsable clusters \cite{MorrisonTaylorClusters}, which typically
support no or minimal matter.  The generic matter fields for a given gauge
group $G$ that we describe here thus address what is in some sense a
second-order question, where the gauge group $G$ itself may require some
fine-tuning of the geometry and we inquire about the expectations for generic
matter given that first-order fine tuning.  In terms of the full 6D
supergravity moduli space, the branches with a tuned gauge group are
themselves of smaller dimensionality than the generic branches with only
(supersymmetrically) non-Higgsable gauge group factors.  How this notion of
genericity for either gauge groups or matter applies for four-dimensional
theories, from F-theory or otherwise, is much less clear, although at the
geometric level there are certain gauge groups that are similarly generic
associated with 4D non-Higgsable clusters \cite{MorrisonTaylor4DClusters}; we
discuss questions of generic matter for 4D theories a bit further in
\cref{sec:further-4d}.

Finally, another important issue that we have not addressed here is the role
of strongly coupled matter, associated in the F-theory picture with $(4, 6)$
non-Kodaira type singularities at codimension two in the base of the elliptic
fibration.  These are associated with superconformal field theories (SCFTs)
coupled to the gravity theory giving rise to ``conformal matter''
\cite{DelZottoEtAlConformal,ApruzziEtAlConformal}.  Studies of the global
structure of the space of 4D F-theory models show that such strongly coupled
conformal matter appears quite broadly throughout the string landscape
\cite{HalversonLongSungAlg,TaylorWangLandscape}, and may in fact provide a
natural mechanism for realizing the standard model within 4D F-theory
\cite{TaylorWangVacua,TianWangEString}.  Incorporating strongly coupled
conformal matter into our understanding of generic gauge groups and matter in
a systematic way presents promising avenues for further research.

%%%%%%%%%%%%%%%%%%%%%%%%%%%%%%%%%%%%%%%%%%%%%%%%%%%%%%%%%%%%%%%%%%%%%%%%%%%%%%
%%%%%%%%%%%%%%%%%%%%%%%%%%%%%%%%%%%%%%%%%%%%%%%%%%%%%%%%%%%%%%%%%%%%%%%%%%%%%%
%%%%%%%%%%%%%%%%%%%%%%%%%%%%%%%%%%%%%%%%%%%%%%%%%%%%%%%%%%%%%%%%%%%%%%%%%%%%%%
\section{Anomaly conditions in 6D supergravity}\label{sec:anomalies}

We write the gauge group in the form
\begin{equation}
G = \prod_{\kappa = 1}^r G_\kappa \times \prod_{i = 1}^{s} \U(1)_i\,,
\end{equation}
where the $G_\kappa$ are simple nonabelian gauge group factors. In general, we
will use lowercase Greek letters to index simple nonabelian gauge group
factors, and lowercase Roman letters to index $\U(1)$ factors. The numbers of
simple nonabelian and abelian gauge group factors are respectively denoted $r$
and $s$.
% Uppercase Roman letters are used to index the irreducible representations of
% the gauge group under which the hypermultiplets transform. Hypermultiplets
% in the representation $R_I$ transform in the representation $R_{I, \kappa}$
% of $G_\kappa$ and have $\U(1)_i$ charge $q_{I, i}$. That is,
% \begin{equation}
% R_I = \bigotimes_{\kappa = 1}^r R_{I, \kappa}
% 	\otimes \bigotimes_{i = 1}^s q_{I, i}\,.
% \end{equation}
Hypermultiplets in the irreducible representation $R$ transform in the
representation $R_\kappa$ of $G_\kappa$ and have $\U(1)_i$ charge $q_{R, i}$.
That is, we indicate the individual factors of a representation via
\begin{equation}
R = \bigotimes_{\kappa = 1}^r R_\kappa
	\otimes \bigotimes_{i = 1}^s q_{R, i}\,.
\end{equation}
Note that a hypermultiplet we refer to as being in representation $R$ actually
contains fields transforming in both $R$ and $\bar{R}$, as the matter
representations in a 6D theory must be quaternionic. For representations $R$
that are themselves quaternionic, we may have ``half-hypermultiplets'' with
half the field content, only transforming in the representation $R$.

% The dimension $M_I$ of the representation $R_I$ is given by
% \begin{equation}
% M_I = \prod_\kappa d_{R_{I, \kappa}}\,,
% \end{equation}
% where $d_{R_\kappa}$ is the dimension of the representation $R_\kappa$ of the
% gauge group factor $G_\kappa$. We use $M_I^\kappa$ and $M_I^{\kappa
% \lambda}$ to respectively denote the number of $G_\kappa$ and $G_\kappa
% \times G_\lambda$ representations in $R_I$, given by
% \begin{equation}
% M_I^\kappa = \prod_{\mu \ne \kappa} d_{R_{I, \mu}}\,,
% 	\quad M_I^{\kappa \lambda} =
% 		\prod_{\mu \ne \kappa, \lambda} d_{R_{I, \mu}}\,.
% \end{equation}
The dimension $d_R$ of the representation $R$ is given by
\begin{equation}
d_R = \prod_\kappa d_{R_\kappa}\,,
\end{equation}
where $d_{R_\kappa}$ is the dimension of the representation $R_\kappa$ of the
gauge group factor $G_\kappa$. We use $d_R^\kappa$ and $d_R^{\kappa
\mu}$ to respectively denote the number of $G_\kappa$ and $G_\kappa
\times G_\mu$ representations $R_\kappa$ and $R_\kappa \otimes R_\mu$
in $R$, given by
\begin{equation}
d_R^\kappa = \prod_{\mathclap{\lambda \ne \kappa}} d_{R_\lambda}\,,
	\quad d_R^{\kappa \mu} =
		\prod_{\mathclap{\lambda \ne \kappa, \mu}} d_{R_\lambda}\,.
\end{equation}

% We use $x$ with various subscripts to indicate the number of hypermultiplets
% transforming in a given representation. Thus, the multiplicity $x_R$ is the
% number of hypermultiplets transforming in the representation $R$ of $G$. The
% multiplicity $x_{R_\kappa}$ indicates the number of hypermultiplets
% transforming in the representation $R_\kappa$ of the factor $G_\kappa$, which
% accounts for the dimensions of the representations of other group factors in
% every representation $\tilde{R}$ containing $R_\kappa$, i.e.

We use $x$ with various superscripts and subscripts to indicate the number of
hypermultiplets transforming in a given representation. Subscripts indicate
representations, and superscripts indicate to which gauge factor these
representations belong. Thus, for example, $x^\kappa_R$ indicates the number
of hypermultiplets transforming in the representation $R$ of the factor
$G_\kappa$, $x^{\kappa, \mu}_{R, S}$ indicates the number of hypermultiplets
transforming in the representation $R \otimes S$ of the product $G_\kappa
\times G_\mu$, and $x^{i, j, k, \ell}_{q, r, s, t}$ indicates the number of
hypermultiplets transforming in the representation $q \otimes r \otimes s
\otimes t$ of the abelian product $\U(1)_i \times \U(1)_j \times \U(1)_k
\times \U(1)_\ell$. For ease of notation, we will omit the superscripts when
they would include all factors of the gauge group, so that, for example,
$x_{q, r} := x^{1, 2}_{q, r}$ for $G = \U(1)^2$. Because we use $R_\kappa$ to
explicitly indicate a representation of factor $G_\kappa$ and $q_i$ to
indicate a charge of factor $\U(1)_i$, we will also use a compressed notation
that omits the superscripts when they can be inferred from the subscripts,
e.g., $x_{R_\kappa, R_\mu} := x^{\kappa, \mu}_{R_\kappa, R_\mu}$.

These multiplicities are defined so as to account for the dimensions of the
representations of the other group factors in every representation $\tilde{R}$
of $G$ that contains the relevant representation, i.e.,
\begin{equation}
x_{R_\kappa} = \sum_{\mathclap{\tilde{R} \supset R_\kappa}}
	x_{\tilde{R}} d_{\tilde{R}}^\kappa\,.
\end{equation}
Similarly, we have
\begin{equation}
\begin{aligned}
x_{R_\kappa, R_\mu} &= \sum_{\mathclap{\tilde{R} \supset
		R_\kappa \otimes R_\mu}}
	x_{\tilde{R}} d_{\tilde{R}}^{\kappa \mu}\,, \\[3pt]
x_{R_\kappa, q_i} &= \sum_{\mathclap{\tilde{R} \supset
		R_\kappa \otimes q_i}}
	x_{\tilde{R}} d_{\tilde{R}}^\kappa\,, \\[3pt]
x_{R_\kappa, q_i, q_j} &= \sum_{\mathclap{\tilde{R} \supset
		R_\kappa \otimes q_i \otimes q_j}}
	x_{\tilde{R}} d_{\tilde{R}}^\kappa\,, \\[3pt]
x_{q_i, q_j, q_k, q_\ell} &= \sum_{\mathclap{\tilde{R} \supset
		q_i \otimes q_j \otimes q_k \otimes q_\ell}}
	x_{\tilde{R}} d_{\tilde{R}}\,.
\end{aligned}
\end{equation}
It is important to note that these multiplicities are related to one another
when indices are duplicated, e.g.,
\begin{equation}
x^{i, i, k, \ell}_{q, r, s, t} = \delta_{q r} x^{i, k, \ell}_{q, s, t}\,,
\quad \text{(no summation)}\,.
\end{equation}

The numbers of massless vector multiplets and hypermultiplets are denoted $V$
and $H$, respectively, and are given by
% \begin{equation}
% V := V_\text{NA} + V_\text{A} := \sum_\kappa d_{\Adj_\kappa} + s\,,
% 	\quad H := \sum_I M_I\,,
% \end{equation}
\begin{equation}
V = V_\text{NA} + V_\text{A} = \sum_\kappa d_{\Adj_\kappa} + s\,,
	\quad H = \sum_R x_R d_R\,,
\end{equation}
where $V_\text{NA}$ is the number of nonabelian vector multiplets in the
theory, $V_\text{A} = s$ is the number of abelian vector multiplets, and
$\Adj_\kappa$ is the adjoint representation of the gauge group factor
$G_\kappa$. The number of tensor multiplets is denoted by $T$.

We will use $\trace$ to denote the trace in the fundamental representation%,
% $\Trace$ to denote the trace in the adjoint representation,
and $\trace_R$ to denote the trace in an arbitrary representation $R$.

The terms in the anomaly polynomial with no abelian field strength factors
give rise to the conditions for cancellation of gravitational, gauge, and
mixed gauge--gravitational anomalies for the nonabelian factors, which in the
notation of \cite{KumarMorrisonTaylorGlobalAspects} are
% \begin{subequations}
% \label{eq:nonabelAC}
% \begin{align}
% 273 &= H - V + 29 T\,, \label{eq:nonabelACgrav} \\
% a \cdot a &= 9 - T\,, \label{eq:nonabelACa} \\
% a \cdot b_\kappa &= -\frac{1}{6} \lambda_\kappa
% 	\left(\sum_I M_I^\kappa A_{I, \kappa} - A_{\Adj_\kappa}\right)\,,
% 	\label{eq:nonabelACA} \\
% 0 &= \sum_I M_I^\kappa B_{I, \kappa} - B_{\Adj_\kappa}\,,
% 	\label{eq:nonabelACB} \\
% b_\kappa \cdot b_\kappa &= \frac{1}{3} \lambda_\kappa^2
% 	\left(\sum_I M_I^\kappa C_{I, \kappa} - C_{\Adj_\kappa}\right)\,,
% 	\label{eq:nonabelACC} \\
% b_\kappa \cdot b_\mu &= \lambda_\kappa \lambda_\mu \sum_I
% 	M_I^{\kappa \mu} A_{I, \kappa} A_{I, \mu}\,, \quad \kappa \ne \mu\,.
% 	\label{eq:nonabelACmix}
% \end{align}
% \end{subequations}
\begin{subequations}
\label{eq:nonabelAC}
\begin{align}
273 &= H - V + 29 T\,, \label{eq:nonabelACgrav} \\
a \cdot a &= 9 - T\,, \label{eq:nonabelACa} \\
a \cdot b_\kappa &= -\frac{1}{6} \lambda_\kappa
	\left(\sum_{R_\kappa} x_{R_\kappa} A_{R_\kappa}
		- A_{\Adj_\kappa}\right)\,,
	\label{eq:nonabelACA} \\
0 &= \sum_{R_\kappa} x_{R_\kappa} B_{R_\kappa}- B_{\Adj_\kappa}\,,
	\label{eq:nonabelACB} \\
b_\kappa \cdot b_\kappa &= \frac{1}{3} \lambda_\kappa^2
	\left(\sum_{R_\kappa} x_{R_\kappa} C_{R_\kappa}
		- C_{\Adj_\kappa}\right)\,,
	\label{eq:nonabelACC} \\
b_\kappa \cdot b_\mu &= \lambda_\kappa \lambda_\mu
	\sum_{\mathclap{R_\kappa, R_\mu}} x_{R_\kappa, R_\mu} A_{R_\kappa}
	A_{R_\mu}\,, \quad \kappa \ne \mu\,.
	\label{eq:nonabelACmix}
\end{align}
\end{subequations}
%Here, we have included the gravitational condition \cref{eq:nonabelACgrav}.
The \emph{anomaly coefficients} $a$ and $b_\kappa$ are $\SO(1, T)$
vectors in the string charge lattice $\Gamma$ of the 6D theory, and
the notation $x \cdot y$ denotes the (integer-valued) $\SO(1,
T)$-invariant product $\Omega_{\alpha \beta} x^\alpha y^\beta$ on
$\Gamma$, where $\Omega$ is an invariant symmetric bilinear form in
$\SO(1, T)$ associated with the Dirac pairing between string
charges. Note that the indices $\kappa, \mu$ in \cref{eq:nonabelACmix}
must be distinct. For each representation $R$ of a given gauge group
factor, the group theory coefficients $A_R$, $B_R$, and $C_R$ are
defined by
\begin{equation}
\trace_R F^2 = A_R \trace F^2\,,
	\quad \trace_R F^4 = B_R \trace F^4 + C_R \left(\trace F^2\right)^2\,.
\end{equation}
% We have adopted the shorthand notation
% \begin{equation}
% A_{I, \kappa} := A_{R_{I, \kappa}}\,,
% 	\quad B_{I, \kappa} := B_{R_{I, \kappa}}\,,
% 	\quad C_{I, \kappa} := C_{R_{I, \kappa}}\,.
% \end{equation}
These group theory coefficients can be computed by hand for any given gauge
group factor $\SU(N)$ in a manner discussed in \cite{KumarTaylorBound} and
\cite{KumarParkTaylorT=0}, among other sources, and can also be determined
systematically \cite{TurnerABCE}. Note that for gauge groups without a quartic
Casimir, like $\SU(2)$, $\SU(3)$, and the exceptional groups, there is no
coefficient $B_R$ and the anomaly equation \labelcref{eq:nonabelACB} does not
constrain the theory.

The $\lambda_\kappa$ are normalization factors associated with the simple
nonabelian Lie groups (given by $\lambda_\kappa = 2 c_\kappa^\vee /
A_{\Adj_\kappa}$, where $c^\vee$ is the dual Coxeter number), given in
\cref{tab:normFactors}.

\begin{table}[h!]
\centering

\[\setlength{\arraycolsep}{5pt}
\begin{array}{cccccccccc} \toprule
& \gA_n & \gB_n & \gC_n & \gD_n & \gE_6 & \gE_7 & \gE_8 & \gF_4 & \gG_2 \\
	\midrule
\lambda & 1 & 2 & 1 & 2 & 6 & 12 & 60 & 6 & 2 \\ \bottomrule
\end{array}
\]

\caption{Normalization factors for the simple Lie groups.}
\label{tab:normFactors}
\end{table}

The terms in the anomaly polynomial with abelian field strength factors yield
the $\U(1)$ and mixed abelian--nonabelian anomaly equations, given by
\cite{Erler6DAnom,ParkTaylorAbelian,ParkAnomalies}
% \begin{subequations}
% \label{eq:abelAC}
% \begin{align}
% a \cdot b_{i j} &= -\frac{1}{6} \sum_I M_I q_{I, i} q_{I, j}\,,
% 	\label{eq:abelACsqr} \\
% 0 &= \sum_I M_I^\kappa E_{I, \kappa} q_{I, i}\,, \label{eq:abelACE} \\
% b_\kappa \cdot b_{i j} &= \lambda_\kappa
% 	\sum_I M_I^\kappa A_{I, \kappa} q_{I, i} q_{I, j}\,, \label{eq:abelACA} \\
% b_{i j} \cdot b_{k \ell} + b_{i k} \cdot b_{j \ell} + b_{i \ell} \cdot b_{j k}
% 	&= \sum_I M_I q_{I, i} q_{I, j} q_{I, k} q_{I, \ell}\,.
% 	\label{eq:abelACquar}
% \end{align}
% \end{subequations}
\begin{subequations}
\label{eq:abelAC}
\begin{align}
a \cdot b_{i j} &= -\frac{1}{6} \sum_{\mathclap{q_i, q_j}}
	x_{q_i, q_j} q_i q_j\,,
	\label{eq:abelACsqr} \\
0 &= \sum_{\mathclap{R_\kappa, q_i}} x_{R_\kappa, q_i} E_{R_\kappa} q_i\,,
	\label{eq:abelACE} \\
b_\kappa \cdot b_{i j} &= \lambda_\kappa \sum_{\mathclap{R_\kappa, q_i, q_j}}
	x_{R_\kappa, q_i, q_j} A_{R_\kappa} q_i q_j\,,
	\label{eq:abelACA} \\
b_{i j} \cdot b_{k \ell} + b_{i k} \cdot b_{j \ell} + b_{i \ell} \cdot b_{j k}
	&= \sum_{\mathclap{q_i, q_j, q_k, q_\ell}}
	x_{q_i, q_j, q_k, q_\ell} q_i q_j q_k q_\ell\,.
	\label{eq:abelACquar}
\end{align}
\end{subequations}
The $b_{i j}$ are $\SO(1, T)$ vectors in $\Gamma$, where, under the additional
mild assumption that the 6D supergravity theory can be compactified on any
spin manifold with any smooth gauge field configuration, we have the
additional condition $b_{i i} \in 2 \Gamma$
\cite{MonnierMooreParkQuantization}, and the fourth group theory coefficient
$E$ is defined by
\begin{equation}
\trace_R F^3 = E_R \trace F^3\,.
\end{equation}
Note that the indices in \cref{eq:abelAC} need not be distinct.

We refer to \cref{eq:nonabelAC,eq:abelAC} collectively as the anomaly
cancellation (AC) equations.

%%%%%%%%%%%%%%%%%%%%%%%%%%%%%%%%%%%%%%%%%%%%%%%%%%%%%%%%%%%%%%%%%%%%%%%%%%%%%%
%%%%%%%%%%%%%%%%%%%%%%%%%%%%%%%%%%%%%%%%%%%%%%%%%%%%%%%%%%%%%%%%%%%%%%%%%%%%%%
%%%%%%%%%%%%%%%%%%%%%%%%%%%%%%%%%%%%%%%%%%%%%%%%%%%%%%%%%%%%%%%%%%%%%%%%%%%%%%
\section{Generic matter}\label{sec:generic}

In this section we describe generic matter in 6D supergravity from several
perspectives.  We begin with gauge groups containing only up to three $\U(1)$
factors and only nonabelian $\SU(N)$ factors, and discuss at the end of the
section how more $\U(1)$ factors and other nonabelian factors can be analyzed
in a similar fashion. Note that the analysis in this section is based only on
the structure of 6D supergravity theories and is independent of F-theory or
any other UV completion.

%%%%%%%%%%%%%%%%%%%%%%%%%%%%%%%%%%%%%%%%%%%%%%%%%%%%%%%%%%%%%%%%%%%%%%%%%%%%%%
%%%%%%%%%%%%%%%%%%%%%%%%%%%%%%%%%%%%%%%%%%%%%%%%%%%%%%%%%%%%%%%%%%%%%%%%%%%%%%
\subsection{Generic matter representations for $G = \SU(N_1) \times \dots
\times \SU(N_r) \times \U(1)^s$, $s \le 3$}\label{sec:generic-u1-3}

We begin by tabulating in \cref{tab:generic} the set of fields that we
identify as living in generic matter representations for models of the form $G
= \SU(N_1) \times \dots \times \SU(N_r) \times \U(1)^s$, for $s \le 3$. In the
following parts of this section, we describe in detail the sense in which this
set of representations can be considered ``generic'' from the points of view
of moduli space dimension and anomaly cancellation. In each case we identify a
canonical subset of the fields that matches the number of anomaly cancellation
conditions.  In all cases with no more than three abelian gauge factors, when
there is a solution to the anomaly equations containing only the canonical set
of generic matter fields, the dimensionality of the associated branch of the
6D supergravity theory should be greater than (or equal to) that of any other
branch with the same discrete structure of tensor fields, gauge groups,
anomaly coefficients, and string charge lattice.  As discussed further below,
when there are multiple abelian or nonabelian factors, the canonical set
indicated in the table is not uniquely determined; for multiple abelian
factors, different subsets of the full set of generic matter fields can be
realized in models with different signs of anomaly coefficients. The generic
matter fields in \cref{tab:generic} arise with non-negative multiplicities in
solutions to the anomaly equations with small values of the anomaly
coefficients; for nonabelian groups, the corresponding dimension of the moduli
space branch for a given gauge theory is larger than for models with larger
choices of anomaly coefficients. We illustrate the simplest examples of
generic matter in \cref{sec:generic-examples}, and discuss in
\cref{sec:dimension-issues} some aspects of how the definition of generic
matter based on the dimension of moduli space branches relates to the choice
of anomaly coefficients.

\begin{table}[h!]
\centering

\[\setlength{\arraycolsep}{10pt}
\begin{array}{ccc} \toprule
\text{Canonical Representations} & \text{Number} &
	\text{Other Generic Representations} \\ \midrule
\bm{1}_{\mathrlap{0}}                                & 1                   &
	\\[0.6em]
\yng{1}_{\mathrlap{\, 0}}                            & r                   &
	\\[0.6em]
\yng{1,1}_{\mathrlap{\, 0}}                          & r_{\ge 4}           &
	\\[0.6em]
\Adj_{\mathrlap{\, 0}}                               & r                   &
	\\[0.6em]
\left(\yng{1}\,, \bar{\yng{1}}\right)_{\mathrlap{0}} & \binom{r}{2}
	& \left(\yng{1}\,, \yng{1}\right)_{\mathrlap{0}}
		\\[0.6em]
\yng{1}_{\mathrlap{\, 1}}                            & r s                 &
	\\[0.6em]
\yng{1}_{\mathrlap{\, -1}}                           & (r_3 + r_{\ge 4}) s &
	\\[0.6em]
\yng{1}_{\mathrlap{\, (1, 1)}}                       & r \binom{s}{2}      &
	\\[0.6em]
\bm{1}_{\mathrlap{1}}                                & s                   &
	\\[0.6em]
\bm{1}_{\mathrlap{2}}                                & s                   &
	\\[0.6em]
\bm{1}_{\mathrlap{(1, 1)}}                           & \binom{s}{2}        &
	\\[0.6em]
\bm{1}_{\mathrlap{(1, -1)}}                          & \binom{s}{2}        &
	\\[0.6em]
\bm{1}_{\mathrlap{(2, -1)}}                          & 2 \binom{s}{2}
	& \bm{1}_{\mathrlap{(2, 1)}}
	\\[0.6em]
\bm{1}_{\mathrlap{(1, 1, -1)}}                       & 3 \binom{s}{3}
	& \bm{1}_{\mathrlap{(1, 1, 1)}} \\ \bottomrule
\end{array}
\]

\caption{Generic matter representations for gauge groups $G = \SU(N_1) \times
\dots \times \SU(N_r) \times \U(1)^s$ with $s \le 3$, along with the number
of distinct (canonical) representations of each type. Here, $r_2$, $r_3$, and
$r_{\ge 4}$ are the number of $\SU(N)$ factors in $G$ with $N = 2$, $N = 3$,
and $N \ge 4$, respectively, with $r = r_2 + r_3 + r_{\ge 4}$. Note that for
some fields charged under multiple gauge factors, the fields in the first
column are canonical choices and there are other ``locally equivalent''
generic matter representations, as discussed in \cref{sec:generic-exchanges};
such further generic matter representations are listed in the third column.}
\label{tab:generic}
\end{table}

There is extra freedom in the set of generic matter fields when multiple gauge
factors are involved, associated with the distinction between representations
and their conjugates.  This kind of ambiguity in generic matter arises in
particular in distinguishing the $\left(\syng{1}, \bar{\syng{1}}\right)$ and
$\left(\syng{1}, \syng{1}\right)$ representations of $\SU(N) \times \SU(M)$,
as well as the $(2, 1)$ vs. $(2, -1)$ representations of $\U(1)^2$ and the
$(1, 1, -1)$ vs. $(1, 1, 1)$ representations of $\U(1)^3$. In the the first of
these cases, the matter fields are locally indistinguishable (since they are
equivalent under conjugating one of the gauge factors and they contribute
equally to anomalies) and the branches of the moduli space with these
alternate charges have equal dimension to the generic branch listed in
\cref{tab:generic}. In the $\U(1)^2$ and $\U(1)^3$ cases, as discussed in
\cref{sec:abelian-2,sec:abelian-3}, the situation is more subtle;
different choices of signs of anomaly coefficients give rise to different
combinations of generic matter fields that lie on branches of equal dimension.

We use in \cref{tab:generic} and in the rest of the paper a compressed
notation for sets of representations, in which $\SU(N)$ representations are
denoted with Young diagrams, $\U(1)$ charges are denoted with subscripts
(except in some cases where there are no nonabelian factors), and only
nontrivial representations are listed. In this way, we refer to a set of
representations of a general given form. For example, the notation
$\syng{1}_{\, (1, -1)}$ actually designates the set
\begin{equation}
\yng{1}_{\, (1, -1)} =
	\Big\{\left(\bm{1}, \dots, \bm{1}, \yng{1}, \bm{1}, \dots,
	\bm{1}\right)_{(0, \dots, 0, 1, 0, \dots, 0, -1, 0, \dots, 0)}\Big\}\,,
\end{equation}
while $\left(\syng{1}, \bar{\syng{1}}\right)_0$ designates the set
\begin{equation}
\left(\yng{1}, \bar{\yng{1}}\right)_0 =
	\Big\{\left(\bm{1}, \dots, \bm{1}, \yng{1}, \bm{1}, \dots, \bm{1},
	\bar{\yng{1}}, \bm{1}, \dots, \bm{1}\right)_{(0, \dots, 0)}\Big\}\,.
\end{equation}
These sets contain all representations with the nontrivial charges listed in
any factor \emph{and in any permutation}.
% There will be cases where we want to
% consider only the listed permutation of multiple nontrivial $\U(1)$ charges,
% in which case we will wrap the charges in braces $\{\}$ instead of parentheses
% $()$. Thus, for example, for an $\SU(N) \times \U(1)^3$ theory, the set
% $\bm{1}_{(1, 2)}$ contains the individual $\U(1)^3$ representations $(0, 1,
% 2)$, $(0, 2, 1)$, $(1, 0, 2)$, $(2, 0, 1)$, $(1, 2, 0)$, and $(2, 1, 0)$,
% while the set $\bm{1}_{\{1, 2\}}$ contains only the $\U(1)^3$ representations
% $(1, 2, 0)$, $(1, 0, 2)$, and $(0, 1, 2)$. \wati{something weird about this
% and not symmetrical. Discuss?}
Recall that a hypermultiplet we refer to as being in representation $R$ is in
fact in the representation $R \oplus \bar{R}$, and so a negation of all
$\U(1)$ charges and conjugation of all nonabelian reps does not result in a
new representation.

%%%%%%%%%%%%%%%%%%%%%%%%%%%%%%%%%%%%%%%%%%%%%%%%%%%%%%%%%%%%%%%%%%%%%%%%%%%%%%
%%%%%%%%%%%%%%%%%%%%%%%%%%%%%%%%%%%%%%%%%%%%%%%%%%%%%%%%%%%%%%%%%%%%%%%%%%%%%%
\subsection{Examples: generic $\U(1)$ and $\SU(2)$ matter}
\label{sec:generic-examples}

To illustrate the basic ideas we begin with a simple example: generic matter
when the gauge group is just an abelian $\U(1)$ group. In this case, the
anomaly equations are very simple, and read
\begin{equation}
\label{eq:U1AC}
\begin{aligned}
-6 a \cdot \tilde{b} &= \sum_{q > 0} x_q q^2\,, \\ %\label{eq:U1ACsqr} \\
3 \tilde{b} \cdot \tilde{b} &= \sum_{q > 0} x_q q^4\,. %\label{eq:U1ACquar}
\end{aligned}
\end{equation}
Here, $\tilde{b} := b_{1 1}$ is the single anomaly coefficient for the $\U(1)$
gauge factor, and $x_q$ is the number of fields of charge $q$.

In this case, the generic matter charges from \cref{tab:generic} are $q = 1,
2$.  There are two independent anomaly equations constraining the charged
matter, so for any anomaly coefficients $a, \tilde{b}$ there is a unique
solution to the pair of equations
\begin{subequations}
\label{eq:u1-2}
\begin{align}
-6 a \cdot \tilde{b} &= x_1 + 4 x_2\,, \label{eq:u1-2-1} \\
3 \tilde{b} \cdot \tilde{b}  &= x_1 + 16 x_2\,. \label{eq:u1-2-2}
\end{align}
\end{subequations}
We can easily show explicitly that whenever there is a solution to these
equations with non-negative charge multiplicities $x_1, x_2$, any other
solution to the anomaly equations that contains higher charges will have fewer
uncharged scalar degrees of freedom, indicating that the branches of
supergravity moduli space with higher, ``exotic,'' $\U(1)$ charges will have
lower dimensionality. This can be seen by noting that the contribution to the
anomaly equations from a single field of charge $q$ is equivalent to that of a
linear combination of $y_1, y_2$ fields of charge $1, 2$, respectively,
determined by the equations
\begin{equation}
\begin{aligned}
q^2 &= y_1 + 4 y_2\,, \\
q^4 &= y_1 + 16 y_2\,.
\end{aligned}
\end{equation}
The solution to these equations is $y_1 = -q^2 (q^2 - 4) / 3, y_2 = q^2
(q^2 - 1) / 12$.  Because the total number of hypermultiplets is fixed by
the gravitational anomaly condition \labelcref{eq:nonabelACgrav}, this
means that the following combinations of fields are equivalent under
anomalies:
\begin{equation}
\label{eq:u1-equivalence}
(q) + \frac{q^2 \left(q^2 - 4\right)}{3} \times (1)
	\longleftrightarrow
	\frac{q^2 \left(q^2 - 1\right)}{12} \times (2) +
	\left(\frac{q^4}{4} - \frac{5 q^2}{4} + 1\right) \times (0)\,.
\end{equation}
The number of uncharged scalars on the right-hand side is always positive for
$q > 2$, so starting with the generic matter solution with only charges $q =
1, 2$ and then ``exchanging'' these charges for any larger charge necessarily
reduces the number of uncharged scalars.  Since any solution to the anomaly
equations can be found by a finite combination of such exchanges, this proves
that the dimension of the moduli space on the branch with generic $\U(1)$
matter is larger than that of any branch with larger charges, when a solution
with generic matter exists. Furthermore, as we discuss in more detail in
\cref{sec:dimension-issues}, when the anomaly coefficient $\tilde{b}$ is
relatively small (subject to the condition that there is a solution of the
anomaly equations with any set of charges), there is always a solution of
\cref{eq:u1-2} with non-negative multiplicities $x_1, x_2$. Note that in this
analysis, the equivalence \labelcref{eq:u1-equivalence} is simply used as a
formal way of relating different solutions to the anomaly equations. In many
cases, however, we expect that there are ``matter transitions'' in which these
changes in spectrum can be realized physically without changing the gauge
group \cite{AndersonGrayRaghuramTaylorMiT}. We discuss ``anomaly
equivalences'' like \cref{eq:u1-equivalence} further in
\cref{sec:generic-exchanges}.

The same kind of analysis can be carried out if the gauge group is $\SU(2)$.
Again, there are only two anomaly equations constraining the charged matter
since $\SU(2)$ has no quartic Casimir; these are
\begin{equation}
\label{eq:su2-2}
\begin{aligned}
-6 a \cdot b &= x_{\, \tyng{1}} + 4 (x_{\, \tyng{2}} - 1)\,, \\
	%\label{eq:su2-2-1} \\
3 b \cdot b &= \frac{1}{2} x_{\, \tyng{1}} + 8 (x_{\, \tyng{2}} - 1)\,,
	%\label{eq:su2-2-2}
\end{aligned}
\end{equation}
with $b := b_{\SU(2)}$. An arbitrary rep
$\underbrace{\ytableaushort{\none{\none[\dotsm]}\none}*{1}*{2+1}}_k$ of
$\SU(2)$ has $A_R = \binom{k + 2}{3}$ and $C_R = \binom{k + 2}{3} \frac{3 k (k
+ 2) - 4}{10}$.  Expressing these as a linear combination of $A_{\, \tyng{1}}
= 1, C_{\, \tyng{1}} = \frac{1}{2}$ and $A_{\, \tyng{2}} = 4, C_{\,
\tyng{2}} = 8$ leads to an anomaly equivalence
\begin{equation}
\label{eq:su2-equivalence}
\begin{aligned}
\underbrace{\ytableaushort{\none{\none[\dotsm]}\none}*{1}*{2+1}}_k
	&+ \frac{(k + 4) (k + 2) (k + 1) k (k - 2)}{30} \times \yng{1} \\
	&\longleftrightarrow
	\binom{k + 3}{5} \times \yng{2}
	+ \frac{(k + 4) (k + 3) (k + 1) (k - 1) (k - 2)}{24} \times
        \bm{1}\,,
\end{aligned}
\end{equation}
where the number of uncharged scalars on the right-hand side is again always
positive for $k > 2$.  So again, the dimension of the branch of moduli space
with only fundamental and adjoint representations is larger than the branches
containing any other combination of matter fields for a pure $\SU(2)$ theory.
And again, as discussed in \cref{sec:dimension-issues}, when $b$ is small,
there is always a good solution of the anomaly equations with only fundamental
and adjoint representations.

A similar analysis holds in an $\SU(2) \times \U(1)$ theory, where we expect
that generic matter will only involve matter charged under the fundamental or
adjoint of $\SU(2)$ that is neutral under the $\U(1)$ ($\syng{1}_{\, 0}$,
$\syng{2}_{\, 0}$), matter that is neutral under $\SU(2)$ and has charge $q =
1$ or $2$ under the $\U(1)$ factor ($\bm{1}_1$, $\bm{1}_2$), and matter that
lives in the fundamental representation of $\SU(2)$ and has charge $q = 1$
under the $\U(1)$ ($\syng{1}_{\, 1}$). In this case, there are five
independent anomaly equations constraining the charged matter (the four
considered above and one equation of the form \labelcref{eq:abelACA}), and
correspondingly there are five charged generic matter fields.  A similar
calculation to those above shows that an exchange that increases the number of
fields in some other representation
$\underbrace{\ytableaushort{\none{\none[\dotsm]}\none}*{1}*{2+1}}_k{}_q$ by
one decreases the number of uncharged scalars by
\begin{equation}
\frac{(k + 4) (k + 3) (k + 1) (k - 1) (k - 2)}{24}
	+ \left(\frac{q^4}{4} - \frac{5 q^2}{4}\right)
	+ \frac{k (k + 1) (k + 2) q^2}{6}\,,
\end{equation}
which is positive for $k > 2$, for $q > 2$, for $k = 2$ and $q > 0$, or for $k
= 1$ and $q > 1$.

All these cases provide simple examples of situations where we can explicitly
show that the branch of moduli space with only generic matter types has
greater dimension than any branch with other ``exotic'' matter types, and that
these generic matter types can appear when the group theory anomaly
coefficients $b$ are not too large. In the subsequent parts of this section,
we explore the various perspectives on generic matter in more detail.

% While we do
% not have a completely general proof for all the gauge groups considered here
% that generalizes these arguments showing that the generic matter branch of
% moduli space (when it exists) has the greatest dimension, we believe this is
% the case for all the generic matter fields listed in \cref{tab:generic}; an
% explicit proof for a large class of gauge groups is given in
% \cref{sec:generic-proof}.

%%%%%%%%%%%%%%%%%%%%%%%%%%%%%%%%%%%%%%%%%%%%%%%%%%%%%%%%%%%%%%%%%%%%%%%%%%%%%%
%%%%%%%%%%%%%%%%%%%%%%%%%%%%%%%%%%%%%%%%%%%%%%%%%%%%%%%%%%%%%%%%%%%%%%%%%%%%%%
\subsection{Anomaly constraints and generic matter content}
\label{sec:generic-AC}

Generalizing the examples just discussed, given a fixed gauge group $G$, and
fixing the discrete parameters $T$ and $a, b_\kappa, b_{i j} \in \Gamma$, if
we consider matter charged under a number of distinct representations equal to
the number of nontrivial AC equations, then we can generally find a unique
solution for the multiplicity $x_R$ of each representation $R$.  In this
sense, any set of anomaly-inequivalent generic matter charges realized on a
maximal-dimensional moduli space branch should have a cardinality equal to the
number of nontrivial AC equations. We now show that this is true for the
canonical subsets of the generic matter charges listed in \cref{tab:generic}.

For $\SU(N)$, we have $B_R = 0$ for $N < 4$ and $E_R = 0$ for $N < 3$. Suppose
that
\begin{equation}
G = \SU(N_1) \times \dots \times \SU(N_r) \times \U(1)^s
\end{equation}
with $s \le 3$, and let $r_2$, $r_3$, $r_{\ge 4}$ be the number of $\SU(N)$
factors in $G$ with $N = 2$, $N = 3$, and $N \ge 4$, respectively, with $r =
r_2 + r_3 + r_{\ge 4}$. Examining the AC equations in the order presented
(noting that \cref{eq:nonabelACa} simply relates $a$ and $T$, and so does not
contribute), we see then that the number of nontrivial AC equations is
\begin{equation}
n_\text{AC}(r_2, r_3, r_{\ge 4}, s) = 1 + r + r_{\ge 4} + r + \binom{r}{2}
	+ \mchoose{s}{2} + (r_3 + r_{\ge 4}) s + r \mchoose{s}{2}
	+ \mchoose{s}{4}\,,
\end{equation}
where
\begin{equation}
\mchoose{n}{k} = \binom{n + k - 1}{k}
\end{equation}
gives the number of multisets of length $k$ on $n$ symbols. Note that the term
$\mchoose{s}{4}$ coming from \cref{eq:abelACquar} includes $\binom{s}{4} = 0$,
because $s \le 3$.

As desired, the counting in $n(r, s)$ exactly matches the total number of
representations in the canonical sets of generic matter fields in
\cref{tab:generic}, noting that
\begin{equation}
\mchoose{n}{k} = \sum_{j = 1}^k \binom{k - 1}{j - 1} \binom{n}{j}\,.
\end{equation}
% Thus, the representations in \cref{tab:generic} are generic from the
% standpoint of the AC equations.

%%%%%%%%%%%%%%%%%%%%%%%%%%%%%%%%%%%%%%%%%%%%%%%%%%%%%%%%%%%%%%%%%%%%%%%%%%%%%%
%%%%%%%%%%%%%%%%%%%%%%%%%%%%%%%%%%%%%%%%%%%%%%%%%%%%%%%%%%%%%%%%%%%%%%%%%%%%%%
\subsection{Anomaly-equivalent exchanges and ambiguities in generic matter}
\label{sec:generic-exchanges}

As we have seen in some simple examples in \cref{sec:generic-examples}, there
are many situations in which distinct combinations of matter representations
will yield solutions to the AC equations \labelcref{eq:nonabelAC,eq:abelAC}
with the same anomaly coefficients on the left-hand side.  For example,
consider the $k = 3$ case of \cref{eq:su2-equivalence}.  In a $G = \SU(2)$
theory, the fundamental representation has $(A_{\, \tyng{1}}, C_{\, \tyng{1}})
= (1, 1 / 2)$, the adjoint has $(A_{\, \tyng{2}}, C_{\, \tyng{2}}) = (4, 8)$,
and the triple-symmetric has $(A_{\, \tyng{3}}, C_{\, \tyng{3}}) = (10, 41)$.
Because these factors appear in the AC equations in sums over all
hypermultiplets, we see that the exchange
\begin{equation}
\label{eq:exchangeExample}
\yng{3} + 14 \times \yng{1} \longleftrightarrow
	6 \times \yng{2} + 14 \times \bm{1}
\end{equation}
relates distinct solutions to the AC equations with the same $b_{\SU(2)}$. The
singlet representations on the right side are necessary to balance the
dimensions and satisfy the gravitational anomaly condition
\labelcref{eq:nonabelACgrav}. Two solutions to the AC equations related in
this way are referred to as \emph{anomaly-equivalent}. As mentioned earlier,
in many cases there are paths in the supergravity moduli space that connect
anomaly-equivalent matter spectra through \emph{matter transitions}
\cite{AndersonGrayRaghuramTaylorMiT}, though this need not always be the case.

In \cref{tab:generic}, we have chosen a canonical set of generic matter
representations with a cardinality that matches the number of AC equations.
Thus, every spectrum that is allowed by anomaly cancellation will be
anomaly-equivalent to some spectrum containing only the matter representations
listed in this table (with the caveat that the resulting spectrum may have
negative multiplicities for some representations).  In many cases, exchanges
will change the number of uncharged scalar moduli fields. As seen in
\cref{eq:exchangeExample}, increasing the number of triple-symmetric
representations in an $\SU(2)$ theory requires reducing the number of moduli.
In this sense, the triple-symmetric representation is non-generic (or
``exotic''). More generally, from \cref{eq:su2-equivalence}, we see that the
fundamental and adjoint representations of $\SU(2)$ are generic in the sense
that any other anomaly-equivalent combination of representations will involve
a smaller number of uncharged scalar moduli. We prove that the representations
in \cref{tab:generic} are generic in the sense that exchanges that result in
models containing only these representations can never decrease (but may
increase) the number of uncharged scalars in \cref{sec:generic-proof}.

It is important to note, however, that as the size of the gauge group
increases, the set of matter representations that are generic in the sense of
maximizing the dimension of the moduli space branch also increases, and can
exceed the number of AC equations.  In particular, when there are multiple
gauge factors, there can be different anomaly-equivalent combinations of
matter fields that have the same number of uncharged scalar moduli.  In the
simplest cases, there are certain combinations of matter representations under
the factors of the gauge group that are ``locally'' indistinguishable in the
sense that they can be related by conjugating one of the factors in the gauge
group in a way that does not affect the anomaly conditions.  For example,
given a gauge group $\SU(N) \times \SU(M)$, the matter representations
$\left(\syng{1}, \syng{1}\right)$ and $\left(\syng{1}, \bar{\syng{1}}\right)$
are precisely equivalent under anomalies, since $A_R, B_R, C_R$ are unchanged
under conjugation of the representation $R$.  In fact, by choosing the
conjugate realization of the gauge group $\SU(M)$, we can globally replace all
matter in every representation $R$ with matter in the representation
$\bar{R}$, so theories in which the complete spectrum is related by $R
\leftrightarrow \bar{R}$ are globally equivalent. This leads to some ambiguity
in the choice of generic matter for product gauge groups.  One way to
interpret this ambiguity is that, as we have done in \cref{tab:generic}, we
can choose a canonical set of generic matter representations whose cardinality
matches the number of AC equations, and then there are other anomaly-free
spectra in which these canonical choices can be freely exchanged for other
generic matter representations that are anomaly-equivalent but not in the
canonical set.  Another way of framing this is simply to observe that the
number of generic matter fields is larger than the number of AC equations, so
that with sufficiently large product groups the AC equations do not uniquely
determine the generic matter content of the theory. The situation is slightly
more complicated for models with multiple $\U(1)$ factors, where for different
choices of anomaly coefficients, different subsets of the set of generic
matter fields are realized, in each case having cardinality equal to the
number of AC equations; in these cases the different matter fields are
similarly locally related by conjugation but contribute differently to the
anomaly conditions. This is discussed in more detail for theories with two or
three $\U(1)$ factors in \cref{sec:abelian-2,sec:abelian-3}. As we
discuss in \cref{sec:further-u1-4}, with four or more $\U(1)$ factors the
number of generic matter fields with the same dimensionality of the moduli
space branch increases still further.
%, beyond the simple local conjugations just mentioned.

In general, we will sometimes use the term ``generic model'' to describe a
model living on a maximal-dimensional moduli space branch; except for the
ambiguity between fundamental--fundamental and fundamental--antifundamental
representations of $\SU(N) \times \SU(M)$ product groups, and similar
ambiguities discussed in \cref{sec:further-u1-4} that arise in the presence of
four or more $\U(1)$ factors, we expect that such generic models will always
have a number of anomaly-inequivalent matter representations equal to the
number of anomaly equations.

%%%%%%%%%%%%%%%%%%%%%%%%%%%%%%%%%%%%%%%%%%%%%%%%%%%%%%%%%%%%%%%%%%%%%%%%%%%%%%
%%%%%%%%%%%%%%%%%%%%%%%%%%%%%%%%%%%%%%%%%%%%%%%%%%%%%%%%%%%%%%%%%%%%%%%%%%%%%%
\subsection{Generic matter and anomaly coefficients}
\label{sec:dimension-issues}

One subtlety in using the dimension of moduli space branches to define generic
matter is that for different anomaly coefficients, different combinations of
matter fields may be possible.  To unambiguously define the generic matter
representations we are interested in, we want to further focus on those
representations that arise when the anomaly coefficients $b$ associated with
the gauge group are relatively small.  One could motivate this additional
component in the precise definition of generic matter by some general sense in
which small anomalies are more generic than large ones.  A more satisfactory
justification for focusing on small $b$ may be the fact that, at least for
nonabelian groups, the branches of moduli space for a given gauge group, among
those arising from all possible compatible anomaly coefficients, are largest
when the $b$s are small. While we do not attempt to prove rigorously for all
the various nonabelian gauge groups considered here that the
highest-dimensional branches of moduli space considered across all possible
anomaly coefficients are always associated with small anomaly coefficients, we
do show this in some of the simplest examples and also confirm in many cases
that the matter content defined as generic in \cref{tab:generic} is realized
with non-negative multiplicities at small $b$. It is also the case that the
models with small anomaly coefficients and generic matter can generally be
easily constructed in F-theory, while the explicit construction of models with
larger anomaly coefficients is generally more difficult or impossible. In this
subsection, we address this additional condition in the context of the
simplest cases considered in \cref{sec:generic-examples}.

To give a sense of how the small $b$ condition affects the anomaly equations,
we begin with a somewhat qualitative discussion.  For simplicity, we focus in
this discussion on the abelian anomaly equations and charges, though similar
arguments hold in the nonabelian part of the theory.  The abelian anomaly
equations \labelcref{eq:abelACsqr} that are linear in $b$ are quadratic in the
charges associated with the representations on the RHS, while the abelian
anomaly equations that are quadratic in $b$ are quartic in the charges.  (The
corresponding statement for $\SU(2)$, for example, is that the anomaly
coefficients $A_k, C_k$ for the $k$-index symmetric representation scale as
the third and fifth powers of $k$ respectively, as noted in
\cref{sec:generic-examples}).  Thus, as the charges of the matter
representations increase, the ratio between the RHS of the equations quadratic
in $b$ and those linear in $b$ increases roughly as $q^2$. For fixed $a$,
then, we expect that small $b$ will generally be associated with anomaly
solutions with non-negative multiplicities for small charges.  As $b$
increases, the typical charges of the solutions will increase.  Thus, the
types of matter that we are identifying as generic here can also be understood
as those with relatively small charges.  The exotic matter representations are
those with larger charges, which arise for larger values of $b$. Note that for
many gauge groups there are simple solutions with minimal anomaly coefficients
$b$ in which only the representations with the very smallest charges have
nonzero multiplicities (for example, for a $\U(1)$ theory with $b = 6$ the
spectrum contains 108 fields of charge $q = 1$ and no other charges); we are
interested in more generic situations where $b$ is large enough so that the
number of representations with nonzero multiplicities is large enough to match
the number of AC equations.

%%%%%%%%%%%%%%%%%%%%%%%%%%%%%%%%%%%%%%%%%%%%%%%%%%%%%%%%%%%%%%%%%%%%%%%%%%%%%%
\subsubsection{$\SU(2)$ generic matter and anomaly coefficients}
\label{sec:dimension-issues-su2}

To illustrate these issues, perhaps the simplest example is the case of
theories with a single $\SU(2)$ factor and no tensor multiplets ($T = 0$).  In
this case, the gravitational anomaly coefficient is $-a = 3$, and the anomaly
equations restricted to the generic fundamental and adjoint matter fields are
\begin{equation}
\label{eq:su2-T0-2}
\begin{aligned}
18 b &= x_{\,  \tyng{1}} + 4 (x_{\, \tyng{2}} - 1)\,, \\
	%\label{eq:su2-T0-2-1} \\
6 b^2 &= x_{\, \tyng{1}} + 16 (x_{\, \tyng{2}} - 1)\,,
	%\label{eq:su2-T0-2-2}
\end{aligned}
\end{equation}
where $b := b_{\SU(2)}$ is a non-negative integer giving the anomaly
coefficient for the $\SU(2)$ factor. It is straightforward to solve these
equations, giving
\begin{equation}
\label{eq:simplest-T0-su2}
\begin{aligned}
x_{\, \tyng{1}} &= 2 b (12 - b)\,, \\
x_{\, \tyng{2}} &= \frac{(b - 1) (b - 2)}{2}\,.
\end{aligned}
\end{equation}
Thus, when $b \leq 12$ there is a good solution
(i.e., a solution with non-negative multiplicities for these generic
matter fields) of the anomaly equations
restricted to these two fields.  Furthermore, the number of charged matter
fields is
\begin{equation}
H_\text{charged} = 2 x_{\, \tyng{1}} + 3 x_{\, \tyng{2}}
	= \frac{6 + 87 b - 5 b^2}{2}\,.
\end{equation}
This number is smallest for small $b$; computing the dimension of the branch
of moduli space, which is given from \cref{eq:nonabelACgrav} by
$H_\text{uncharged} = 276 - H_\text{charged}$, for $b \le 12$ gives
\begin{equation}
\label{eq:su2-12-models}
H_\text{uncharged} (b = 1, 2, \dots, 12)
	= 232, 196, 165, 139, 118, 102, 91, 85, 84, 88, 97, 111\,.
\end{equation}
This dimension is clearly largest at small values of $b$.
  When $b = 13$, there is no solution with generic matter; we naively
would have $x_{\, \tyng{1}}= -26$ and $x_{\,
\tyng{2}} = 66$.  This is anomaly-equivalent through \cref{eq:exchangeExample}
to a model with spectrum
\begin{equation}
b = 13\colon \quad (x_{\, \tyng{1}}, x_{\, \tyng{2}}, x_{\, \tyng{3}})
	= (2, 54, 2)\,, \quad H_\text{uncharged} = 102\,.
\end{equation}
A similar analysis up to $b = 24$ gives models with fields in these three
representations, and $H_\text{uncharged} \leq 102$.  At this point, further
representations such as $\syng{4}$ would need to be included for a
non-negative spectrum. Thus, we see that for the gauge group $\SU(2)$ and
theories with no tensor multiplets, the largest-dimensional branches of moduli
space are associated with small values of $b$ and are realized by the generic
matter content described in \cref{tab:generic}.

Continuing to consider the case of gauge group $\SU(2)$ and $T = 0$, it is
interesting to note that for certain values of $b$ there are anomaly-free
non-negative spectra with positive values only for $x_{\, \tyng{2}}, x_{\,
\tyng{3}}$; for example, at $b = 14$ we have the spectrum
\begin{equation}
b = 14\colon \quad (x_{\, \tyng{1}}, x_{\, \tyng{2}}, x_{\, \tyng{3}})
	= (0, 54, 4)\,, \quad H_\text{uncharged} = 98\,.
\end{equation}
For this choice of anomaly coefficient, this spectrum maximizes the dimension
of the branch of moduli space.  Thus, if we did not consider the range of
anomaly coefficients $b$ or make an absolute comparison between the dimensions
of branches for different values of $b$, it might seem natural to define
generic matter as depending on $b$, so that the representations $\syng{2},
\syng{3}$ would be ``generic'' matter for $b = 14$.  From this point of view,
\cref{eq:exchangeExample} could be viewed as increasing the dimensionality of
the moduli space branch by trading away all fundamentals to end up at the
maximal-dimensional branch with only representations $\syng{2}, \syng{3}$.  We
choose to focus on the types of matter that arise at small $b$ for several
reasons: first, this gives a more universal definition of generic matter;
second, this definition matches with the branches of moduli space of largest
dimension compared across anomaly coefficients for nonabelian groups such as
$\SU(2)$; and third, this matches most naturally with the structures we find
from F-theory, as described later in the paper.

%%%%%%%%%%%%%%%%%%%%%%%%%%%%%%%%%%%%%%%%%%%%%%%%%%%%%%%%%%%%%%%%%%%%%%%%%%%%%%
\subsubsection{$\U(1)$ generic matter and anomaly coefficients}
\label{sec:dimension-issues-u1}

Now let us consider the case of generic matter for models with a gauge group
$\U(1)$, again restricting to the case $T = 0$, where $-a = 3$ and $\tilde{b}
:= b_{1 1}$ is a non-negative integer.  In this case, there is an additional
subtlety coming from the fact that for the smallest values of $\tilde{b}$
there are no solutions at all to the anomaly equations with non-negative
spectra. In particular, we can see immediately by taking the difference of the
equations \labelcref{eq:U1AC} that
\begin{equation}
\label{eq:u1PositiveEven}
3 \tilde{b} (\tilde{b} - 6) = \sum_{q > 1} x_q q^2 (q^2 - 1)\,,
\end{equation}
so there can only be a non-negative spectrum for any set of charges when
$\tilde{b} \geq 6$.  Thus, when we assert that generic matter representations
should be associated with the matter fields arising in the moduli space branch
of highest dimension when the anomaly coefficients are small, we must include
the condition that the anomaly coefficients be at least big enough that there
is a possible non-negative spectrum.

With this condition, it is straightforward to compute the $\U(1)$ spectrum
with charges $q = 1, 2$ when $\tilde{b}$ is small but not less than 6.  The
result is
\begin{equation}
\begin{aligned}
x_1 &= \tilde{b} (24 - \tilde{b})\,, \\
x_2 &= \frac{\tilde{b} (\tilde{b} - 6)}{4}\,.
\end{aligned}
\end{equation}
This spectrum is simply that associated with Higgsing the $\SU(2)$ models in
\cref{eq:simplest-T0-su2}, when there is at least one adjoint matter field
(see \cite{TaylorTurnerU1} for further related analysis). From this, it is
clear that there is an acceptable spectrum of the $\U(1)$ model with these
charges whenever $\tilde{b}$ is even and $6 \leq \tilde{b} \leq 24$ (note
that, from \cref{eq:u1PositiveEven}, there cannot be any solutions to the
anomaly equations when $\tilde{b}$ is odd).

Note that for the $\U(1)$ gauge group, the dimension of the branches of moduli
space does not decrease with increasing $\tilde{b}$ in the same way that it
does for $\SU(2)$ and other nonabelian gauge groups.  This is related to the
fact that for $\U(1)$ there are an infinite number of possible solutions of
the anomaly equations with arbitrarily large charges \cite{TaylorTurnerU1},
unlike the finite number of matter spectra allowed for a fixed nonabelian
gauge group, which follows from the fact that the dimensions of nonabelian
group representations increase so that only a finite set of distinct
representations are allowed for a theory with a given nonabelian gauge group,
from the gravitational anomaly condition \labelcref{eq:nonabelACgrav}.

%%%%%%%%%%%%%%%%%%%%%%%%%%%%%%%%%%%%%%%%%%%%%%%%%%%%%%%%%%%%%%%%%%%%%%%%%%%%%%
\subsubsection{Models with $T > 0$}\label{sec:dimension-issues-T>0}

In the examples we have considered thus far, we have focused on the case with
no tensor multiplets ($T = 0$).  The observation that the generic matter
fields are realized with non-negative multiplicities in solutions to the
anomaly equations for small values of the anomaly coefficients $b$ also holds
more generally, though the analysis is a bit more involved.

Part of the challenge in dealing with models at larger values of $T$
is that the constraints on the anomaly coefficient $a$ and the
positivity cone in the string charge lattice are not as well
understood from the low-energy point of view
\cite{KumarMorrisonTaylorGlobalAspects}. Here, we briefly consider the
analysis for the gauge group $\SU(2)$ at $T = 1$ using the simplest positivity
cones that are known to come from F-theory; a further analysis
for larger values of $T$ is included in \cref{sec:appendix-t}.

When $T > 0$, the string charge lattice is a unimodular lattice $\Gamma$
\cite{SeibergTaylorLattices} and $-a$ is a characteristic vector in $\Gamma$
\cite{MonnierMooreParkQuantization} that satisfies $a \cdot a = 9 - T$.  The
anomaly coefficients $b_{i i}, b_{\kappa}$ for each abelian and nonabelian
factor lie in a positivity cone in $\Gamma$ (this provides the proper sign for
the gauge kinetic terms).  For $T = 1$, the two simplest choices of inner
products associated with $\Gamma$ and positivity cones that come from F-theory
are
\begin{equation}
\label{eq:even-case}
(x, y) \cdot (x', y') = x y' + y x'\,; \quad x, y \geq 0
\end{equation}
and
\begin{equation}
\label{eq:odd-case}
(x, y) \cdot (x', y') = x x' - y y'\,; \quad  x, y + x \geq 0\,.
\end{equation}
We consider these cases in turn. (The structure of these two unimodular
lattices $\Gamma$ is described in more detail in \cref{sec:6D-structure}.)

In the first, ``even'' case \labelcref{eq:even-case}, we can choose a basis
where $-a = (2, 2)$.  The $\SU(2)$ anomaly equations for a gauge factor with
anomaly coefficient $b = (b_0, b_1)$ are then
\begin{equation}
\label{eq:su2-even-2}
\begin{aligned}
-6 a \cdot b = 12 (b_0 + b_1) &= x_{\, \tyng{1}} + 4 (x_{\, \tyng{2}} - 1)\,,
	\\ %\label{eq:su2-even-2-1} \\
6 b \cdot b = 12 b_0 b_1 &= x_{\, \tyng{1}} + 16 (x_{\, \tyng{2}} - 1)\,.
	%\label{eq:su2-even-2-2}
\end{aligned}
\end{equation}
Solving for generic $\SU(2)$ matter, we have
\begin{equation}
\label{eq:simplest-T1-su2-even}
\begin{aligned}
x_{\, \tyng{1}} &= 16 (b_0 + b_1) - 4 b_0 b_1\,, \\
x_{\, \tyng{2}} &= 1 + b_0 b_1 - b_0 - b_1\,.
\end{aligned}
\end{equation}

\begin{figure}[h!]
\centering

\pic{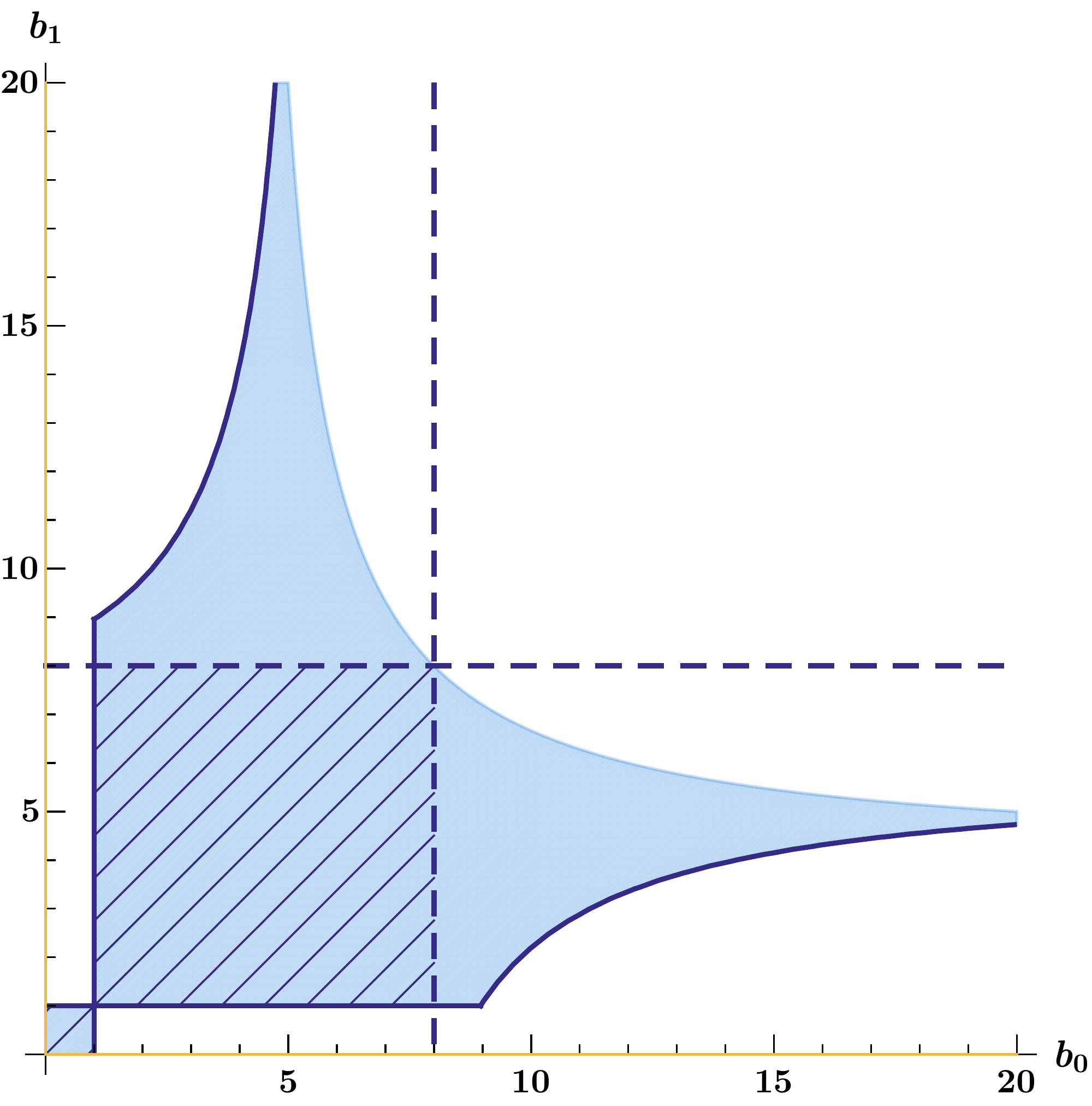}

\caption{Plot of the $(b_0, b_1)$-plane for an $\SU(2)$ model with $T
= 1$, even string charge lattice, and positivity cone $b_0, b_1 \geq 0$.
Choices of $b = (b_0, b_1)$ within the positivity cone (bounded by the yellow
curves) for which the AC equations have non-negative solutions with only
generic matter are shaded in solid blue. The solid curves denote $g = 0$
($b_0, b_1 = 1$) and the gravitational bound \labelcref{eq:nonabelACgrav} on
generic matter, which becomes a bound on $b_0, b_1$ using
\cref{eq:simplest-T1-su2-even}. The dashed curves are $b_0, b_1 = 8$. The
solid and dashed curves bound the hashed region where $g \ge 0$ (a constraint
necessary to have any good solutions to the AC conditions), the gravitational
bound is satisfied, and $b_0, b_1 \le 8$ (the small $b$ constraint). The
hashed region lies entirely within the solid shaded region, showing that all
such choices of $b$ yield non-negative solutions of the AC equations with only
generic matter. As described in \cref{sec:generic-swampland}, the models
within the hashed region have a simple F-theory construction using a Tate
tuning, while the models in the shaded but not hashed region would require a
more sophisticated F-theory Weierstrass model, which may or may not exist.}
\label{fig:su2-t=1-even}
\end{figure}

As shown in \cref{fig:su2-t=1-even}, these equations have non-negative
solutions whenever $b_0, b_1 \le 8$, except at $b_0 = 0, b_1 > 0$ and $b_0 >
0, b_1 = 0$. However, there are no solutions at all to the AC conditions for
choices of $b$ where $b_i = 0, b_{1 - i} > 1$: as discussed in
\cref{sec:appendix-t}, any valid model must have
\begin{equation}
g = \frac{b \cdot (b + a)}{2} + 1 \ge 0\,.
\end{equation}
We see that when $b_i = 0$, $g = 1 - b_{1 - i}$, and so there are no valid
solutions in these cases. Alternatively, we see that for solutions in terms of
generic matter, $x_{\, \tyng{2}} = g$, so any choice of $b_i$ for which the
generic matter solution has a negative multiplicity of adjoints has no good
solution for any set of representations. Thus, there are valid generic matter
models for all $b_0, b_1 \le 8$ in the positivity cone for which there are
any valid solutions.

We can use a similar approach for the ``odd'' case \labelcref{eq:odd-case}.
Here, we can choose a basis where $-a = (3, -1)$. The $\SU(2)$ anomaly
equations then become
\begin{equation}
\label{eq:su2-odd-2}
\begin{aligned}
-6 a \cdot b = 6 (3 b_0 + b_1) &= x_{\, \tyng{1}} + 4 (x_{\, \tyng{2}} - 1)\,,
	\\ %\label{eq:su2-odd-2-1} \\
6 b \cdot b = 6 \left(b_0^2 - b_1^2\right)
	&= x_{\, \tyng{1}} + 16 (x_{\, \tyng{2}} - 1)\,,
	%\label{eq:su2-odd-2-2}
\end{aligned}
\end{equation}
yielding
\begin{equation}
\label{eq:simplest-T1-su2-odd}
\begin{aligned}
x_{\, \tyng{1}} &= 24 b_0 + 8 b_1 - 2 b_0^2 + 2b_1^2\,, \\
x_{\, \tyng{2}} &= \frac{1}{2} \left(b_0^2 - b_1^2 - 3 b_0 - b_1 + 2\right)\,.
\end{aligned}
\end{equation}
As above, a choice of $b$ that yields any good solution of the AC equations
will have
\begin{equation}
x_{\, \tyng{2}} = g \ge 0
\end{equation}
for the solution with generic matter.

\begin{figure}[h!]
\centering

\pic{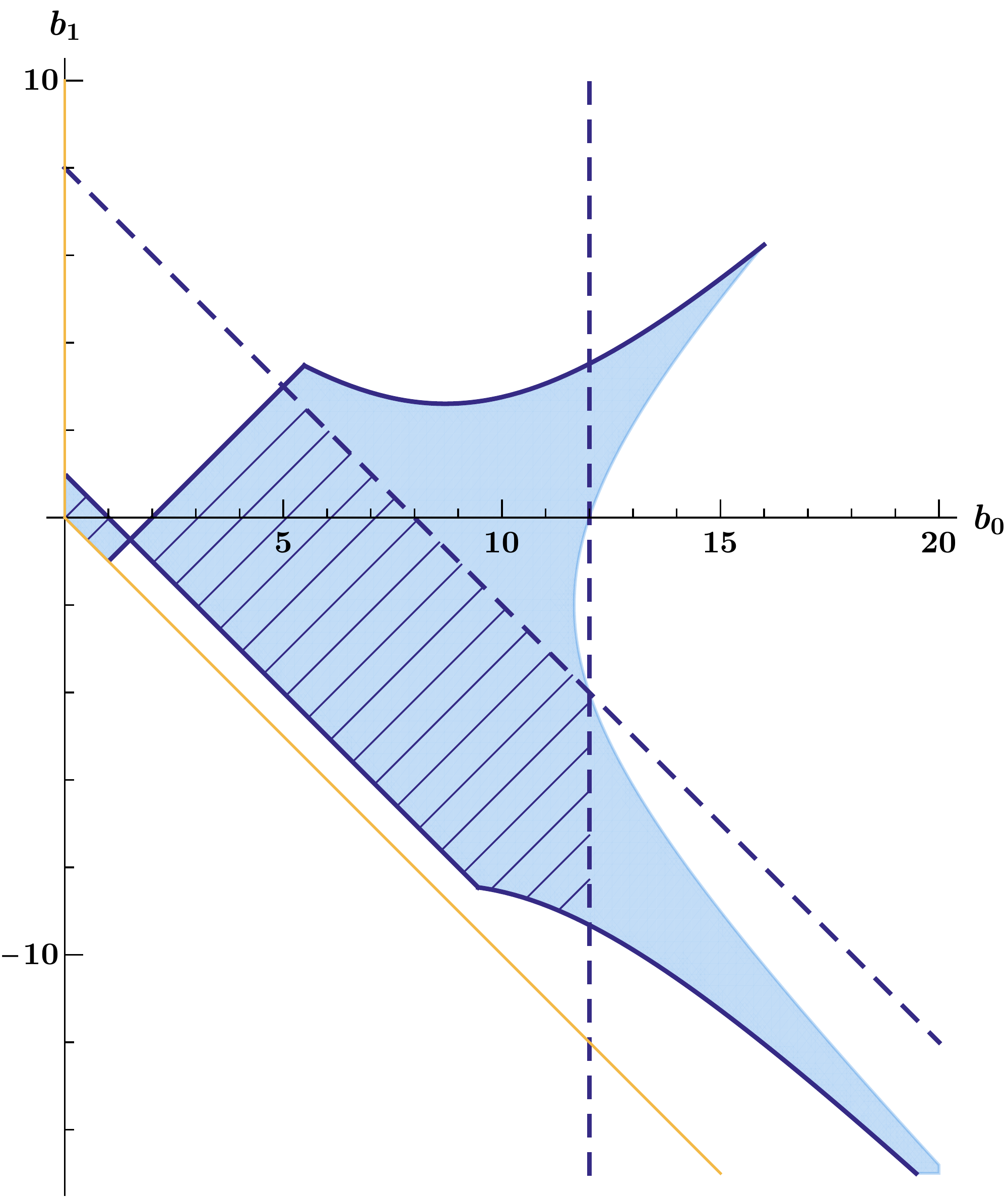}

\caption{Plot of the $(b_0, b_1)$-plane for an $\SU(2)$ model with $T = 1$,
odd string charge lattice, and positivity cone $b_0, b_1 + b_0 \geq 0$.
Choices of $b = (b_0, b_1)$ within the positivity cone (bounded by the yellow
curves) for which the AC equations have non-negative solutions with only
generic matter are shaded in solid blue. The solid curves denote $g = 0$ ($b_0
+ b_1 = 1$, $b_0 - b_1 = 2$) and the gravitational bound
\labelcref{eq:nonabelACgrav} on generic matter, which becomes a bound on $b_0,
b_1$ using \cref{eq:simplest-T1-su2-odd}. The dashed curves are $b_0 = 12, b_0
+ b_1 = 8$. The solid and dashed curves bound the hashed region (again
indicating the models with an F-theory Tate tuning) where $g \ge 0$ (a
constraint necessary to have any good solutions to the AC conditions), the
gravitational bound is satisfied, and $b_0 \le 12, b_0 + b_1 \le 8$ (the
small $b$ constraint). The hashed region lies entirely within the solid
shaded region, showing that all such choices of $b$ yield non-negative
solutions of the AC equations with only generic matter.}
\label{fig:su2-t=1-odd}
\end{figure}

As shown in \cref{fig:su2-t=1-odd}, the multiplicities
\labelcref{eq:simplest-T1-su2-odd} are positive for all choices of $b$ with
$b_0 \le 12$ and $b_0 + b_1 \le 8$ that satisfy $g \ge 0$. Thus, we see that
generic matter is realized with non-negative multiplicities for appropriate
small values of $b$ in the positivity cone on the odd lattice as well.

In \cref{sec:appendix-t}, we use a different approach to demonstrate these
results for higher $T > 0$ in a way that does not rely on a choice of
positivity cone.

%%%%%%%%%%%%%%%%%%%%%%%%%%%%%%%%%%%%%%%%%%%%%%%%%%%%%%%%%%%%%%%%%%%%%%%%%%%%%%
%%%%%%%%%%%%%%%%%%%%%%%%%%%%%%%%%%%%%%%%%%%%%%%%%%%%%%%%%%%%%%%%%%%%%%%%%%%%%%
\subsection{Two abelian factors: $\U(1)^2$}
\label{sec:abelian-2}

When the gauge group contains multiple abelian $\U(1)$ factors, the structure
of generic matter becomes more complicated.  In particular, the number of
generic matter fields becomes larger than the number of anomaly cancellation
equations, and different choices of anomaly coefficients give rise to
different subsets of the set of generic matter fields.  In the case of two
$\U(1)$ factors, the fields
\begin{equation}
\label{eq:2-field-pairs}
(1, \pm 1)\,, \quad (2, \pm 1)\,, \quad (1, \pm 2)
\end{equation}
are all included in the set of  generic matter fields.  In any valid
``generic'' model  (i.e., with a maximal-dimensional branch of moduli space
for given relatively small anomaly coefficients), however, at most four of
these six fields have nonzero multiplicities. This matches with the number
expected from the anomaly cancellation conditions. The way in which these
fields appear, however, is somewhat different than in the previously examined
cases.

%%%%%%%%%%%%%%%%%%%%%%%%%%%%%%%%%%%%%%%%%%%%%%%%%%%%%%%%%%%%%%%%%%%%%%%%%%%%%%
\subsubsection{Choices of generic matter spectra for $\U(1) \times \U(1)$
theories}\label{sec:abelian-2-generic}

The fields in \cref{eq:2-field-pairs} are related in pairs by conjugating one
of the $\U(1)$ factors, similar to the story with $\left(\syng{1},
\syng{1}\right)$ and $\left(\syng{1}, \bar{\syng{1}}\right)$ matter
fields for the group $\SU(N) \times \SU(M)$. When abelian $\U(1)$ factors are
involved, however, even matter fields that are locally equivalent under a
conjugation of a gauge factor can be anomaly-inequivalent. For example, with a
gauge group of $\SU(N) \times \U(1)$, the representations $\syng{1}_{\, 1}$
and $\syng{1}_{\, -1}$ are locally equivalent, in that one representation
switches to the other under a conjugation of either gauge factor; however,
these representations are actually distinguished by the anomaly equation
\labelcref{eq:abelACE}, and hence are not anomaly-equivalent and are
separately listed as generic matter representations in the canonical set in
\cref{tab:generic} that matches the number of AC equations. Similarly, for
$\U(1)^2$ the representations $(1, 1)$ and $(1, -1)$ are distinguished by the
anomaly equations \labelcref{eq:abelAC}, as are the representations $(2, \pm
1)$, etc.

The abelian and mixed abelian--nonabelian anomaly equations
\labelcref{eq:abelAC} have the feature that they are invariant under a
simultaneous change of signs
\begin{equation}
\label{eq:sign-symmetry}
\begin{aligned}
q_i &\rightarrow -q_i\,, \\
b_{i j} &\rightarrow -b_{i j}\,, \quad \forall j \neq i
\end{aligned}
\end{equation}
for any fixed choice of $i$.  Thus, in a theory with $G = \U(1) \times
\U(1)$, a valid spectrum of matter fields $x_{p, q}$ that solve the anomaly
equations with a fixed anomaly coefficient $b_{1 2}$ is related to another
solution with the anomaly coefficient $b'_{1 2} = -b_{1 2}$, where the
multiplicities of the matter fields in the sign-flipped solution are
\begin{equation}
\label{eq:flip-12}
b_{1 2} \rightarrow b'_{1 2} = -b_{1 2}
\quad \Rightarrow \quad
x'_{p, q} = x_{p, -q}\,.
\end{equation}

While all six fields in \cref{eq:2-field-pairs} can arise as generic matter
fields, for any fixed choice of anomaly coefficients we expect from the number
of AC equations that at most four of these fields will have positive
multiplicities in a valid generic model with a maximal dimension of the moduli
space branch. Indeed, this is the case. While generic $\U(1) \times \U(1)$
models contain fields of both sign choices $(1, \pm 1)$, the solutions with
the largest number of uncharged scalars (at small $b$) have only one each
taken from the pairs $(2, \pm 1)$ and $(1, \pm 2)$.  In \cref{tab:generic}, we
have made a canonical choice of the fields with both signs negative, but the
other three choices of sign combinations are also possible.

One way to understand this is from anomaly equivalences. From the anomaly
equations, we can find the anomaly equivalence
\begin{equation}
\label{eq:21-exchange}
(2, 1) + (2, -1) + 6 \times (0, 1) + 8 \times (1, 0) \longleftrightarrow
	2 \times (2, 0) + 4 \times (1, 1) + 4 \times (1, -1) + 6 \times (0, 0)\,.
\end{equation}
 Thus, we can always exchange a pair of charges $(2, 1) + (2, -1)$ for other
generic matter (assuming a sufficient number of fields of charges $(0, 1)$ and
$(1, 0)$) and increase the number of uncharged scalars. A similar statement
clearly holds for $(1, 2) + (-1, 2)$ matter. When we only have $(2, 1)$ matter
fields and no fields of charge $(2, -1)$, however, we cannot make further
exchanges without decreasing the number of uncharged scalars.  Thus, a model
with generic matter is expected to have only one of the types of charge $(2,
1)$ or $(2, -1)$, and only one of the types of charge $(1, 2)$ or $(-1, 2)$.

In fact, the sign choices of the allowed matter fields are determined directly
by the structure of the anomaly coefficients.  As described in more detail in
\cref{sec:two-u1-appendix}, solving the eight relevant anomaly equations for
the canonical choice of generic $\U(1) \times \U(1)$ matter content in
\cref{tab:generic} gives
\begin{equation}
\label{eq:2-1mults}
\begin{aligned}
x_{2, -1} &= -b_{1 2} \cdot \left(a + \frac{1}{2} b_{1 1}\right)\,, \\
x_{-1, 2} &= -b_{1 2} \cdot \left(a + \frac{1}{2} b_{2 2}\right)\,.
\end{aligned}
\end{equation}
These multiplicities are non-negative when the RHS of both equations is
non-negative. This combination of signs for the RHS of \cref{eq:2-1mults} and
the subset of generic matter fields containing $(2, -1), (-1, 2)$ represents
one consistent class of models associated with specific signs for the anomaly
coefficients.  Because there is no positivity constraint on the anomaly
coefficient $b_{1 2}$, we can consider another closely related model, as
discussed above, in which the sign of this coefficient is flipped; this
corresponds to an equivalent model in which one of the $\U(1)$ factors is
conjugated, giving a model with a spectrum related to the original spectrum
through \cref{eq:flip-12}.  In particular, the resulting model has generic
matter fields of types $(2, 1)$ and $(1, 2)$.  Indeed, explicitly solving the
eight relevant anomaly equations for the generic matter content with this
choice of fields gives
\begin{equation}
\label{eq:21mults}
\begin{aligned}
x_{2, 1} &= b_{1 2} \cdot \left(a + \frac{1}{2} b_{1 1}\right)\,, \\
x_{1, 2} &= b_{1 2} \cdot \left(a + \frac{1}{2} b_{2 2}\right)\,.
\end{aligned}
\end{equation}
When $T = 0$, we can show that in fact these two types of matter spectra are
the only possible combinations consistent with the anomaly constraints. From
\cref{eq:u1-2} (noting that these equations will hold separately for each
$\U(1)$ factor), we see that for each $\U(1)$ factor we have
\begin{equation}
x^i_2 = \left(\frac{1}{2} b_{i i}\right) \cdot
	\left(a + \frac{1}{2} b_{i i}\right) \geq 0\,.
\end{equation}
For theories with $T = 0$, the anomaly coefficients $a, b_{i j}$ are integers,
and $-a = 3$, so \cref{eq:u1-2-1} implies that $b_{i i} \geq 0$. This implies
that for each $i$, $a + b_{i i} / 2 \geq 0$. As a result, the RHS of
\cref{eq:2-1mults} always both have the same sign when $T = 0$, as do the RHS
of \cref{eq:21mults}.

When $T > 0$, however, the terms on the RHS of the two equations
\labelcref{eq:2-1mults} can have opposite signs.  In this case, we get a mixed
spectrum, with either charges $(2, 1), (-1, 2)$ or $(1, 2), (2, -1)$.  In
these cases, the expressions for the multiplicities of the non-negative charges
of these types are given by one equation from \cref{eq:2-1mults} and one
equation from \cref{eq:21mults}, as appropriate. We describe an explicit
example of such a spectrum below.

\subsubsection{Generic $\U(1) \times \U(1)$ matter from Higgsing nonabelian
theories}\label{sec:abelian-2-higgsing}

One illuminating perspective on these spectra can be understood from the point
of view of Higgsing a nonabelian theory.  As described in
\cite{CveticKleversPiraguaTaylorU1U1}\footnote{In the model in
\cite{CveticKleversPiraguaTaylorU1U1}, the sign of $b_{1 2}$ is such that the
spectrum contains $(2, 1), (1, 2)$ charged matter, but the embedding
construction there is essentially equivalent to that used in the discussion
here to realize our choice here of canonical $\U(1) \times \U(1)$ matter.},
the canonical $\U(1) \times \U(1)$ generic matter types from
\cref{tab:generic} can be realized when a theory with nonabelian gauge group
$G_{(2)} = \SU(2) \times \SU(2) \times \SU(3)$ and generic matter is broken by
Higgsing pairs of bifundamental $(\syng{1}, \bm{1}, \syng{1})$ and $(\bm{1},
\syng{1}, \syng{1})$ fields to give the $\U(1) \times \U(1)$
model.\footnote{Note that in this and the following sections on $\U(1)^s$
models for $s = 3, s > 3$, we choose to use fundamental--fundamental $\SU(N)
\times \SU(M)$ matter as canonical rather than fundamental--antifundamental
matter to simplify the structure of the Higgsing and embedding formulae; a
different choice of signs for the embedding would match with the
fundamental--antifundamental canonical matter choice in \cref{tab:generic}.}

In particular, we can embed the two $\U(1)$ factors into $G_{(2)}$ as
\begin{equation}
\label{eq:2-embedding}
\begin{aligned}
U(1)_1 & \rightarrow
	\mat[p]{1 & 0 \\ 0 & -1}_1
		+ \mat[p]{0 & 0 \\ 0 & 0}_2
		+ \mat[p]{1 & 0 & 0 \\ 0 & -1 & 0 \\ 0 & 0 & 0}_3\,, \\
U(1)_2 & \rightarrow
	\mat[p]{0 & 0 \\ 0 & 0}_1
		+ \mat[p]{1 & 0 \\ 0 & -1}_2
		+ \mat[p]{0 & 0 & 0 \\ 0 & 1 & 0 \\ 0 & 0 & -1}_3\,.
\end{aligned}
\end{equation}
Here, the subscript on each matrix indicates which factor of $G_{(2)}$ it acts
in. With this embedding, we see that the generic matter types for $G_{(2)}$
descend to the canonical generic matter types for $\U(1) \times \U(1)$.  For
example, the adjoint of $\SU(3)$ contains fields that will give charges $(2,
-1)$ and $(-1, 2)$, as do the bifundamental fields $(\syng{1}, \bm{1},
\syng{1})$ and $(\bm{1}, \syng{1}, \syng{1})$.

If we denote the anomaly coefficients in the nonabelian $G_{(2)}$ model by $A
= b_{\SU(2)_1}, B = b_{\SU(2)_2}, C = b_{\SU(3)}$, then it is straightforward
to deduce that
\begin{equation}
\begin{aligned}
b_{1 1} &= 2 (A + C) \\
b_{2 2} &= 2 (B + C) \\
b_{1 2} & = -C\,.
\end{aligned}
\end{equation}
With the canonical choice of generic matter and the above embedding, we can
then interpret the multiplicities \labelcref{eq:2-1mults} in terms of the
original nonabelian model.  In particular,
\begin{equation}
x_{2, -1} = C \cdot (a + A + C) = C \cdot A + 2 g_C - 2\,,
\end{equation}
and a similar relation holds for $x_{-1, 2}$, where $g_C = C \cdot (a + C) / 2
+ 1$. This immediately matches with the analysis above: $C \cdot A$ is the
number of $(\syng{1}, \bm{1}, \syng{1})$ fields (which give one $(2, -1)$
field each), $g_C$ is the number of adjoints (which give two $(2, -1)$ fields
each), and two bifundamental fields need to be used for the Higgsing.

From this point of view, many consistent generic $\U(1) \times \U(1)$ spectra
can be realized by Higgsing a nonabelian model with gauge group $G_{(2)}$. The
alternative spectra with $(2, 1)$ and $(1, 2)$ matter can be realized
similarly by using \cref{eq:sign-symmetry}, which can be done explicitly by
simply flipping the sign of one of the $\U(1)$ factors in the embedding
\labelcref{eq:2-embedding} and changing the sign of $b_{1 2}$.  Not all
generic $\U(1) \times \U(1)$ models can be unHiggsed in this fashion, however.
In some cases, there may not be enough uncharged scalar matter.  Another more
interesting class of cases arises when we have the mixed matter types such as
$(2, 1)$ and $(1, -2)$.  We conclude this section with a simple example of
such a case.

Consider the case of $T = 1$ with the odd string lattice and positivity cone
\labelcref{eq:odd-case}.  We choose
\begin{equation}
b_{1 1} = (6, 0)\,, \quad
b_{2 2} = (8, -4)\,, \quad
b_{1 2} = (0, 1)\,.
\end{equation}
We see that in this case, we expect the spectrum
\begin{equation}
\begin{aligned}
x_{2, -1} &= -b_{1 2} \cdot \left(a + \frac{1}{2} b_{1 1}\right) = 1\,, \\
x_{1, 2}  &=  b_{1 2} \cdot \left(a + \frac{1}{2} b_{2 2}\right) = 3\,.
\end{aligned}
\end{equation}
This cannot come from a Higgsed nonabelian model as described above, however.
Choosing, for example, $C = b_{1 2}$ and $A = b_{1 1} / 2 - C$, we see that
the nonabelian model would have no $\SU(3)$ adjoint matter and only one
$(\syng{1}, \bm{1}, \syng{1})$ bifundamental field, so there would not be
sufficient matter available for the Higgsing. Nonetheless, as we discuss
further in \cref{sec:generic-f-theory}, it seems likely that this $\U(1)
\times \U(1)$ spectrum can be realized in F-theory.

%%%%%%%%%%%%%%%%%%%%%%%%%%%%%%%%%%%%%%%%%%%%%%%%%%%%%%%%%%%%%%%%%%%%%%%%%%%%%%
\subsubsection{Generic $\U(1) \times \U(1)$ matter at small $b$, $T = 0$}
\label{sec:abelian-2-small-b}

We believe that for all sufficiently small choices of $b$ that have good
solutions to the AC conditions, a subset of generic matter representations
with cardinality equal to the number of nontrivial AC conditions is realized
with non-negative multiplicities. We prove that this is the case for $T = 0$
models here, and briefly discuss the more general case in
\cref{sec:two-u1-appendix}.

\begin{figure}[h!]
\centering

\pic{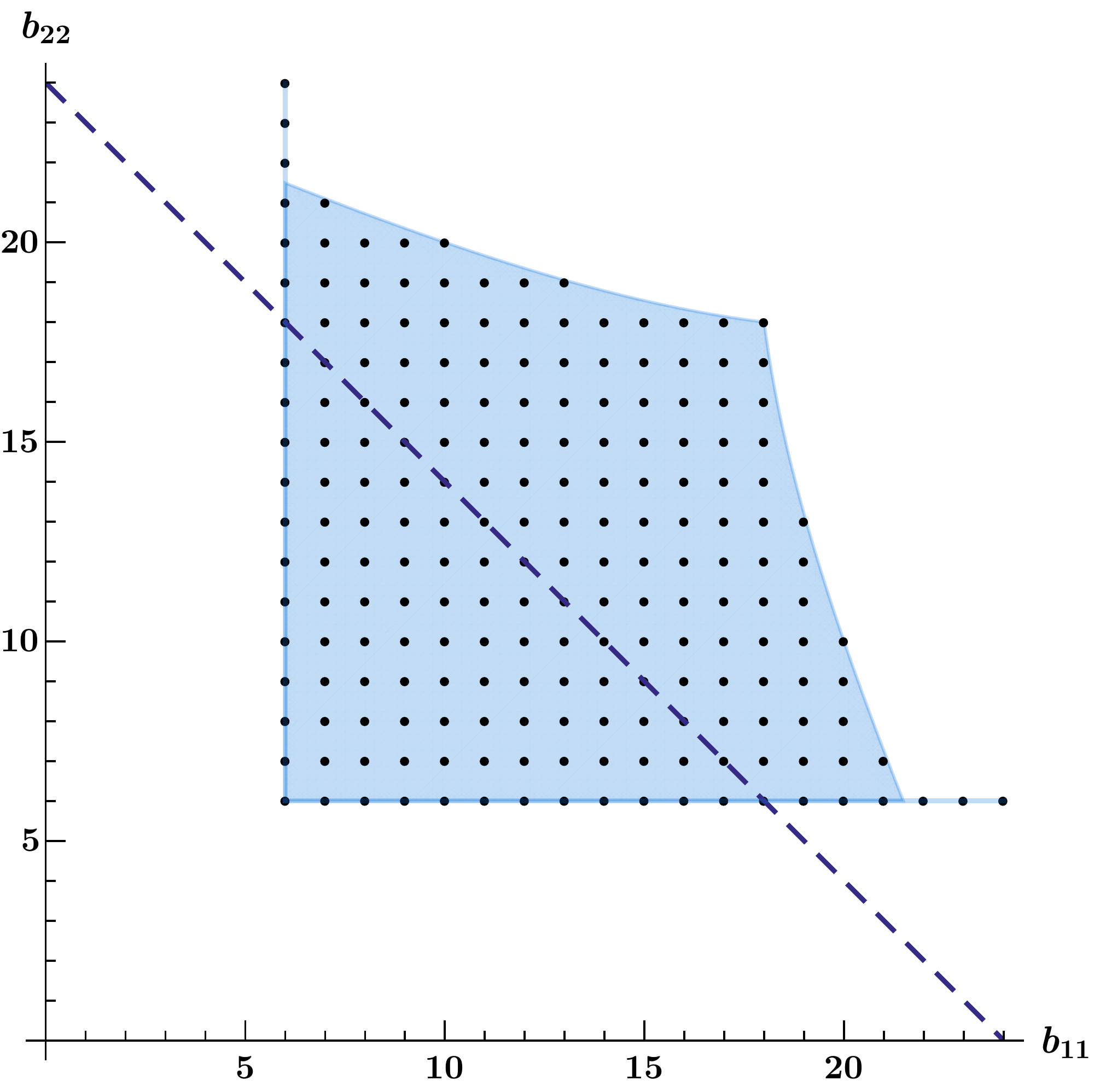}

\caption{Plot of the $(b_{1 1}, b_{2 2})$-plane for a $\U(1) \times \U(1)$
model with $T = 0$. Choices of $b_{1 1}, b_{2 2}$ for which the AC equations
have non-negative solutions with only generic matter (either with the $(2,
-1), (-1, 2)$ or the $(2, 1), (1, 2)$ pair of charge combinations) are shaded
in solid blue. Integer choices of $b_{i i}$ for which there is such a solution
with integer $b_{i j}$ are superimposed in black. Below the dashed curve $b_{1
1} + b_{2 2} = 24$, all choices of $b_{i i}$ that yield any good solutions of
the AC equations lie within the blue shaded region, and within this region,
there is a good solution with one of the two above subsets of generic matter
for all $\abs{b_{1 2}} \le \min(b_{1 1}, b_{2 2}) / 2$.}
\label{fig:u1-2-t=0}
\end{figure}

We have already seen in \cref{sec:abelian-2-generic} that for any choice of
$b_{i j}$ that yields good solutions of the AC equations, at most one of the
pairs $(2, -1), (-1, 2)$ and $(2, 1), (1, 2)$ can have nonzero multiplicities
in a generic model on a maximal-dimensional branch of moduli space. Thus, we
only need to inspect the multiplicities for the remaining generic matter
representations. The full set of equations determining these multiplicities is
given in \cref{sec:two-u1-appendix}. For $T = 0$, we find that for all $b_{i
i}$ with $b_{1 1} + b_{2 2} \le 24$ that yield good solutions of the AC
equations, there exists a good solution with generic matter for one of the two
choices of $(2, \pm 1), (\pm 1, 2)$ for all $\abs{b_{1 2}} \le \min(b_{1 1},
b_{2 2}) / 2$. This result is shown in \cref{fig:u1-2-t=0}.

%%%%%%%%%%%%%%%%%%%%%%%%%%%%%%%%%%%%%%%%%%%%%%%%%%%%%%%%%%%%%%%%%%%%%%%%%%%%%%
%%%%%%%%%%%%%%%%%%%%%%%%%%%%%%%%%%%%%%%%%%%%%%%%%%%%%%%%%%%%%%%%%%%%%%%%%%%%%%
\subsection{Three abelian factors: $\U(1)^3$}
\label{sec:abelian-3}

Turning to $\U(1)^3$, in addition to the canonical fields in
\cref{tab:generic}, matter fields of the form $(1, 1, 1)$ are locally
equivalent to those with charges $(1, 1, -1)$, and can also play the role of
generic matter fields. Similarly to the case of charges $(2, 1)$ and $(2,
-1)$, we find that there is an exchange
\begin{equation}
\label{eq:111-exchange}
\begin{aligned}
&(1, 1, 1) + (-1, 1, 1) + (1, -1, 1) + (1, 1, -1) \\
&\quad + 4 \times (1, 0, 0) + 4 \times (0, 1, 0) + 4 \times (0, 0, 1) \\
&\qquad\qquad\qquad\quad \longleftrightarrow \\
&\qquad\qquad\qquad\qquad\quad 2 \times (1, 1, 0) + 2 \times (1, 0, 1)
	+ 2 \times (0, 1, 1) + 2 \times (1, -1, 0) \\
&\qquad\qquad\qquad\qquad\qquad + 2 \times (1, 0, -1) + 2 \times (0, 1, -1)
	+ 4 \times (0, 0, 0)\,,
\end{aligned}
\end{equation}
so that, assuming there are a sufficient number of fields of charges $(1, 0,
0)$, $(0, 1, 0)$, and $(0, 0, 1)$, we can always remove quartets of charges
$(1, 1, 1) + (-1, 1, 1) + (1, -1, 1) + (1, 1, -1)$ and increase the number of
uncharged scalars. Thus, we expect that models with generic matter and a
maximal number of uncharged scalars (at small $b$ as usual) should only have a
size three subset of these four charges present. Note that the number of
canonical fields listed in the table has cardinality $3 \binom{s}{3}$, which
matches the number of AC equations, while the number of fields of type $(1, 1,
1)$ is $\binom{s}{3}$.

For each pair of $\U(1)$ charges, we can again use the analysis of the
previous section to show that there are only one each of the pairs $(2, \pm 1,
0)$, $(1, \pm 2, 0)$, etc. Furthermore, because the analysis for each pair of
$\U(1)$ factors considered independent of the third charge is unchanged, we
can read off which elements of each of the $(2, \pm 1, 0)$ type pairs are
realized directly from the signs of the terms $b_{i j} \cdot (a + b_{i i} /
2)$, analogous to \cref{eq:2-1mults,eq:21mults}, for each
pair $i, j$.

While the structure of the minimal set of three of the four $(1, \pm 1, \pm
1)$ type fields (recall that $(1, -1, -1)$ is the same type of matter as $(-1,
1, 1)$) is correlated with the structure of the $(2, \pm 1, 0)$ type fields,
in some cases the minimal set of $(1, \pm 1, \pm 1)$ fields is not determined
uniquely by the  $(2, \pm 1, 0)$ type fields or by the signs of the terms
$b_{i j} \cdot (a + b_{i i} / 2)$. To say a little more about this, it is
helpful to distinguish two classes of configurations. To illustrate the first
class, let us start with a case where $-b_{i j} \cdot (a + b_{i i} / 2) \geq
0$ for all six possible pairs $i, j$ (with $b_{i j} = b_{j i}$). In the spirit
of the discussion above for two $\U(1)$ factors, such a $\U(1)^3$ model can be
realized by Higgsing a model with gauge group $G_{(3)} = \SU(2)^3 \times
\SU(3)^3$, where the $\U(1)$ factors are embedded as shown in
\cref{tab:U1-3embed}. Roughly speaking, the $\SU(2)$ factors are associated
with the anomaly coefficient combinations $b_{i i} / 2 + \sum_j b_{i j}$ for
each $i$, and the $\SU(3)$ factors are associated with the $-b_{i j}$, $i \ne
j$.  With such an embedding, it is straightforward to see that the $\SU(3)$
adjoint and bifundamental fields realize $\U(1)$ charges in the canonical set
listed in \cref{tab:generic}.  Thus, when all these signs are consistently
chosen in this way for the anomaly coefficients, we should realize this
canonical set of generic matter fields (for sufficiently small $b$, as usual).
For example, when $T = 0$, this corresponds to the case where all $b_{i j}$
are negative.

\begin{table}[h!]
\centering

\[\setlength{\arraycolsep}{5pt}
\begin{array}{c|ccc}
& \U(1) & \U(1) & \U(1) \\ \hline
\SU(2) & \mat[p]{0 & 0 \\ 0 & 0} & \mat[p]{0 & 0 \\ 0 & 0} &
	\mat[p]{1 & 0 \\ 0 & -1} \rule{0pt}{2.2em} \\[1.3em]
\SU(2) & \mat[p]{0 & 0 \\ 0 & 0} &
	\mat[p]{1 & 0 \\ 0 & -1} & \mat[p]{0 & 0 \\ 0 & 0} \\[1.3em]
\SU(2) & \mat[p]{1 & 0 \\ 0 & -1} &
	\mat[p]{0 & 0 \\ 0 & 0}  & \mat[p]{0 & 0 \\ 0 & 0} \\[1.3em]
\SU(3) & \mat[p]{0 & 0 & 0 \\ 0 & 0 & 0 \\ 0 & 0 & 0} &
	\mat[p]{1 & 0 & 0 \\ 0 & -1 & 0 \\ 0 & 0 & 0} &
	\mat[p]{0 & 0 & 0 \\ 0 & 1 & 0 \\ 0 & 0 & -1} \\[2em]
\SU(3) & \mat[p]{1 & 0 & 0 \\ 0 & -1 & 0 \\ 0 & 0 & 0} &
	\mat[p]{0 & 0 & 0 \\ 0 & 0 & 0 \\ 0 & 0 & 0} &
	\mat[p]{-1 & 0 & 0 \\ 0 & 0 & 0 \\ 0 & 0 & 1} \\[2em]
\SU(3) & \mat[p]{-1 & 0 & 0 \\ 0 &  0 & 0 \\ 0 & 0 & 1} &
	\mat[p]{0 & 0 & 0 \\ 0 & 1 & 0 \\ 0 & 0 & -1} &
	\mat[p]{0 & 0 & 0 \\ 0 & 0 & 0 \\ 0 & 0 & 0}
\end{array}
\]

\caption{Embedding of $\U(1)^3$ into $\SU(2)^3 \times \SU(3)^3$. The generator
of a given $\U(1)$ factor is the sum of the $\SU(N)$ generators in the
corresponding column.}
\label{tab:U1-3embed}
\end{table}

For simplicity, let us continue to consider the case where $T = 0$, so the
anomaly coefficients are integers.  We assume all $b_{i j}$ are nonzero; when
one or more of these coefficients vanish, the configuration is degenerate and
missing some generic matter fields. From the preceding Higgsing construction,
we can consider three other classes of models that are realized by
implementing the change of signs \labelcref{eq:sign-symmetry} for each of the
three choices of $i$.  This will give a total of four possible minimal sets of
generic matter fields that we expect to be associated with moduli space
branches of maximal dimension; these four sets are given by the canonical set
listed in \cref{tab:generic} and the sets realized by flipping the signs of
$q_i$ in all charges for each choice of $i$.  For $T = 0$, where the $b_{i j}$
are integers, these four classes of models correspond to sets of $b_{i j}$
where an even number of these anomaly coefficients are positive.  In analogy
with a spin model where $b_{i j}$ gives the coupling between pairs of spins,
we refer to this as a ``non-frustrated'' configuration.  In these cases, the
type $(1, \pm 1, \pm 1)$ matter choices in generic models are also fixed by
the anomaly coefficient signs. It is interesting to also consider the four
``frustrated'' configurations of $b_{i j}$, where an odd number of the
off-diagonal abelian anomaly coefficients are positive. In these cases, again
the spectrum of $(2, \pm 1)$ type matter is determined from the sign of the
$b_{i j}$ (or in models with $T > 0$ by the signs of $b_{i j} \cdot (a + b_{i
i} / 2)$).  The spectrum of $(1, \pm 1, \pm 1)$ type fields is, however, not
determined uniquely from the signs of the anomaly coefficients in these cases.
Roughly speaking, the choice of $(1, \pm 1, \pm 1)$ type fields follows
whichever non-frustrated configuration is ``closest'' to the given frustrated
configuration.  For example, with anomaly coefficients $b_{i i} = 8, b_{1 2} =
-2, b_{1 3} = -2, b_{2 3} = 1$, we get the canonical set of $(1, \pm 1, \pm
1)$ fields, while with $b_{i i} = 8, b_{1 2} = -2, b_{1 3} = -1, b_{2 3} = 2$,
which have the same signs, we get the set of generic matter fields $(1, 1, 1),
(1, -1, -1), (1, -1, 1)$ associated with positive choices of $b_{1 3}, b_{2
3}$ (which can be related to the canonical set of generic matter fields by
flipping the sign on $q_3$).

%We should also emphasize that in some cases, valid solutions to the AC
%equations exist that cannot be exchanged to solutions with only generic matter
%having positive multiplicities.\andrew{Maybe at least mention this in the
%statement before, as in footnote?} For example, the AC equations for $G =
%\SU(2)$ and $T = 0$ are solved by the model
%\begin{equation}
%102 \times \bm{1} + 2 \times \yng{1} + 54 \times \yng{2} + 2 \times \yng{3}
%\end{equation}
%with $b_{\SU(2)} = 13$, which does not have enough hypermultiplets charged
%under the fundamental to exchange away both of the hypermultiplets charged
%under the triple-symmetric representation. Further examples of such models are
%discussed in \cite{KleversMorrisonRaghuramTaylorExotic,TaylorTurnerU1}, and
%would include for example $\U(1)^2$ models as discussed above with both $(2,
%1)$ and $(2, -1)$ matter but insufficient $(0, 1)$ and $(1, 0)$ multiplets to
%exchange away the non-generic combination $(2, 1) + (2, -1)$ through
%\cref{eq:21-exchange}.  We treat solutions of this type as non-generic, even
%though their equivalent formulation in terms of generic matter would have
%negative multiplicities.

To conclude this section, we consider how the anomaly-equivalent
representations $\left(\syng{1}, \syng{1}, \bm{1}\right)$ and $\left(\syng{1},
\bar{\syng{1}}, \bm{1}\right)$ of $\SU(3)^3 \subset G_{(3)}$ behave under
Higgsing to $\U(1) \times \U(1)$.  As discussed above, with the embedding of
\cref{tab:U1-3embed}, the fundamental--fundamental matter field gives rise to
only the canonical choices of generic matter fields. On the other hand, the
fundamental--antifundamental field gives rise to charges including $(-1, -1,
2)$. (Note that such a charge can also arise from fundamental--fundamental
matter fields when the signs on the embedding are chosen differently). While
this may seem to suggest that $(2, -1, -1)$ type matter fields may arise in
generic $\U(1)^3$ models, this is not the case.  Comparing the complete sets
of fields that arise from the fundamental--fundamental and
fundamental--antifundamental $\SU(3) \times \SU(3)$ representations, we find
that the resulting sets of $\U(1)^3$ fields have an anomaly equivalence
\begin{equation}
\label{eq:211-exchange}
\begin{aligned}
&(2, -1, -1) + (2, 0, 0) + (1, 1, -1) + (1, -1, 1) + (1, -1, 0)\\
&\quad + (1, 0, -1)+(0, 1, 1) + (0, 1, 0) + (0, 0, 1)\\
&\qquad\qquad\qquad\quad \longleftrightarrow \\
&\qquad\qquad\qquad\qquad\quad (2, 0, -1) + (2, -1, 0) +  (-1, 1, 1)
  + (1, 1, -1) +(1, 1, 0)\\
&\qquad\qquad\qquad\qquad\qquad
	+ (1, 0, 1) + 2 \times (0, 1, -1) +  (0, 0, 0) \,.
\end{aligned}
\end{equation}
This illustrates several things: first, it gives a simple understanding of the
anomaly equivalence from $(2, -1, -1)$ to canonical generic matter; second, it
is an example of how Higgsing matter that is anomaly-equivalent to generic
matter can give non-generic matter fields; third, it emphasizes the point that
while they are anomaly-equivalent, the representations $\left(\syng{1},
\syng{1}\right)$ and $\left(\syng{1}, \bar{\syng{1}}\right)$ of $\SU(N) \times
\SU(M)$ behave physically quite differently in many circumstances.

%%%%%%%%%%%%%%%%%%%%%%%%%%%%%%%%%%%%%%%%%%%%%%%%%%%%%%%%%%%%%%%%%%%%%%%%%%%%%%
%%%%%%%%%%%%%%%%%%%%%%%%%%%%%%%%%%%%%%%%%%%%%%%%%%%%%%%%%%%%%%%%%%%%%%%%%%%%%%
\subsection{Generic matter with more than three $\U(1)$ factors}
\label{sec:further-u1-4}

For gauge groups with more than three $\U(1)$ factors, there are still further
ambiguities in what defines generic matter.  For a gauge group $\U(1)^s$ with
$s > 3$, in addition to the generic matter charge types $(0), (1), (2), (1,
\pm 1), (2, \pm 1)$ and $(1, \pm 1, \pm 1)$ that can arise with a gauge group
of $\U(1)^3$, among generic matter charge types we must also include fields
with charges $(1, \pm 1, \pm 1, \pm 1)$ for any subset of four of the $\U(1)$
factors. Based on the structure of $\U(1)^3$ generic matter and on counting of
anomaly cancellation conditions, in any given generic $\U(1)^s$ model we might
expect that for each subset of four $\U(1)$ factors only one of the possible
eight of these charge types will appear for any given set of anomaly
coefficients. This turns out not to be the case, however, as we now describe
in further detail.

From the point of view of counting anomaly constraints, we see that
\cref{eq:abelACquar} gives an additional $\binom{s}{4}$ constraints beyond
those considered in the analysis for up to three $\U(1)$ factors, so including
one new charge type for each set of four $\U(1)$ factors would give a set of
generic matter fields that matches the number of AC equations, as we might
desire for the subset of generic matter types realized in any given generic
model. We can, however, identify anomaly equivalences that exchange any field
of charge type $(1, \pm 1, \pm 1, \pm 1)$ for any other field of another such
charge type under the same set of four $\U(1)$ factors without changing the
number of uncharged scalars, assuming sufficient quantities of all other
generic matter fields.  Restricting attention to the simplest case of
$\U(1)^4$, we have for example the anomaly equivalence
\begin{equation}
\begin{aligned}
&(1, -1, 1, -1) + (-1, 1, 1, 0) + (1, 0, -1, 1) + (1, -1, 0, 1) +
	(0, 1, 1, -1)  \\
&\quad + (1, 0, 0, -1) + (0, 1, -1, 0) + (0, 1, 0, 1) + (1, 0, 1, 0) \\
&\qquad\qquad \longleftrightarrow \\
&\qquad\qquad\qquad (1, -1, -1, 1) + (1, 0, 1, -1) + (1, -1, 1, 0) + (0,
  1, -1, 1) + (-1, 1, 0, 1)\\
&\qquad\qquad\qquad\quad
	+ (0, 1, 0, -1) + (1, 0, -1, 0) + (1, 0, 0, 1) + (0, 1, 1, 0)\,.
\end{aligned}
\label{eq:4-equivalence}
\end{equation}
This anomaly equivalence suggests that there can be multiple
anomaly-equivalent models
%, with the same anomaly coefficients
with different subsets of the generic matter fields; furthermore, we may
expect to see ``generic models'' of maximal moduli space dimension that
contain more distinct matter charges than the appropriate number of AC
equations.  In fact, this seems to indeed be possible, and can be understood
in relation to the anomaly equivalence of the fundamental--fundamental and
fundamental--antifundamental fields for an $\SU(N) \times \SU(M)$ gauge group.

To explore this further, we can take an embedding of $\U(1)^4$ into $G_{(4)} =
\SU(2)^4 \times \SU(3)^6$, analogous to that given in \cref{tab:U1-3embed} for
$\U(1)^3$. The charges resulting from the Higgsing of such a model with
generic matter for the nonabelian gauge group are of the forms $(0)$, $(1)$,
$(2)$, $(1, 1)$, $(1, -1)$, $(2, -1)$, $(1, 1, -1)$, $(1, 1, -1, -1)$. (Note
that other choices of embedding, or including fundamental--antifundamental
fields in the nonabelian theory can give rise to charges type $(1, \pm 1, \pm
1, \pm 1)$ with different sign choices.  Other embedding choices can also give
rise to charges of type $(2, \pm 1, \pm 1)$, but these can be removed in favor
of the generic matter types using the anomaly equivalence
\labelcref{eq:211-exchange}.) While this Higgsing construction gives rise to a
set of $\U(1)^4$ models with only generic matter types, it generally will give
more than one set of $(1, 1, -1, -1)$ type charges.  While we can exchange one
such charge for another through \cref{eq:4-equivalence}, this does not change
the number of uncharged scalar fields, so this construction  gives models with
more distinct generic matter charge types than expected from the number of AC
equations, although these models should be generic in the sense of maximizing
moduli space dimension.  Indeed, from the point of view of this Higgsing
process, the equivalence \labelcref{eq:4-equivalence} can be associated with
the Higgsed version of the anomaly equivalence between
fundamental--fundamental and fundamental--antifundamental $\SU(3) \times
\SU(3)$ representations. This can be further understood, in particular, by
noting that the LHS and RHS of the anomaly equivalence are related by simply
flipping the signs on the last two charges. Thus, we see that at four (or
more) $\U(1)$ factors, just in the same way as for multiple nonabelian
factors, the number of distinct types of generic matter fields realized in any
given generic model can exceed the number of anomaly cancellation equations.

Combining this analysis with our understanding of generic matter with three or
fewer nonzero abelian charges, we can summarize the situation for the generic
matter spectrum of a general $\U(1)^s$ theory: the bi-charged matter under any
pair of $\U(1)$ factors will be determined just as in the $\U(1)^2$ and
$\U(1)^3$ cases from the anomaly coefficients, with in particular the types of
$(2, \pm 1)$ matter being determined by the signs of the terms $b_{i j} \cdot
(a + b_{i i} / 2)$. Similar to the analysis of $\U(1)^3$ models in the
previous subsection, we can consider each $\U(1)^3$ subgroup of $\U(1)^k$ to
be ``frustrated'' or ``non-frustrated'' depending on the signs of the relevant
$b_{i j}$ anomaly coefficients, and the spectrum of $(1, \pm 1, \pm 1)$ matter
will be correlated to these anomaly coefficients for each $\U(1)^3$ subgroup
as discussed above.  For 4-charged matter of the types $(1, \pm 1, \pm 1, \pm
1)$, however, we may generally expect to find models with different
combinations of matter charges even for fixed anomaly coefficients; such
models will be related by anomaly equivalences such as
\cref{eq:4-equivalence}. While for certain combinations of anomaly
coefficients we expect only certain types of $(1, \pm 1, \pm 1, \pm 1)$ to be
possible in generic models, we leave a systematic analysis of the constraints
and possibilities for further work.

%%%%%%%%%%%%%%%%%%%%%%%%%%%%%%%%%%%%%%%%%%%%%%%%%%%%%%%%%%%%%%%%%%%%%%%%%%%%%%
%%%%%%%%%%%%%%%%%%%%%%%%%%%%%%%%%%%%%%%%%%%%%%%%%%%%%%%%%%%%%%%%%%%%%%%%%%%%%%
\subsection{Other nonabelian factors}\label{sec:other-nonabel}

The analysis of other nonabelian factors is closely parallel to that of
$\SU(N)$. While we have not carried out this analysis at the same level of
detail as with $\SU(N)$, we can enumerate the types of representations we
expect to find as generic for the other compact simple Lie groups. Namely, for
$\Sp(N)$ we expect to find fundamentals, adjoints, and antisymmetric
representations, as for $\SU(N)$, while for $\SO(N)$ we expect to find
fundamentals, spinors, and adjoints. The exceptional groups have no quartic
Casimir, and so behave like $\SU(2)$ and $\SU(3)$; thus, we expect only
fundamentals and adjoints to appear generic for these groups. (For the
exceptional groups the relevant fundamental representations are the
smallest-dimensional nontrivial representations, i.e., the $\bm{7}$ of $\gG_2$,
$\bm{27}$ of $\gE_6$, etc.). Note that for $\gE_8$, there are not distinct
fundamental and adjoint representations, and so there appears to be a counting
mismatch with the AC conditions; however, we do not expect to find matter in
6D supergravity theories with gauge group $\gE_8$, and so this is not an
issue.

We expect for each of these nonabelian gauge factors that the representations
listed play the same role as the analogous representations for $\SU(N)$, with
in particular matter charged under multiple fields including a given
nonabelian factor involving the fundamental in the same way as for $\SU(N)$.
Thus, for example, for $\gG_2 \times \SU(2)$, generic matter includes
bi-charged matter in the $(\bm{7}, \bm{2})$ representation (which is
self-conjugate, and so can appear in half-hypermultiplets).

%%%%%%%%%%%%%%%%%%%%%%%%%%%%%%%%%%%%%%%%%%%%%%%%%%%%%%%%%%%%%%%%%%%%%%%%%%%%%%
%%%%%%%%%%%%%%%%%%%%%%%%%%%%%%%%%%%%%%%%%%%%%%%%%%%%%%%%%%%%%%%%%%%%%%%%%%%%%%
%%%%%%%%%%%%%%%%%%%%%%%%%%%%%%%%%%%%%%%%%%%%%%%%%%%%%%%%%%%%%%%%%%%%%%%%%%%%%%
\section{Generic matter in F-theory and the string swampland}
\label{sec:generic-f-theory}

From the point of view of F-theory, the generic matter representations
described in \cref{sec:generic-u1-3} arise naturally through the simplest
Weierstrass model constructions of both nonabelian and abelian gauge groups.
Other ``exotic'' matter representations require more complicated constructions
that involve more singular geometries and tuning additional moduli.  The
framework of F-theory also provides explicit constructions of ``matter
transitions,'' which exchange one kind of matter fields for another
anomaly-equivalent set of matter fields without changing the gauge group. We
summarize here how the models constructed in F-theory fit into the framework
of generic matter defined earlier.  In \cref{sec:generic-f-theory-sun}, we
describe F-theory models with $\SU(N)$ gauge factors only; F-theory models
with other nonabelian gauge factors besides $\SU(N)$ can be constructed in a
similar fashion to the simplest $\SU(N)$ models (using Tate tunings), and
similarly give rise to generic matter fields. In
\cref{sec:generic-f-theory-u1}, we describe F-theory models with one or more
$\U(1)$ factors.

F-theory gives rise to most 6D supergravity theories that can be realized in
any version of string theory.\footnote{Some exceptions include models that can
be realized in the ``frozen'' phase of F-theory \cite{BhardwajEtAlFrozen}, or
which include exotic matter that cannot be realized in conventional F-theory
\cite{LudelingRuehleDuals,CveticHeckmanLinExoticDiscrete}.} In
\cref{sec:further-swampland}, we describe how generic matter provides a
helpful paradigm with which to explore the ``swampland'' of 6D models that
appear acceptable from anomaly cancellation and other known quantum
consistency conditions, but which are not realized in F-theory.

%%%%%%%%%%%%%%%%%%%%%%%%%%%%%%%%%%%%%%%%%%%%%%%%%%%%%%%%%%%%%%%%%%%%%%%%%%%%%%
%%%%%%%%%%%%%%%%%%%%%%%%%%%%%%%%%%%%%%%%%%%%%%%%%%%%%%%%%%%%%%%%%%%%%%%%%%%%%%
\subsection{Tate and Weierstrass models for $\SU(N)$ gauge groups}
\label{sec:generic-f-theory-sun}

We review briefly the basic elements of F-theory that are involved in
constructing a 6D theory with a gauge group $\SU(N)$.  For a more extensive
background on F-theory see the original papers
\cite{VafaF-theory,MorrisonVafaI,MorrisonVafaII} or the reviews
\cite{TaylorTASI,WeigandTASI}.

An F-theory model of a 6D supergravity theory is described by an elliptically
fibered Calabi--Yau threefold encoded in a Weierstrass model
\begin{equation}
y^2 = x^3 + f x + g\,,
\end{equation}
where $f, g$ are functions (or more properly sections of certain line bundles)
on a complex surface base $B$. The geometry of the Calabi--Yau threefold
encodes the physics of the corresponding 6D theory; the number of tensor
multiplets is given by $T = h^{1, 1}(B) - 1$, the 6D string charge lattice is
given by $\Gamma = H^{1, 1}(B, \Z)$, and the anomaly coefficient $a$ is the
element in $\Gamma$ associated with the canonical class $K_B$ of the base.
The gauge group and matter fields are encoded in singularities of the elliptic
fibration.

The F-theory model gives an $\SU(N)$ gauge group over a divisor (complex
codimension one locus) $\sigma = 0$ when the discriminant
\begin{equation}
\Delta = 4 f^3 + 27 g^2
\end{equation}
vanishes to order $N$ over $\sigma$, while either $f$ or $g$ does not vanish
over $\sigma$.  In general, this condition requires a precise cancellation of
the terms in the discriminant at each order in $\sigma$.  The simplest way
that these cancellations can be imposed is through a ``Tate form'' model
\cite{BershadskyEtAlSingularities,KatzEtAlTate}
\begin{equation}
\label{eq:Tate}
y^2 + a_1 x y z + a_3 y z^3 = x^3 + a_2 x^2 z^2 + a_4 x z^4 + a_6 z^6\,.
\end{equation}
When \cref{eq:Tate} is converted into Weierstrass form by completing the
square and clearing the quadratic term in $x$ (and setting $z = 1$), an
$\SU(N)$ singularity is realized over the divisor $\sigma = 0$ when the $a_n$
coefficients vanish to the orders $[a_1, a_2, a_3, a_4, a_6] = \left[0, 1,
\floor{N / 2}, \floor{(N + 1) / 2}, N\right]$.

This simple Tate form \labelcref{eq:Tate} for $\SU(N)$ gives rise to a 6D
theory where the $\SU(N)$ anomaly coefficient is the class in $\Gamma = H^{1,
1}(B, \Z)$ associated with the divisor $\sigma$ ($b = [\sigma]$), and $\SU(N)$
matter fields transform under the fundamental, adjoint, and two-index
antisymmetric matter representations---precisely the generic matter types
listed in \cref{tab:generic}.  This can be seen geometrically in the F-theory
picture from the fact that adjoint representations arise from the genus of a
smooth divisor, and the fundamental and two-index antisymmetric matter
representations arise from simple codimension-two singularities associated
with points on $\sigma$ where $\Delta$ vanishes to orders higher than $N$
through a simple singularity enhancement described using the Kodaira
classification as $A_{N - 1} \rightarrow A_N, A_{N - 1} \rightarrow D_N$,
respectively. The Tate form construction of $\Sp(N)$ and $\SO(N)$ gauge groups
proceeds in a similar fashion, and for exceptional groups the Tate
construction is even simpler, involving only conditions on the orders of
vanishing of $f$ and $g$.

Any other kind of matter representation for $\SU(N)$ (or any other nonabelian
gauge factor) in F-theory requires some additional more special tuning of the
Weierstrass coefficients $f, g$.  There is only one case in which it is known
that this can be achieved through a Tate-type tuning: it was shown in
\cite{HuangTaylorLargeHodge} that a special Tate tuning of $\SU(6)$ can be
realized when the $a_n$ coefficients vanish to orders $[0, 2, 2, 4, 6]$
(instead of the usual Tate $\SU(6)$ tuning $[0, 1, 3, 3, 6]$).  In this case,
at the locus $a_1 = \Delta = 0$ there arise higher singularities associated
with the three-index $\bm{20}$ representation of $\SU(6)$.  Since $a_4$
generally has at least as many degrees of freedom at order $\sigma^3$ as $a_3$
does at order $\sigma^2$, and $a_2$ generally also has degrees of freedom at
order $\sigma$, it turns out that this exotic Tate tuning always reduces the
number of degrees of freedom in the model, in accord with the principle that
exotic matter always reduces the number of degrees of freedom (uncharged
matter hypermultiplets) in the theory.

A systematic approach to tuning $\SU(N)$ Weierstrass models by arranging an
order-by-order cancellation in the terms of the expansion of $\Delta$ in terms
of expansions of $f, g$ as power series in $\sigma$ was carried out in
\cite{MorrisonTaylorMaS}.  In that paper, the assumption was made that
$\sigma$ is a smooth divisor, and that the ring of functions on $\sigma$ is a
Unique Factorization Domain (UFD). From this point of view, the three-index
antisymmetric tensor representations of not only $\SU(6)$ but also $\SU(7)$
and $\SU(8)$ can be realized explicitly as tuned Weierstrass models.  In each
case, it can be seen explicitly that more degrees of freedom must be tuned
away (associated with a decrease in the number of uncharged hypermultiplets)
to realize the three-index antisymmetric tensor representations.

A further study of exotic matter was carried out in
\cite{KleversMorrisonRaghuramTaylorExotic}.  In that paper it was shown
explicitly how the two-index symmetric representation of $\SU(N)$ and the
three-index symmetric representation of $\SU(2)$ can be realized through a
non-UFD tuning of $\SU(N)$ on a singular divisor $\sigma$. Again, in these
situations, tuning the curve $\sigma$ so that it is singular and the spectrum
includes exotic matter representations removes degrees of freedom encoded in
uncharged hypermultiplets. In that paper, it is also argued that these are the
only exotic representations of $\SU(N)$ that can be realized in conventional
F-theory. Furthermore, for other gauge factors including in particular the
exceptional gauge algebras, the same argument shows that no other non-generic
matter parameterization should be possible in conventional F-theory other than
for $\SU(N)$ and closely related $\Sp(N)$ cases.

Note that, as discussed in \cite{KleversMorrisonRaghuramTaylorExotic},
although exotic matter representations can generally only be achieved by
fine-tuning from a model with only generic matter, there are cases in which
models with generic matter are ruled out by anomaly cancellation
considerations that would imply that some matter representations have negative
multiplicity, as mentioned in \cref{sec:generic-exchanges}. In such
situations, there can be F-theory models that realize a given gauge group on a
divisor $\sigma$ of a certain class with exotic matter, where no generic
matter model is possible.  In general, this occurs precisely where the anomaly
coefficient $b$ becomes too large for generic matter to be possible with
non-negative multiplicities.

As an example of how F-theory realizes generic $\SU(N)$ matter through Tate
tunings at small $b$, we consider the simplest case of the base $\bP^2$,
corresponding to $T = 0$.  In this case, homology classes in $H^{1, 1}(\bP^2)$
are given by integer multiples of the generating class $H$, so anomaly
coefficients are simply denoted by integers. The anomaly coefficient $-a = 3$
corresponds to the anticanonical class $-K = 3 H$.  A Tate tuning of $\SU(2)$
is possible up to $b = 12$, giving the generic matter $\SU(2)$ models listed
in \cref{eq:su2-12-models}.  For $b > 12$, any F-theory model must be
described by a more general class of Weierstrass model that does not take the
Tate form; in particular, as described in
\cite{KleversMorrisonRaghuramTaylorExotic}, the divisor $\sigma$ must have a
triple-point singularity for each (half-hyper) matter multiplet in the
$\syng{3}$ representation.  While there is no completely systematic
construction of such models, some F-theory models of this type with $b > 12$
are explicitly constructed in \cite{KleversTaylor3Sym}; for example (see
Table~3.22 in that paper) there is a model with $b = 13$ that has 6
triple-symmetric matter fields, associated with 12 triple-point singularities
in the locus $\sigma = 0$.  The F-theory construction of such exotic matter
models is discussed further in \cref{sec:further-swampland}.

For $\SU(N_1) \times \dots \times \SU(N_r)$ theories, standard F-theory
constructions using the Tate construction for each factor yield the generic
representations of the form $\bm{1}$, $\syng{1}$, $\Adj$, $\syng{1,1}$,
$\left(\syng{1}, \bar{\syng{1}}\right)$, and $\left(\syng{1},
\syng{1}\right)$. Some more exotic representations, such as $\left(\syng{1,1},
\syng{1}\right)$, can also be constructed with a reduction in the number of
uncharged scalars, though the Weierstrass models for such exotic
multiply-charged have not been systematically described.

It is interesting to consider how the anomaly-equivalent generic matter
representations $\left(\syng{1}, \bar{\syng{1}}\right)$ and $\left(\syng{1},
\syng{1}\right)$ arise in simple F-theory models.  While determining which of
these representations arises at a given singularity involves somewhat subtle
global aspects of a given Weierstrass model \cite{MorrisonTaylorMaS}, it is
straightforward to see that many general Tate models for $\SU(N) \times
\SU(M)$ will include both types of representations, using a simple Higgsing
argument.  Consider, for simplicity, models with no tensor multiplets,
corresponding to F-theory on $\bP^2$.  We can perform a Tate tuning of
$\SU(N)$ on a smooth curve of degree $d$, which has self-intersection $n =
d^2$ and genus $g = (d - 1) (d - 2) / 2$.  The resulting $\SU(N)$ model has
generic matter content
\begin{equation}
g \times {\Adj} + [16 (1 - g) + (8 - N) n] \times \syng{1}
	+ (n + 2 - 2 g) \times \syng{1,1}\,.
\end{equation}
We can break $\SU(N) \rightarrow \SU(N - k) \times \SU(k)$ by Higgsing on a
pair of antisymmetric matter fields.  The resulting model has the spectrum we
expect from a Tate tuning of the product gauge group on a pair of distinct
divisors of degree $d$.  Under this Higgsing process, the decomposition of the
$\syng{1,1}$ representations of $\SU(N)$ includes bifundamental
representations $\left(\syng{1}, \syng{1}\right)$, while the decomposition of
the adjoints of $\SU(N)$ includes ($2 \times$) fundamental--antifundamental
matter representations $\left(\syng{1}, \bar{\syng{1}}\right)$. Thus, the
resulting $\SU(N - k) \times \SU(k)$ model contains, among other fields, the
matter representations
\begin{equation}
2 g \times \left(\syng{1}, \bar{\syng{1}}\right)
	+ (d^2 - 2 g) \left(\syng{1}, \syng{1}\right)\,.
\end{equation}
As expected from anomaly cancellation, the total number of fields in these two
representations sums to $d^2$, which is the number of points where the two
degree $d$ curves intersect; however, the spectrum includes $d^2 - 3 d + 2$
fundamental--antifundamental matter fields and $3 d - 2$ bifundamental fields.
This is a simple example of how F-theory generically gives rise to a mixture
of the different anomaly-equivalent generic matter representations in cases
where there are more generic matter fields than anomaly cancellation
equations.

%%%%%%%%%%%%%%%%%%%%%%%%%%%%%%%%%%%%%%%%%%%%%%%%%%%%%%%%%%%%%%%%%%%%%%%%%%%%%%
%%%%%%%%%%%%%%%%%%%%%%%%%%%%%%%%%%%%%%%%%%%%%%%%%%%%%%%%%%%%%%%%%%%%%%%%%%%%%%
\subsection{$\U(1)$ matter in F-theory}
\label{sec:generic-f-theory-u1}

F-theory constructions of 6D models with abelian gauge groups are not well
understood beyond the simplest models with a single $\U(1)$ factor and the
generic $q = 1, 2$ matter representations.  We summarize briefly here the
situation of what is known. The generic matter analysis for $\U(1)^2$,
$\U(1)^3$, and $\U(1)^s, s > 3$ models carried out above gives new insights
into the structure of F-theory constructions in these cases.

%%%%%%%%%%%%%%%%%%%%%%%%%%%%%%%%%%%%%%%%%%%%%%%%%%%%%%%%%%%%%%%%%%%%%%%%%%%%%%
\subsubsection{One $\U(1)$ factor}\label{sec:one-abelian-f}

A general construction of an F-theory model over a given base $B$ with a
single $\U(1)$ factor was given by Morrison and Park in \cite{MorrisonParkU1}.
In the Morrison--Park model, the only matter fields that arise are the generic
charges $q = 1, 2$. Thus, again, in this case the simplest F-theory
construction gives a model with only generic charges.

There are only a limited set of explicit Weierstrass model constructions known
that give rise to $\U(1)$ theories with charges $q > 2$. A specific set of
models with $q = 3$ was first identified in \cite{KleversEtAlToric}.  This set
was generalized to a more general class of models with $q = 3$ charges and
related to non-UFD singularities in \cite{Raghuram34}; in this paper, Raghuram
also identified a limited class of models with charge $q = 4$.  At this time,
there are no explicit Weierstrass model constructions known for any $\U(1)$
charges $q > 4$, although general arguments from Higgsing known nonabelian
models suggest that there should be F-theory models with $\U(1)$ charges $q =
5, 6$ that arise from Higgsing simple Tate $\SU(N)$ models over the base $B
=\bP^2$ \cite{TaylorTurnerU1} and F-theory models with $\U(1)$ charges up to
$q = 21$ at least that arise from Higgsing more exotic nonabelian models on
other bases \cite{RaghuramTaylor21}. A recent paper \cite{CianciPenaValandro}
gives an alternative construction of F-theory models with charge $q = 4$
matter through a weak coupling limit of type IIB string theory, and identifies
IIB models that should correspond to charge $q = 5, 6$ matter.

%%%%%%%%%%%%%%%%%%%%%%%%%%%%%%%%%%%%%%%%%%%%%%%%%%%%%%%%%%%%%%%%%%%%%%%%%%%%%%
\subsubsection{Two $\U(1)$ factors}\label{sec:two-abelian-f}

For models with two $\U(1)$ factors ($G = \U(1) \times \U(1)$), a rather
general construction was put forth in \cite{CveticKleversPiraguaTaylorU1U1},
following some more specific constructions with more constrained matter
content \cite{BorchmannEtAllSU5U1U1,CveticKleversPiraguaU1s,CveticEtAlSU5U1s,
BorchmannEtAlTops}. The matter content of the general $\U(1)^2$ construction
matches well with the generic $\U(1) \times \U(1)$ matter content described
above, and can be understood as coming from the Higgsing of a nonabelian model
with gauge group $G_{(2)} = \SU(2) \times \SU(2) \times \SU(3)$. Here, the
$\U(1)$ factors are associated with divisors $A C, B C$, where the nonabelian
factors are supported on $A, B, C$; this leads to a natural embedding of the
$\U(1)$ factors in the Cartan subalgebra of the nonabelian group as in
\cref{eq:2-embedding}. The explicit analysis of the $\U(1) \times \U(1)$
Weierstrass model in \cite{CveticKleversPiraguaTaylorU1U1} was shown to give
the generic matter spectrum described in \cref{sec:abelian-2} with charges
$(2, 1)$ and $(1, 2)$. This setup should also allow for the construction of
models with generic matter charges $(2, -1)$ and $(1, -2)$, with different
choices of divisors associated with the anomaly coefficients $b_{i j}$, though
that is not addressed directly in the paper; we leave an investigation of the
details of this analysis for further work. One interesting feature of the
construction in \cite{CveticKleversPiraguaTaylorU1U1} is that the most general
model there automatically includes non-generic charge $(2, 2)$ matter.  A
slightly more restricted version of the model gives only the generic matter
charges in \cref{tab:generic}.

%%%%%%%%%%%%%%%%%%%%%%%%%%%%%%%%%%%%%%%%%%%%%%%%%%%%%%%%%%%%%%%%%%%%%%%%%%%%%%
\subsubsection{Three or more $\U(1)$ factors}\label{sec:three-abelian-f}

At this point there is very little understanding of explicit F-theory
Weierstrass models for theories with more than two $\U(1)$ factors. In
\cite{CveticEtAlU13}, a model was given with three $\U(1)$ factors, but it
provides only a very limited class of spectra and does not contain all
expected generic matter fields or generic models. In \cite{KimuraK3s},
explicit Weierstrass models were given for K3 surfaces of Mordell--Weil rank
up to 4, but the matter spectra of F-theory compactifications on these
surfaces was not explored. There are also some explicit 6D models known with
Mordell--Weil rank up to 8 and no matter charged under the $\U(1)$ factors
(i.e., non-Higgsable $\U(1)$ factors)
\cite{MartiniTaylorSemitoric,MorrisonParkTaylorNHAbelian}. These models are
rather special. We can, however, from the analysis of generic matter, gain
some insight into the structure one might expect for a general F-theory
construction of generic $\U(1)^s$ models.

Following the structure of the general $\U(1)^2$ F-theory model, it was
conjectured in \cite{CveticKleversPiraguaTaylorU1U1} that a general class of
$\U(1)^s$ models could be constructed from the Higgsing of a model with gauge
group
\begin{equation}
G = \SU(2)^s \times \SU(3)^{\binom{s}{2}} \times \SU(4)^{\binom{s}{3}}
	\times \dots \times \SU(s)^s \times \SU(s + 1)\,,
\end{equation}
in which each $\U(1)$ has a component in one of the $\SU(2)$ factors, each
pair of $\U(1)$s has a common component in one of the $\SU(3)$ factors, and so
on. From the analysis of generic matter in $\U(1)^s$ models above, however,
and the structure of the anomaly equations, it seems that a broad class of
generic matter models can be realized from Higgsing nonabelian models with the
simpler gauge group
\begin{equation}
G_{(s)} = \SU(2)^s \times \SU(3)^{\binom{s}{2}}\,.
\end{equation}
Here, there is an $\SU(2)$ factor associated with each $\U(1)$ factor, and an
$\SU(3)$ factor associated with each pair of $\U(1)$ factors. We can always
construct an embedding of $\U(1)^s$ into this gauge group analogous to the
embedding of $\U(1)^3$ into $G_{(3)} = \SU(2)^3 \times \SU(3)^3$ given in
\cref{tab:U1-3embed}.  Higgsing a $G_{(s)}$ model with generic matter to
$\U(1)^s$ using such an embedding gives a spectrum with charges of the generic
matter types described in the previous section.

While it is straightforward to construct an F-theory model with generic matter
and gauge group $G_{(s)}$ with any desired (relatively small) anomaly
coefficients using the Tate construction, unfortunately it is not clear given
such a Weierstrass model how to perform a Higgsing.  The existence of the
nonabelian model guarantees that there should be a Weierstrass model for the
resulting Higgsed $\U(1)^s$ model, which with an appropriate embedding will
have generic matter fields as described in
\cref{sec:abelian-3,sec:further-u1-4}.  Unfortunately we do not know how to
compute this model explicitly.  One approach would be to simultaneously tune
Morrison--Park type $\U(1)$ factors on the relevant combinations of $\SU(2)$
and $\SU(3)$ divisors, but it is not clear even in the case of $\U(1)^2$ how
to derive the general model of \cite{CveticKleversPiraguaTaylorU1U1} from this
point of view. Indeed, understanding high-rank abelian gauge groups associated
with large rank Mordell--Weil groups in the F-theory geometry is an open
problem that deserves further attention; hopefully the structure of generic
matter analyzed here may help provide some insights into these problems.

%%%%%%%%%%%%%%%%%%%%%%%%%%%%%%%%%%%%%%%%%%%%%%%%%%%%%%%%%%%%%%%%%%%%%%%%%%%%%%
%%%%%%%%%%%%%%%%%%%%%%%%%%%%%%%%%%%%%%%%%%%%%%%%%%%%%%%%%%%%%%%%%%%%%%%%%%%%%%
\subsection{Generic matter, string universality, and the swampland}
\label{sec:further-swampland}

The concept of ``generic'' matter representations in 6D supergravity theories
is a useful tool in analyzing the relationship between low-energy constraints
on supergravity theories and UV complete models that come from F-theory or
other string compactifications.  It was conjectured in
\cite{KumarTaylorStringUni6D} that ``string universality'' may hold for 6D
$\cN = (1, 0)$ supergravity theories, meaning that every massless 6D spectrum
of tensor fields, gauge fields, and matter representations that can be
consistently coupled to quantum gravity is realized in string theory.
Currently, there is still a significant gap in understanding whether this is
true; in particular, even for theories with no tensor multiplets there is an
infinite family of $\U(1)$ charge spectra that are consistent with anomalies
and other known low-energy consistency conditions but are not realized in
F-theory \cite{TaylorTurnerU1}. The set of theories that satisfy known quantum
consistency conditions but that are not realized by known string theory
constructions has been dubbed the ``swampland''
\cite{VafaSwamp,OoguriVafaSwamp}.\footnote{Note that different authors use the
term swampland differently.  We refer to the swampland as containing any model
that does not violate known quantum consistency conditions but does not have a
known explicit string construction.  Thus, the swampland by this definition is
time-dependent and can shrink as new constraints and new string constructions
are identified.} At least for theories with generic matter types, it seems
that some simple assumptions about the structure of the low-energy theory
suffice to limit the swampland almost completely.  This essentially focuses
general questions about string universality and the swampland to more specific
questions about exotic matter and some subtle issues related to the positivity
cone of the low-energy theory.  We summarize here briefly the general setup
and explain how the swampland is substantially reduced for theories with
generic matter.

%%%%%%%%%%%%%%%%%%%%%%%%%%%%%%%%%%%%%%%%%%%%%%%%%%%%%%%%%%%%%%%%%%%%%%%%%%%%%%
\subsubsection{Structure of 6D supergravity theories}
\label{sec:6D-structure}

Elaborating further on the basic framework described in \cref{sec:anomalies},
a 6D $\cN = (1, 0)$ supergravity theory is described by some basic data for
the low-energy theory.  Any such theory has one gravity (super)multiplet, $T$
tensor multiplets, a gauge group $G$ associated with $\dim(G)$ gauge
multiplets, and $H$ matter hypermultiplets that transform under various
representations of the group $G$.  In addition to this structure, there is a
signature-$(1, T)$ integer lattice $\Gamma$, often referred to as the
``anomaly lattice,'' which must be unimodular \cite{SeibergTaylorLattices}.
There is a gravitational anomaly coefficient $a \in \Gamma$, which satisfies
$a \cdot a = 9 - T$ by the gravitational anomaly condition. For each
nonabelian factor of the gauge group there is an anomaly coefficient $b_\kappa
\in \Gamma$, and for each pair of abelian factors there is an anomaly
coefficient $b_{i j} \in \Gamma$.  These anomaly coefficients satisfy the
anomaly constraints \labelcref{eq:nonabelAC,eq:abelAC}, which lead to various
further integrality constraints
\cite{KumarTaylorStringUni6D,MonnierMooreParkQuantization}, including the
condition that for a theory with $\U(1)$ gauge factors the anomaly
coefficients $b_{i i}$ are even, in the sense that $b_{i i} \in 2 \Gamma$.
Finally, there is a positivity cone in the anomaly lattice, corresponding to
the K\"ahler moduli space of the theory. All of these quantities have a direct
interpretation in the geometry of F-theory: the lattice $\Gamma$ is the
homology lattice of the base surface $B$, the positivity cone is the cone of
effective divisors, $a$ corresponds to the canonical class, $b$ corresponds to
the divisor class of the seven-branes supporting a given gauge factor, etc.

For each value of $T$, then, we can classify theories beginning with a choice
of $a$ and lattice $\Gamma$ with positivity cone.  It was shown in
\cite{MonnierMooreParkQuantization} (based on some minimal assumptions) that
$a$ is a characteristic vector for $\Gamma$, meaning that $a \cdot x + x \cdot
x \in 2 \Gamma$ for any $x \in \Gamma$.  For small values of $T$, there are
only a few possible choices of lattice and $a$, and the characteristic vector
condition suffices to limit the possibilities to those that can arise from
F-theory.  In particular, for $T = 0$ the lattice $\Gamma$ is uniquely defined
as $\Gamma = (1)$, and $-a = 3$ is the unique choice for $a$ (with the sign
fixed by the condition that $-a$ lies in the positivity cone, which by
convention we take to be the set of positive values in $\Gamma$).  For $T =
1$, there are two possible unimodular lattices: the odd lattice with inner
product
\begin{equation}
\Omega_1 = \mat[p]{1 & 0 \\ 0 & - 1}
\end{equation}
and the even lattice with inner product
\begin{equation}
\label{eq:even-lattice}
\Omega_0 = U = \mat[p]{0 & 1 \\ 1 & 0}\,.
\end{equation}
We denote the corresponding string charge lattices by $\Gamma_1, \Gamma_0$.
For $\Gamma_1$, the only possible choice for $a$ so that $a \cdot a = 9 - T =
8$ is $-a = -a_1 = (3, -1)$, up to symmetries. For $\Gamma_0$, there are two
possibilities: $-a = -a_0 = (2, 2)$ and $-a = -a_0' = (4, 1)$. The second
choice is, however, not a characteristic vector, so is not allowed in a
consistent low-energy theory. For $T = 0, 1$, these allowed combinations of
the anomaly lattice and $a$ are precisely those allowed by F-theory. In fact,
if we assume that the anomaly lattice for any higher value of $T$ has the
inner product $\diag(+1, -1, -1, \dots, -1)$, then it seems that the
characteristic vector condition uniquely determines the value $-a = (3, -1,
-1, \dots, -1)$.\footnote{Thanks to Noam Elkies for helpful discussions on
this point.}

%%%%%%%%%%%%%%%%%%%%%%%%%%%%%%%%%%%%%%%%%%%%%%%%%%%%%%%%%%%%%%%%%%%%%%%%%%%%%%
\subsubsection{Positivity cone and anomaly lattice related swampland issues}
\label{sec:cone-lattice}

There are several aspects of the anomaly lattice $\Gamma$ and the
positivity cone that are not fully understood from the low-energy
point of view.  In particular, it is not clear how the low-energy
theory constrains the positivity cone, and while in F-theory any
vector in the positivity cone that has an inner product with itself $x
\cdot x < -2$ must support a gauge group $\SU(3)$ or larger
\cite{MorrisonTaylorClusters}, this has not been proven purely from
consistency considerations arising from coupling the low-energy theory to
quantum gravity.  These issues lead to various families of swampland theories
that cannot be realized in F-theory but do not violate any proven quantum
consistency conditions on the low-energy theory.  It seems likely that these
parts of the swampland can be removed by considering consistency conditions on
the world-volume of strings that couple to the tensor fields
\cite{KumarTaylorStringUni6D} (see, e.g., \cite{LockhartVafaSCPart,
HaghighatEtAlM,GaddeEtAlChains,DelZottoEtAlTopStrings,DelZottoLockhartBPS} for
some recent progress in understanding these string world-volume theories).
There are thus various important open swampland questions related to the
anomaly lattice and positivity cone.  Here, however, we assume that the
lattice and positivity cone are of a form compatible with F-theory, and
consider various gauge groups and matter content in that context.

%%%%%%%%%%%%%%%%%%%%%%%%%%%%%%%%%%%%%%%%%%%%%%%%%%%%%%%%%%%%%%%%%%%%%%%%%%%%%%
\subsubsection{The swampland for generic matter}
\label{sec:generic-swampland}

Restricting attention to anomaly lattices $\Gamma$ and positivity cones
compatible with F-theory, and focusing only on theories with the matter fields
we have identified as generic, the only questions about theories in the
swampland arise from three issues:
\begin{enumerate}[label={(\alph*)}]
\item \label{item:largeB} large anomaly coefficients $b$,
\item \label{item:rep-ambiguities} ambiguities between
  fundamental--fundamental/fundamental--antifundamental matter for products of
  nonabelian factors and the related ambiguity associated with the anomaly
  equivalence \labelcref{eq:4-equivalence} for theories with more than three
  $\U(1)$ factors,
\item \label{item:lack} a lack of explicit Weierstrass models for generic
models with multiple abelian factors.
\end{enumerate}
In essentially all of these cases, the appearance of swampland models is
associated with our incomplete understanding of how to construct Weierstrass
models describing elliptic Calabi--Yau threefolds with sufficiently
complicated structure. We now briefly review how the known classes of F-theory
constructions described in
\cref{sec:generic-f-theory-sun,sec:generic-f-theory-u1} match with the set of
allowed low-energy theories with generic matter, and where the limitations of
these constructions lead us to the swampland.

Recall that for generic matter for any combination of nonabelian and abelian
gauge fields, we have found that the anomaly constraints uniquely identify the
set of allowed matter representations given the gauge fields and associated
anomaly vectors $b_\kappa, b_{i j}$, up to the ambiguities in
\cref{item:rep-ambiguities} above.  Thus, we can define an apparently
consistent low-energy theory by choosing a set of gauge group factors
$G_\kappa, \U(1)_i$ and associated values of $b_\kappa, b_{i j}$ that are
sufficiently small that the anomaly equations are satisfied by a subset of the
set of generic matter fields.  In cases with multiple nonabelian factors, or
more than three abelian factors, there may be a family of theories associated
with choosing different multiplicities of the anomaly-equivalent sets of
generic matter fields associated with \cref{item:rep-ambiguities} above. The
swampland question in this context is: for which models in this class does
there exist an explicit F-theory construction through a Weierstrass model, or
more indirectly an F-theory construction of a model with a larger gauge group
that can be Higgsed to the desired model.  (There are currently no
non-F-theory constructions of models in this class that cannot also be
realized by conventional F-theory.)

Many of the theories in this general class can be directly constructed using
the Tate form \labelcref{eq:Tate}, as long as the anomaly coefficients are not
too large.  While there is no direct known Weierstrass construction for
generic models with more than two $\U(1)$ factors, for small values of the
anomaly coefficients as described in \cref{sec:abelian-3,sec:further-u1-4} we
can realize models with more $\U(1)$ factors by Higgsing a Tate model with the
appropriate number of $\SU(2)$ and $\SU(3)$ factors. These approaches
essentially lead to F-theory models for any low-energy theory that does not
suffer from issues \labelcref{item:largeB,item:rep-ambiguities}, though for
multiple $\U(1)$ factors the details of the Higgsing construction and
associated embeddings have not been worked out in detail for all generic
matter combinations.  To illustrate the relevant issues more explicitly, we
focus first on models with a single gauge factor, where only issue
\labelcref{item:largeB} is relevant.

As the simplest example, as discussed in \cite{TaylorTurnerU1}, when the gauge
group is $\SU(2)$ and $T = 0$, there are anomaly-consistent theories for each
value $1 \le b \le 12$, and all of these theories can be directly constructed
as Weierstrass models using the Tate form, so there is no associated swampland
for models with only generic matter (except for one subtlety regarding the
case $b = 12$, which we return to below).  The situation is less clear at
larger values of $T$.  A Tate $\SU(2)$ construction is possible when $b \leq
-4 a$, meaning that $-4 a - b$ is in the positivity cone.  The condition for
generic matter, however, is that $b \cdot b \leq -4 a \cdot b$, which is a
weaker condition.  Thus, there are some cases that admit generic matter
solutions to the anomaly equations, for which a Tate construction is not
possible.

As an example, consider the case of the even lattice
\labelcref{eq:even-lattice}, with positivity cone as defined in
\cref{eq:even-case}; this corresponds to the class of F-theory models
compactified on the Hirzebruch surface $\F_0 = \bP^1 \times \bP^1$. In this
case, $-a = (2, 2)$.  The anomaly coefficient $b = (2, 9)$ (which lies in the
shaded but not the hatched region in \cref{fig:su2-t=1-even}) is too large for
a Tate $\SU(2)$ construction, but is compatible with a spectrum $x_{\,
\tyng{1}}= 104, x_{\, \tyng{2}} = 8$ (with 15 uncharged scalars needed to
saturate the gravitational anomaly bound).  This model is thus currently in
the swampland since it satisfies the known quantum consistency conditions (see
\cref{fig:su2-t=1-even}) but does not at this time have a known F-theory
construction.

While there is no Tate form Weierstrass model for this $\SU(2)$ theory on
$\F_0$, it is possible that the more general class of non-UFD constructions
analyzed in \cite{KleversMorrisonRaghuramTaylorExotic} can realize this model.
Specifically, this could occur if the genus $8$ curve describing the divisor
$\sigma$ in the class $b$ where the $\SU(2)$ ($A_1$) singularity is localized
is taken to be singular, so that some of the eight adjoints become localized;
in this case the non-UFD ring of functions on $\sigma$ can lead to nontrivial
cancellations giving an $\SU(2)$ Weierstrass model that is not in Tate form.
We leave a more explicit analysis of this and related situations for future
work. Note that the model with anomaly coefficient $b = (1, 9)$ is not
consistent as it violates the gravitational anomaly bound; since the
corresponding curve would be rational (genus $0$), the non-UFD construction
possibility is not an option in this case.

For theories with a single $\SU(N)$ factor more generally, similar issues can
arise giving possible swampland models when the combination $b N$ is too large
for a Tate realization. In general, a Tate realization of a theory with gauge
group $\SU(N)$ will be possible when $b N \le -8 a$. When this condition is
violated (but the Kodaira bound $ b N \le -12 a$ is still satisfied), there
may be Weierstrass constructions even when there is no Tate construction, as
in the above $\SU(2)$ example, but there is no general methodology known or
general condition for when such a Weierstrass model will exist. Another,
slightly more subtle issue can arise when tuning maximal even or odd rank
$\SU(N)$ factors through Tate \cite{KumarMorrisonTaylor6DSUGRA,
MorrisonTaylorMaS, JohnsonTaylorEnhanced}. The simplest example of this is
found when trying to construct an $\SU(23)$ or $\SU(24)$ group on a divisor $b
= 1$ in $\bP^2$. In this case, the Tate tuning of $\SU(23)$ automatically
forces a gauge group $\SU(24)$. So, the $T = 0, b = 1$ model with gauge group
$\SU(23)$ is in the swampland; it is not clear if there is a non-Tate F-theory
realization of such a model, but it seems unlikely since unlike in the
$\SU(2)$ case mentioned above the genus of the $b = 1$ curve is zero, so there
is no opportunity for non-UFD structure. Similar issues arise when tuning, for
example, an $\SU(15)$ on $\F_1$ or $\F_2$.

An even more subtle issue is relevant in the $\SU(24)$ case.  In this case,
the Tate tuning gives a model with gauge group $\SU(24) / \Z_2$. This model
has only matter in the two-index antisymmetric tensor representation, which is
invariant under the $\Z_2$ center.  In fact, this seems to be a general
pattern: in any model with no massless matter transforming under a central
component of the gauge group, the resulting F-theory model has a gauge group
that is quotiented by that central component. Since every known
F-theory model that we are aware of satisfies this condition, it is
natural to make the hypothesis that this is universally true for all
F-theory constructions.

For example, with $T = 0, b = 12$ the $\SU(2)$ model described earlier has
only $54$ adjoint fields and no fundamentals, and the gauge group is actually
$\SO(3) = \SU(2) / \Z_2$ \cite{MorrisonTaylorSections,TaylorTurnerU1}.  This
is reminiscent of the completeness hypothesis \cite{BanksSeibergSymm,
HarlowOoguriSymms} stating that any theory coupled to quantum gravity must
contain matter transforming in all nontrivial charges under the gauge group,
but this condition applies only to massless fields and is therefore much
stronger.  It would be interesting to understand whether this condition is
truly a universal constraint; for now, the models like those with a gauge
group $\SU(24)$ and only two-index antisymmetric matter lie in the swampland.

Turning to models with a only single $\U(1)$ factor, again, generic matter
content of charges $q = 1, 2$ is produced in many cases by the Morrison--Park
model \cite{MorrisonParkU1}.  As discussed in \cite{TaylorTurnerU1}, in the
simplest cases where $T = 0$, with the allowed values of $a$ and positivity
cones compatible with F-theory geometry, the $\U(1)$ anomaly constraints
precisely match those for the existence of a Morrison--Park model, and there
is no swampland (with the exception of the model with only charge $2$ matter
that arises from Higgsing the $\SU(2), b = 12$ model mentioned above, which
violates the general condition discussed there). For larger values of $T$, the
anomaly constraints are weaker but, as in the $\SU(2)$ case discussed above,
have a seemingly parallel form to the constraints for a Morrison--Park model
that would be interesting to understand better. An example of a swampland
model here would be the Higgsing of the $b = (2, 9), T = 1$ $\SU(2)$ model
mentioned above; presumably this model can be realized in F-theory if and only
if the corresponding $\SU(2)$ model can also be realized.

Now considering theories with a gauge group containing multiple
nonabelian factors, for a gauge group $G = \prod_{\kappa}
\SU(N_\kappa)$ in general a Tate realization will be possible if
 $\sum_\kappa b_\kappa N_\kappa \le -8 a$, and again there will be cases where
this condition is violated but the Kodaira bound $\sum_\kappa b_\kappa
N_\kappa \le -12 a$ is still satisfied where there may be exotic non-UFD or
other F-theory constructions. A further issue that arises here is that the
Tate construction will give some specific combination of
fundamental--fundamental and fundamental--antifundamental matter for each
product of nonabelian factors; there will be other anomaly-equivalent models
with different multiplicities after exchanging these equivalent
representations, and we do not currently have a general approach to analyzing
or constructing the sets of models with arbitrary distributions of these
anomaly-inequivalent representations.  This contributes to parts of the
swampland related to issue \labelcref{item:rep-ambiguities} above.

Finally, the F-theory construction of models with more than two $\U(1)$
factors is still not well understood (issue \labelcref{item:lack} above).  For
two $\U(1)$ factors, the general construction in
\cite{CveticKleversPiraguaTaylorU1U1} gives a general class of models with
generic matter types.  Presumably, like the Morrison--Park model discussed
above, there will be cases at $T > 0$ where no Tate tuning is possible but we
expect a valid spectrum, which will contribute to the swampland of $\U(1)^2$
models.  As discussed in \cref{sec:two-abelian-f}, the original analysis of
these models made specific positivity choices for the anomaly coefficients,
and a complete story would involve generalizing this analysis to all sign
choices, though it seems likely this will lead to a consistent construction of
all the different classes of $(2, \pm 1)$ matter spectra.  For more than two
$\U(1)$ factors, there is no general explicit model known, so the most general
approach available is to implicitly construct $\U(1)^3$ and higher rank
$\U(1)^s$ models by Higgsing generic nonabelian $G_{(s)} = \SU(2)^s \times
\SU(3)^{\binom{s}{2}}$ models.  This will give a broad class of generic
$\U(1)^s$ models, but as for the lower rank models discussed more explicitly,
there will likely be similar further components of the swampland.  The
swampland for $\U(1)^4$ and higher models will also contain components from
different distributions through the anomaly equivalence
\labelcref{eq:4-equivalence}, associated with issue
\labelcref{item:rep-ambiguities} above.

%%%%%%%%%%%%%%%%%%%%%%%%%%%%%%%%%%%%%%%%%%%%%%%%%%%%%%%%%%%%%%%%%%%%%%%%%%%%%%
\subsubsection{Exotic matter and the 6D swampland}
\label{sec:exotic-swampland}

The concept of generic matter is thus helpful in organizing analysis of
questions related to string universality and the swampland for 6D supergravity
theories.  Going beyond generic matter, the question of which exotic matter
types are allowed in F-theory involves much more complex questions of
algebraic geometry.  For exotic matter charged under nonabelian gauge groups,
a systematic analysis of three-index antisymmetric (``genus 0''
\cite{KumarParkTaylorT=0}) $\SU(N)$ matter was carried out in
\cite{MorrisonTaylorMaS,AndersonGrayRaghuramTaylorMiT}, and a general
methodology for understanding ``higher genus'' matter representations in terms
of singularities over divisors that themselves are singular was developed in
\cite{KleversMorrisonRaghuramTaylorExotic}.  While in many cases there are
F-theory models with exotic singularity structures that match with
anomaly-free low-energy theories with certain exotic matter content, in other
cases it is not known whether F-theory models exist, and in still other cases
it is known that F-theory models cannot exist and there are apparently
consistent models in the swampland. It may be that some exotic matter is
consistent and can be realized in string theory but not conventional F-theory;
incorporating such exotic matter such as $E_7 \times \SU(2)$ bifundamental
matter into F-theory may involve the ``T-brane'' world-volume fields on
7-branes \cite{CveticHeckmanLinExoticDiscrete}, and other exotic matter may
appear in the ``frozen phase'' of F-theory \cite{BhardwajEtAlFrozen}. For
abelian exotic matter, the story is even less clear.  As discussed earlier,
explicit models for F-theory constructions with abelian charges $q = 3, 4$
were constructed in \cite{KleversEtAlToric,Raghuram34,CianciPenaValandro}, and
Higgsing nonabelian constructions can give certain F-theory models with
charges up to $q = 21$, but there still exists an infinite swampland of
higher-charge $\U(1)$ models that have no F-theory realizations, even though
there is no clear understanding of what the finite maximum $\U(1)$ charge
allowed in F-theory constructions can be.

%%%%%%%%%%%%%%%%%%%%%%%%%%%%%%%%%%%%%%%%%%%%%%%%%%%%%%%%%%%%%%%%%%%%%%%%%%%%%%
\subsubsection{Swampland summary}
\label{sec:swampland-summary}

Summarizing our discussion of the swampland and string universality, there are
open questions at the level of the positivity cone, some detailed questions
about which generic matter models can be realized for large gauge groups and
at larger values of $b, T$, and questions about which exotic nonabelian and
abelian matter representations can be realized in F-theory.  The analysis of
generic matter that we have carried out here provides a useful framework in
which to organize further research in these directions.

%%%%%%%%%%%%%%%%%%%%%%%%%%%%%%%%%%%%%%%%%%%%%%%%%%%%%%%%%%%%%%%%%%%%%%%%%%%%%%
%%%%%%%%%%%%%%%%%%%%%%%%%%%%%%%%%%%%%%%%%%%%%%%%%%%%%%%%%%%%%%%%%%%%%%%%%%%%%%
%%%%%%%%%%%%%%%%%%%%%%%%%%%%%%%%%%%%%%%%%%%%%%%%%%%%%%%%%%%%%%%%%%%%%%%%%%%%%%
\section{Generic matter with global gauge group structure and 4D physics}
\label{sec:further}

%%%%%%%%%%%%%%%%%%%%%%%%%%%%%%%%%%%%%%%%%%%%%%%%%%%%%%%%%%%%%%%%%%%%%%%%%%%%%%
%%%%%%%%%%%%%%%%%%%%%%%%%%%%%%%%%%%%%%%%%%%%%%%%%%%%%%%%%%%%%%%%%%%%%%%%%%%%%%
\subsection{Generic matter and the global structure of the gauge group}
\label{sec:further-global}

In our discussions of generic matter for $\SU(N)$ and $\U(1)$ gauge groups, we
have primarily assumed that there are no subtleties in the global structure of
the gauge group.  In particular, we have assumed that the semisimple part of
the gauge algebra is associated with a simply connected gauge group and that
the gauge group is connected.  One can consider more complicated cases where
the nonabelian part of the gauge group is not simply connected, or where the
gauge group has discrete structure and is not connected. The global structure
of the gauge group played a role in the last section in the discussion of
swampland models with no massless matter charged under a central component of
the gauge group. A complete analysis of these more general cases is beyond the
scope of this paper, but we make some basic remarks here about a few aspects
of these questions.

One useful tool in considering both low-energy 6D theories and their F-theory
constructions is the Higgsing process, which can connect theories with
different gauge groups.\footnote{Actually, since there is no superpotential in
six dimensions, 6D supergravity theories with different numbers of tensor
multiplets, different gauge groups, and different matter representations
generally all live on branches of a single moduli space, connected through
Higgs transitions, tensionless string transitions (which trade a tensor
multiplet for 29 scalar multiplets in the simplest situations), and matter
transitions \cite{AndersonGrayRaghuramTaylorMiT}. For convenience, however, we
often refer to branches of the theory with different massless spectra as
different models or ``theories.'' See also \cref{foot:singleMod}.} We have
used Higgsing processes in many places in this paper to describe generic
$\U(1)^s$ models from the Higgsing of generic nonabelian models with gauge
groups $G_{(s)} = \SU(2)^s \times \SU(3)^{\binom{s}{2}}$, for $s = 1, 2, 3,
\dots$. For nonabelian groups such as $\SU(N)$, a theory with generic matter
content will still have generic matter content when a Higgs transition occurs
from giving a vacuum expectation value (VEV) to a pair of fields in the
fundamental representation, for example.  In this case, the new gauge group is
$\SU(N - 1)$, and the other matter fields in the fundamental of $\SU(N)$, for
example, decompose as $\bm{N} \rightarrow \bm{(N - 1)} + \bm{1}$ under $\SU(N
- 1)$.

It is interesting to consider, however, what happens when we take an
$\SU(N + 1)$ theory with generic matter including at least one adjoint
multiplet, and Higgs on an adjoint field with the VEV $\diag(1, 1,
\dots, 1, -N)$.  In this case, the new gauge group is $\SU(N) \times \U(1) /
\Z_N$, where the discrete group $\Z_N$ has elements of the form $\diag(\omega,
\omega, \dots, \omega) \times \omega^{-1} \in \SU(N) \times \U(1)$, with
$\omega$ an $N$th root of unity.  That these elements of the discrete center
of $\SU(N) \times \U(1)$ correspond to trivial elements of the original group
$\SU(N + 1)$ can be seen from the fact that the $\SU(N)$ factor embeds
naturally in the first $N$ components of $\SU(N + 1)$, while the $\U(1)$
factor is generated by $\diag(1, 1, \dots, 1, -N)$. Under such a Higgsing, the
``bifundamental'' type fields that are charged under both $\SU(N)$ and $\U(1)$
take forms that differ from the generic form when the product group $\SU(N)
\times \U(1)$ is simply connected.  In particular, fields that take a given
representation of $\SU(N)$ lead to $\U(1)$ charges that can only differ by
shifts through multiples of $N$. When the $\U(1)$ charges are labeled in units
of $1 / N$, this corresponds to unit shifts of the $\U(1)$ charges. For
example, a fundamental of $\SU(N + 1)$ leads to a fundamental of $\SU(N)$ with
$\U(1)$ charge 1, which may naturally be described as charge $1 / N$, and a
single hypermultiplet that is uncharged under $\SU(N)$ and has $\U(1)$ charge
$N$, naturally described as charge $1$ in the units where the fundamental has
charge $1 / N$.  It is interesting to note that it was observed that this same
shift property seems to be generic for certain F-theory realizations of
$\SU(N) \times \U(1)$ gauge groups \cite{CveticLinU1} (see also
\cref{foot:LingMirjam}), though it is not known that all F-theory models with
such a gauge group can be thought of as coming from a Higgsing of a larger
gauge symmetry such as $\SU(N + 1)$. Note that there are multiple possible
actions of the $\Z_N$ on the product group. For instance, when $N$ is prime,
we can have a gauge group $\SU(N) \times \U(1) / \Z_N$ where the discrete
group consists of elements of the form $\diag(\omega, \omega, \dots, \omega)
\times \omega^{-\bar{k}} \in \SU(N) \times \U(1)$, with $\omega$ an $N$th root
of unity and $\bar{k}$ the modular multiplicative inverse of $k$, which allows
representations such as $\syng{1}_{\, k}$.

The intuition gained from Higgsing $\SU(N + 1) \to \SU(N) \times \U(1) /
\Z_N$ naturally leads us to the question of determining the full set of
generic matter for the latter gauge group. Under this Higgsing, the generic
matter of $\SU(N + 1)$ decomposes as
\begin{equation}
\label{eq:higgsToQuotient}
\begin{aligned}
\yng{1} &\rightarrow \yng{1}_{\, 1 / N} + \bm{1}_1\,, \\
\yng{1,1} &\rightarrow \yng{1,1}_{\, 2 / N} + \yng{1}_{\, -1 + 1 / N}\,, \\
\Adj &\rightarrow \Adj_0 + 2 \times \yng{1}_{\, 1 + 1 / N} + \bm{1}_0\,,
\end{aligned}
\end{equation}
where we have chosen to label the $\U(1)$ charges in units of $1 / N$. The AC
conditions tell us that we expect eight generic representations in this case,
and indeed, if we add $\bm{1}_2$ to the representations on the right-hand side
of \cref{eq:higgsToQuotient}, this collection of representations appears to be
generic.\footnote{We have not proven this in complete generality, but have
checked other allowed representations with small charges and confirmed that
all the representations we checked can be exchanged for these generic
representations with an increase in moduli space dimension.} Note that these
representations look very similar to the generic representations for $\SU(N)
\times \U(1)$, in that they agree in the $N \to \infty$ limit. More generally,
the discrete subgroup $\Z_N$ can be embedded in the center of $\SU(N) \times
\U(1)$ in multiple ways, as noted above, and the choice of embedding
determines the offset of the $\U(1)$ charges from those for the $\SU(N) \times
\U(1)$ generic matter, as shown in \cref{tab:genericQuotient}.

\begin{table}[h!]
\centering

\[\setlength{\arraycolsep}{15pt}
\begin{array}{cc} \toprule
\SU(N) \times \U(1)         & \SU(N) \times \U(1) / \Z_N         \\
\text{Generic Matter}       & \text{Generic Matter}              \\ \midrule
\bm{1}_{\mathrlap{0}}       & \bm{1}_{\mathrlap{0}}              \\[0.8em]
\bm{1}_{\mathrlap{1}}       & \bm{1}_{\mathrlap{1}}              \\[0.8em]
\bm{1}_{\mathrlap{2}}       & \bm{1}_{\mathrlap{2}}              \\[0.8em]
\yng{1}_{\mathrlap{\, 0}}   & \yng{1}_{\mathrlap{\, k / N}}      \\[0.8em]
\yng{1}_{\mathrlap{\, 1}}   & \yng{1}_{\mathrlap{\, 1 + k / N}}  \\[0.8em]
\yng{1}_{\mathrlap{\, -1}}  & \yng{1}_{\mathrlap{\, -1 + k / N}} \\[0.8em]
\yng{1,1}_{\mathrlap{\, 0}} & \yng{1,1}_{\mathrlap{\, 2 k / N}}  \\[1em]
\Adj_{\mathrlap{\, 0}}      & \Adj_{\mathrlap{\, 0}}             \\
	\bottomrule
\end{array}
\]

\caption{Generic matter representations for gauge groups $\SU(N_1) \times
\U(1)$ and $\SU(N_1) \times \U(1) / \Z_N$. Here, $k \in \Z$ is determined by
the embedding of the $\Z_N$ in the center of $\SU(N) \times \U(1)$, and we
have $-N / 2 < k \le N / 2$. If $k$ is not relatively prime to $N$, then the
quotient reduces to a quotient by the relevant subgroup of $\Z_N$, which is
the trivial group for $k = 0$.}
\label{tab:genericQuotient}
\end{table}

Another set of questions involves 6D theories with discrete gauge groups.
These have been a subject of much recent research
(\cite{BraunGenus1,MorrisonTaylorSections,MayrhoferEtAlGlobal,
AndersonEtAlNoSection,CveticEtAlZ3,KimuraEnhanced}, see \cite{WeigandTASI,
CveticLinTASI} for reviews and further references).  We make only a few brief
comments here. Starting from a $\U(1)$ theory with generic matter, we have
only charges $q = 1, 2$.  Higgsing such a theory on fields of charge $q = 2$
leads to a theory with discrete gauge group $\Z_2$ and charges $q = 1$.  Thus,
there is a natural sense in which the discrete gauge group $\Z_2$ fits into
the generic class of 6D supergravity theories.  On the F-theory side, however,
and from the point of view of counting uncharged scalar hypermultiplets, it is
not clear why a $\Z_2$ discrete gauge group is more ``generic'' in any
meaningful sense than a theory with a $\Z_3$ discrete gauge group.  We leave a
further exploration of these questions to future research.

%%%%%%%%%%%%%%%%%%%%%%%%%%%%%%%%%%%%%%%%%%%%%%%%%%%%%%%%%%%%%%%%%%%%%%%%%%%%%%
%%%%%%%%%%%%%%%%%%%%%%%%%%%%%%%%%%%%%%%%%%%%%%%%%%%%%%%%%%%%%%%%%%%%%%%%%%%%%%
\subsection{Generic matter in 4D supergravity theories and F-theory models}
\label{sec:further-4d}

From the point of view of the low-energy supergravity theory, much of
the structure we have used in 6D to classify generic matter is not
available in four dimensions.  In particular,
for 4D theories with $\cN = 1$ supersymmetry, there is in
general a superpotential that lifts many or most of the uncharged
scalar moduli fields, so the notion of generic matter as being
associated with a larger number of moduli does not hold in the
low-energy 4D theory; one may expect that the number
of flux vacua will be exponentially larger on higher-dimensional
moduli spaces \cite{DenefLesHouches}, but the details of this are a bit
harder to quantify explicitly.
In 4D, there is also no
analogous gauge--gravity anomaly, so the matter content in a low
energy theory is less constrained by obvious anomaly constraints.  Furthermore,
the
absence of a purely gravitational anomaly in 4D means that there
is not as clear an upper bound on the number of fields in the theory
in 4D, as opposed to 6D where \cref{eq:nonabelACgrav} puts a
strict bound on the number and complexity of charged and uncharged
matter fields for a given gauge group.

A clearer indication perhaps for four dimensional theories is that the same
geometric structures arise in constructing 4D $\cN = 1$ supergravity theories
from F-theory as arise in 6D.  In particular, the kinds of singularities that
are most generic in Weierstrass models give rise to common types of matter in
4D and 6D. There are further subtleties related to fluxes, the superpotential,
and chiral matter (see \cite{WeigandTASI} for a recent detailed review), but
there is a sense in which the same types of matter fields that are generic for
6D are also the most generic constructions in 4D F-theory models. Thus, we
would expect that for $\SU(N)$ gauge groups coming from generic F-theory
constructions without exotic singularities and associated matter, the natural
representations would be the fundamental, adjoint, and two-index antisymmetric
representations.  Similarly, for a theory with a $\U(1)$ gauge group we would
expect generically only charges $q = 1, 2$, unless as discussed in the
previous section the $\U(1)$ is part of a larger group like $\SU(N) \times
\U(1) / \Z_N$ that has more complicated global structure.

%%%%%%%%%%%%%%%%%%%%%%%%%%%%%%%%%%%%%%%%%%%%%%%%%%%%%%%%%%%%%%%%%%%%%%%%%%%%%%
%%%%%%%%%%%%%%%%%%%%%%%%%%%%%%%%%%%%%%%%%%%%%%%%%%%%%%%%%%%%%%%%%%%%%%%%%%%%%%
\subsection{Generic matter and the standard model}\label{sec:further-SM}

The structure of generic matter seems to shed some interesting light on a
long-standing question regarding the standard model of particle physics.  The
gauge group of the standard model is generally described as $G_\text{({SM})} =
\SU(3)_\text{c} \times \SU(2)_\text{L} \times \U(1)_Y$, with matter fields
taken by standard convention to have fractional charges under the $\U(1)_Y$
gauge field.  For example, the left-handed quarks in the standard model
transform in the fundamental representations of $\SU(3)$ and $\SU(2)$ with
$\U(1)_Y$ charge $Y = 1 / 6$, while the right-handed up quark transforms in
the fundamental representation of $\SU(3)$ and the trivial representation of
$\SU(2)$, with $\U(1)_Y$ charge $Y = 4 / 6$. If the gauge group of the
standard model observed in nature really has the global structure of
$G_\text{({SM})}$, the natural units of charge would be such that the
left-handed quarks would have a unit charge $6 Y = 1$, while the right-handed
electron would have charge $6 Y = -6$.  Independent of charge normalization,
this global structure of the gauge group would make it appear to be an
accident of nature that the electron has an electromagnetic charge three times
larger than the natural charge units for quarks. An alternative hypothesis is
that the actual global structure of the standard model gauge group is $G =
G_\text{({SM})} / \Z_6$, where the the discrete $\Z_6$ has elements of the
form $\diag(\omega^2) \otimes \diag(\omega^3) \otimes \omega$, with $\omega$ a
sixth root of unity; all the charges of the fundamental particles in the
standard model are invariant under the central $\Z_6$, so there is no
empirical evidence for or against this alternative hypothesis. (In fact, there
is also no empirical evidence that the $\U(1)_Y$ factor is compact instead of
a non-compact $\R$ gauge group). See \cite{TongLineOps} for a recent analysis
and further references regarding this ambiguity in the gauge group of the
standard model.

From the point of view of the generic matter representations identified in
this paper, the standard model gauge group without the $\Z_6$ quotient seems
unnatural, in the sense that the matter fields do not fit into the generic
classes of fields listed in \cref{tab:generic}.  In particular, as mentioned
above the right-handed up quark has 4 units of $\U(1)_Y$ charge, the
left-handed leptons have $-3$ units of $\U(1)_Y$ charge, etc.  It is natural
therefore to consider generic matter for the gauge group $\SU(3)_\text{c}
\times \SU(2)_\text{L} \times \U(1)_Y / \Z_6$. The generic matter we find,
along with each corresponding multiplet from the MSSM, is shown in
\cref{tab:genericMSSM}.

\begin{table}[h!]
\centering

\[
\begin{array}{c@{\extracolsep{30pt}}c} \toprule
\text{Generic Matter} & \text{MSSM Multiplet} \\ \midrule
(\bm{1}, \bm{1})_{\mathrlap{0}} & \bm{N}^c                        \\[0.8em]
(\bm{1}, \bm{1})_{\mathrlap{1}} & \bm{E}^c                        \\[0.8em]
(\bm{1}, \bm{1})_{\mathrlap{2}} &                                 \\[0.8em]
\left(\yng{1}, \bm{1}\right)_{\mathrlap{2 / 3}}  & \bar{\bm{U}^c} \\[0.8em]
\left(\yng{1}, \bm{1}\right)_{\mathrlap{-1 / 3}} & \bar{\bm{D}^c} \\[0.8em]
\left(\yng{1}, \bm{1}\right)_{\mathrlap{-4 / 3}} &                \\[0.8em]
\left(\bm{1}, \yng{1}\right)_{\mathrlap{1 / 2}}
	& \bar{\bm{L}} = {\mat[p]{\bar{\bm{N}} & \bar{\bm{E}}}},
		\bm{H}_u, \bar{\bm{H}_d}                                  \\[0.8em]
\left(\bm{1}, \yng{1}\right)_{\mathrlap{3 / 2}}  &                \\[0.8em]
(\Adj, \bm{1})_{\mathrlap{0}}                    &                \\[0.8em]
(\bm{1}, \Adj)_{\mathrlap{0}}                    &                \\[0.8em]
\left(\yng{1}, \yng{1}\right)_{\mathrlap{1 / 6}}
	& \bm{Q} = \mat[p]{\bm{U} \\ \bm{D}}                          \\
	\bottomrule
\end{array}
\]

\caption{Generic matter representations (not including conjugates) for the
gauge group $\SU(3)_\text{c} \times \SU(2)_\text{L} \times \U(1)_Y / \Z_6$,
which include all the left-handed MSSM multiplets. The generic matter for
the group $\SU(3)_\text{c} \times \SU(2)_\text{L} \times \U(1)_Y$ does not
include the MSSM multiplets.}
\label{tab:genericMSSM}
\end{table}

Since our concrete definition of generic matter relies on the structure of 6D
supergravity, this is the context in which we have determined the generic
matter fields in \cref{tab:genericMSSM}.  We have not proven rigorously that
all possible other representations can be exchanged for the fields in this
table, but we have checked this by hand for all reasonably small
representations.  In particular, for the gauge group of the SM,
\cref{eq:nonabelACgrav} tells us that $H \le 285$ \emph{for any $T$} (the
bound becomes more strict for larger $T$) in a 6D supergravity theory. Thus,
at least for 6D theories with this gauge group, it suffices to check that the
representations in \cref{tab:genericMSSM} are generic for all exchanges with
representations of dimension at most $285$. Carrying out this brute force
check, we do find that exchanges to the representations in
\cref{tab:genericMSSM} never decrease the number of uncharged scalars.

The upshot of this analysis is that the matter content of the MSSM consists of
generic matter field types, so long as the global structure of the gauge group
is $\SU(3)_\text{c} \times \SU(2)_\text{L} \times \U(1)_Y / \Z_6$.  Note that
this structure of the gauge group can arise in particular when the standard
model is realized by breaking a GUT group such as $\SU(5)$, $\gE_6$, $\gE_7$,
or $\gE_8$.

% The standard model has the gauge group $\SU(3) \times \SU(2) \times \U(1)$.
% The various quark and lepton fields transform under the fundamental
% representation of $\SU(2)$, and the fundamental and anti-fundamental
% (equivalent to the two-index antisymmetric) representations of $\SU(3)$. While
% the $\U(1)$ charges do not appear immediately to be of the minimal $q = 1, 2$
% type, taking values $q = 1 / 6, 1 / 3, 1 / 2$ in the usual units, along the
% lines of the discussion in \cref{sec:further-global} these can actually be
% understood as the natural generic charges if the global structure of the gauge
% group is $\SU(3) \times \SU(2) \times \U(1) / \Z_6$, as might arise for
% example from a GUT group of $\SU(5)$, $\gE_6$, $\gE_7$, or $\gE_8$. We hope to
% return to this topic in more detail elsewhere.

It is also interesting to consider the question of whether there are
nontrivial 4D chiral matter models that contain only generic matter for the
gauge group $\SU(3)_\text{c} \times \SU(2)_\text{L} \times \U(1)_Y$ (without
the $\Z_6$ quotient). For generic matter, we therefore consider the possible
multiplicities of the fields from \cref{tab:generic} for a chiral theory with
this gauge group. The AC conditions in 4D are linear and cubic in the
hypercharge, rather than quadratic and quartic. These models can indeed have
chiral matter, and so we can ask how the multiplicities of each representation
must be related to those of their conjugates in order to satisfy anomaly
cancellation. We note immediately that fields like the left-handed quark
fields, which transform under all three gauge factors, are not among the set
of generic matter fields.

Furthermore, the only field charged under both the $\SU(2)$ and $\U(1)_Y$
factors has charges $(\bm{1}, \syng{1})_1$, so the difference between left-
and right-handed multiplicities of this field must vanish by the $\U(1)
\SU(2)^2$ anomaly.  This implies that we cannot have any chiral matter that is
charged under both the $\SU(2)_\text{L}$ and the $\U(1)_Y$, including, e.g., a
field like an ``electron.'' There is, however, a nontrivial multi-parameter
family of solutions to the complete set of 4D anomaly equations. Defining
$\Delta x_R := x_R - x_{\bar{R}}$, we find a family of solutions of the form
\begin{equation}
\label{eq:generic-no-quotient}
\begin{gathered}
\Delta x_{(\tyng{1}, \bm{1})_0} = a\,, \quad
	\Delta x_{(\bm{1}, \tyng{1})_0} = b\,, \quad
	\Delta x_{(\Adj, \bm{1})_0} = c\,, \quad
	\Delta x_{(\bm{1}, \Adj)_0} = d\,, \\
\Delta x_{(\tyng{1}, \bm{1})_1} =
	\Delta x_{(\tyng{1}, \bm{1})_{-1}} = e\,, \quad
	\Delta x_{(\tyng{1}, \tyng{1})_0} =  (-a - 2 e) / 2\,, \\
\Delta x_{(\bm{1}, \tyng{1})_1} = \Delta x_{(\bm{1}, \bm{1})_1} =
	\Delta x_{(\bm{1}, \bm{1})_2} = 0\,.
\end{gathered}
\end{equation}
If the standard model gauge group did not have a global structure with the
$\Z_6$ quotient, this would appear to be the most generic type of matter we
would expect from considerations of 6D supergravity and F-theory geometry.
This would be a less phenomenologically rich world than the one we live in,
however, with no fields simultaneously charged under the $\SU(2)$ and
$\U(1)_Y$ gauge factors, like the left-handed quarks and charged leptons. In
the context of F-theory, we might expect to have constructions leading to
either the MSSM with the gauge group having the quotient structure and the
standard MSSM matter fields associated with generic matter from
\cref{tab:genericMSSM}, or the gauge group having the product structure
without the quotient and a spectrum in the family of fields listed in
\cref{eq:generic-no-quotient}. The MSSM with the gauge group having no
quotient taken, however, seems to involve non-generic matter and is likely
disfavored by F-theory or perhaps any other approach to string
compactification.\footnote{\label{foot:LingMirjam} Note that in
\cite{CveticLinU1}, a swampland hypothesis was put forward regarding F-theory
constructions of theories with product groups, and it was suggested that this
implied that the standard model gauge group as constructed by F-theory would
generally have the $\Z_6$ quotient.  That argument is somewhat different from
what we are saying here. The precise statement made in \cite{CveticLinU1}
depends upon the assumption that in a theory with a $\U(1)$ factor, the
lattice of singlet charges associated with fields charged under that $\U(1)$
but no other factors determine the preferred normalization of that $\U(1)$
charge.  For theories with only a single $\U(1)$ and no other gauge factors,
this is a special case of the massless field swampland hypothesis mentioned in
\cref{sec:generic-swampland}.  Under the more general hypothesis that this
statement is true even in the presence of one or more nonabelian factors,
\cite{CveticLinU1} argue that in any F-theory model, all fields that transform
in a given representation of the remaining gauge factors must differ by a
multiple of the normalized $\U(1)$ charge.  Several comments on this result
and the connection to the present work: 1) This argument does not rule out in
any way the existence of F-theory models with a standard model gauge group
(without $\Z_6$ quotient) and spectrum composed of the fields listed in
\cref{eq:generic-no-quotient}, as long as there is at least one multiplet with
charges $(\bm{1}, \bm{1})_1$ (note that this multiplicity need not vanish in
6D theories or for vector-like 4D multiplets).  2) It is not clear that the
assumption quoted above is correct.  In particular, consider a theory with
gauge group $\SU(2) \times \U(1)$ with $x_{\, \tyng{1}_{\, 1}} = 0$ but
$x_{\bm{1}_2}, x_{\, \tyng{1}_{\, 2}} \neq 0$.  This would not be allowed by
this assumption.  Nevertheless, there are anomaly-free models with such
spectra in 6D.  While no F-theory constructions for such models are known,
this is likely because $\syng{1}_{\, 2}$ is a non-generic representation and
would require a fine-tuned exotic Weierstrass model; thus, we see no reason
why such models cannot exist.  3) The statement that any model where the
singlet $\U(1)$ charge can be used as a measuring stick for the
\emph{massless} charged fields must have a gauge group with a quotient
structure is again a special case of the massless field swampland hypothesis
of \cref{sec:generic-swampland}.}

%%%%%%%%%%%%%%%%%%%%%%%%%%%%%%%%%%%%%%%%%%%%%%%%%%%%%%%%%%%%%%%%%%%%%%%%%%%%%%
%%%%%%%%%%%%%%%%%%%%%%%%%%%%%%%%%%%%%%%%%%%%%%%%%%%%%%%%%%%%%%%%%%%%%%%%%%%%%%
%%%%%%%%%%%%%%%%%%%%%%%%%%%%%%%%%%%%%%%%%%%%%%%%%%%%%%%%%%%%%%%%%%%%%%%%%%%%%%
\section{Conclusions}\label{sec:conclusions}

In this paper, we have introduced a notion of ``generic'' matter
representations for different gauge groups in supergravity theories. This
notion is given a specific and quantitative meaning in the context of
six-dimensional supergravity theories, where we define generic matter
representations to be those that arise on the branches of the moduli space of
largest dimension for a given gauge group when the anomaly coefficients are
suitably small.  This definition matches nicely with the anomaly cancellation
conditions in six dimensions, and also matches with the matter representations
that arise through the most direct and straightforward geometric constructions
in the language of F-theory.  While we use six dimensional supergravity to
give a concrete definition to the notion of generic matter, the correspondence
with natural constructions in F-theory suggests that this notion should also
be meaningful for four-dimensional supergravity theories containing gauge
groups and matter fields in various representations.

The notion of generic matter illuminates some outstanding puzzles related to
6D supergravity and F-theory.  As we have described in \cref{sec:generic}, the
structure of generic matter clarifies what kinds of charged matter we expect
in the most generic constructions of F-theory models with multiple abelian
gauge factors, and may be helpful in guiding further research on the
challenging problem of explicitly constructing such models.

Generic matter also provides a useful tool for framing questions about the
string swampland.  As discussed in \cref{sec:generic-f-theory}, when we
restrict to 6D supergravity theories with a string charge lattice and
positivity cone compatible with those known to arise from geometric
constructions, the swampland of apparently consistent theories with no known
realization in F-theory is rather limited for theories with only generic
matter content.  The main questions in this regard can then in large part be
related to questions about exotic matter representations and the construction
of Weierstrass models realizing sufficiently exotic singularities to realize
these representations.

Perhaps the most interesting application of these ideas is in the context of
4D physics.  Naively, even taking the structure of the standard model gauge
group $\SU(3)_\text{c} \times \SU(2)_\text{L} \times \U(1)_Y$ to be fixed, one
could imagine consistent theories with an essentially infinite number of
possible different combinations of light matter fields in different
representations.  Of course, the standard model is known to have one of the
simplest combinations of such fields that satisfies 4D anomaly cancellation
conditions, but the notion of generic matter gives a concrete framework that
motivates why such a ``simplest combination'' may be favored by nature, at
least in the context of a UV completion such as F-theory.  In fact, as we have
found here, the matter representations in the standard model are only generic
if the gauge group has the global structure $\SU(3)_\text{c} \times
\SU(2)_\text{L} \times \U(1)_Y / \Z_6$.  While there is as yet no simple and
direct experimental mechanism for testing this aspect of the global structure
of the gauge group, this distinction is a promising sign that this kind of
analysis may eventually lead us to new insights regarding important and
observable features of physics beyond the standard model.

As mentioned at the end of the introduction, it is important to note that we
are defining generic matter in this paper in terms of a fixed choice of gauge
group.  Thus, the analysis presented here represents a refinement of our
understanding of the space of 6D supergravity theories and F-theory vacua that
goes beyond the more basic question of which gauge groups are most generic or
prevalent in the broader landscape of supergravity or string vacua. As also
noted in the introduction, we have not here discussed strongly coupled
conformal matter associated with gravitationally coupled SCFTs, which
represent another important and interesting arena for study.

\acknowledgments
We would like to thank Noam Elkies, Ling Lin, Greg Moore, and Nikhil Raghuram
for helpful discussions. Thanks to Yinan Wang for comments on a preliminary
version of this manuscript. The authors are supported by DOE grant
DE-SC00012567.

%%%%%%%%%%%%%%%%%%%%%%%%%%%%%%%%%%%%%%%%%%%%%%%%%%%%%%%%%%%%%%%%%%%%%%%%%%%%%%
%%%%%%%%%%%%%%%%%%%%%%%%%%%%%%%%%%%%%%%%%%%%%%%%%%%%%%%%%%%%%%%%%%%%%%%%%%%%%%
%%%%%%%%%%%%%%%%%%%%%%%%%%%%%%%%%%%%%%%%%%%%%%%%%%%%%%%%%%%%%%%%%%%%%%%%%%%%%%
%%%%%%%%%%%%%%%%%%%%%%%%%%%%%%%%%%%%%%%%%%%%%%%%%%%%%%%%%%%%%%%%%%%%%%%%%%%%%%
\appendix

%%%%%%%%%%%%%%%%%%%%%%%%%%%%%%%%%%%%%%%%%%%%%%%%%%%%%%%%%%%%%%%%%%%%%%%%%%%%%%
%%%%%%%%%%%%%%%%%%%%%%%%%%%%%%%%%%%%%%%%%%%%%%%%%%%%%%%%%%%%%%%%%%%%%%%%%%%%%%
%%%%%%%%%%%%%%%%%%%%%%%%%%%%%%%%%%%%%%%%%%%%%%%%%%%%%%%%%%%%%%%%%%%%%%%%%%%%%%
\section{Generic matter for
	$\SU(N_1) \times \dots \times \SU(N_r) \times \U(1)^3$}
\label{sec:generic-proof}

In this appendix, we will prove that the matter representations given in
\cref{tab:generic} are generic in the case of $\prod_i \SU(N_i) \times
\U(1)^3$, in the sense that exchanges from other representations to these
representations never decrease the number of uncharged scalars. We will first
prove this to be true for canonical generic matter for $\SU(N)$ and $\U(1)^3$
individually, after which we can use these results in the proof for $\SU(N)
\times \U(1)^3$ and $\prod_i \SU(N_i) \times \U(1)^3$. We have addressed the
non-canonical generic matter types in the various sections of the main text.

%%%%%%%%%%%%%%%%%%%%%%%%%%%%%%%%%%%%%%%%%%%%%%%%%%%%%%%%%%%%%%%%%%%%%%%%%%%%%%
%%%%%%%%%%%%%%%%%%%%%%%%%%%%%%%%%%%%%%%%%%%%%%%%%%%%%%%%%%%%%%%%%%%%%%%%%%%%%%
\subsection{$\SU(N)$}\label{sec:sun-proof}

Note that, defining
\begin{equation}
\label{eq:defGens}
\begin{aligned}
(T_{1 2})_{i j} &= \delta_{i 1} \delta_{j 1} - \delta_{i 2} \delta_{j 2}\,, \\
(T_{3 4})_{i j} &= \delta_{i 3} \delta_{j 3} - \delta_{i 4} \delta_{j 4}\,, \\
(T_{1 2 3})_{i j} &= \delta_{i 1} \delta_{j 1} + \delta_{i 2} \delta_{j 2}
	- 2 \delta_{i 3} \delta_{j 3}\,,
\end{aligned}
\end{equation}
the group theory coefficients $A_R, B_R, C_R, E_R$ can be computed directly
via
\begin{subequations}
\label{eq:ABCEcomp}
\begin{align}
A_R &= \frac{1}{2} \trace_R T_{1 2}^2\, \label{eq:Acomp} \\
B_R + 2 C_R &= \frac{1}{2} \trace_R T_{1 2}^4\,, \label{eq:Bcomp} \\
C_R &= \frac{3}{4} \trace_R T_{1 2}^2 T_{3 4}^2\,, \label{eq:Ccomp} \\
E_R &= -\frac{1}{6} \trace_R T_{1 2 3}^3\,. \label{eq:Ecomp}
\end{align}
\end{subequations}
In this section, we take the definition of $C_R$ in \cref{eq:Ccomp} at face
value, so that $C_R = 0$ for $\SU(2)$ and $\SU(3)$, while $B_R$ does not
necessarily vanish. This is in contrast to the conventions of
\cref{sec:anomalies}, where we take $B_R = 0$ for $\SU(2)$ and $\SU(3)$, and
would have $C_R = \frac{1}{4} \trace_R T_{1 2}^4$ for these groups.

Consider a theory with gauge group $G = \SU(N)$. An exchange that trades a
hypermultiplet charged under the non-generic representation $R$ for some
combination of hypermultiplets charged under the canonical generic matter
representations in \cref{tab:generic} increases the number of uncharged
scalars by an amount
\begin{equation}
X = d + \left(\frac{3 N + 1}{12}\right) g_R + \frac{N (N - 3)}{6} C_R - N A_R
\end{equation}
where
\begin{equation}
\begin{aligned}
d &= \dim R\,, \\
g_R &= B_R + 2 C_R - A_R\,.
\end{aligned}
\end{equation}
In terms of traces, we can rewrite $X$ in the form
\begin{equation}
\label{eq:sun-proof-trace}
X = d + \trace_R \left\{\frac{N}{8} T_{1 2}^2 \left[T_{1 2}^2
	+ (N - 3) T_{3 4}^2 - 5\right]
	+ \frac{1}{24} T_{1 2}^2 \left(T_{1 2}^2 - 1\right)\right\}\,.
\end{equation}

Recall that the irreducible representations of $\SU(N)$ are in bijection with
Young diagrams with at most $N - 1$ rows. Let $\lambda$ be the partition such
that the Young diagram of shape $\lambda$, which we will call $[\lambda]$,
corresponds to the representation $R$. The basis elements of the
representation $R$ then correspond to semistandard Young tableaux (SSYT) of
shape $\lambda$ with entries in $1, \dots, N$; we denote the set of such Young
tableaux as $\SSYT_N(\lambda)$.

Using this fact, the trace in \cref{eq:sun-proof-trace} can be evaluated as a
sum over $\SSYT_N(\lambda)$:
\begin{equation}
\begin{split}
X = \sum_{T \in \SSYT_N(\lambda)} \Bigg\{1 &+ \frac{N}{8}
	\left(\mu_1 - \mu_2\right)^2 \left[\left(\mu_1 - \mu_2\right)^2
	+ \theta(N - 4) (N - 3) \left(\mu_3 - \mu_4\right)^2 - 5\right] \\
&{}+ \frac{1}{24} \left(\mu_1 - \mu_2\right)^2
	\left[\left(\mu_1 - \mu_2\right)^2 - 1\right]\Bigg\}\,.
\end{split}
\end{equation}
Here, $\mu = \left(\mu_1, \mu_2, \dots, \mu_N\right)$ is the \emph{weight} of
the SSYT $T$, so that the entry $i$ occurs $\mu_i$ times in $T$, and
$\theta$ is the Heaviside step function.

Our goal is to show that $X$ is always non-negative. We proceed by casework.
We want to identify the SSYT for which the corresponding summand is negative,
so that we can ensure they are compensated for by other positive summands and
the resulting sum is never negative. For ease of notation, denote the
contribution from SSYT $T$ of weight $\mu$ by $S_\mu$, so that
\begin{equation}
X = \sum_{\mathclap{T \in \SSYT_N(\lambda)}} S_\mu\,.
\end{equation}
Note that if $\mu_1 = \mu_2$, then $S_\mu = 1$.

We first consider the case that $\mu_3 = \mu_4$; this also covers the
cases where $N < 4$, in which case there is no quartic Casimir and thus no
generator $T_{3 4}$. In this case, we have
\begin{equation}
S_\mu = 1 + \frac{k^2}{8} \left[N \left(k^2 - 5\right)
	+ \frac{k^2 - 1}{3}\right]
\end{equation}
for $\mu_1 = \mu_2 \pm k$, which gives $S_\mu = 1$ for $k = 0$, $S_\mu = 1 -
\frac{N}{2}$ for $k = 1$, $S_\mu = \frac{3}{2} - \frac{N}{2}$ for $k = 2$, and
$S_\mu > 1$ for $k > 2$.

Next, we consider the case $\mu_3 = \mu_4 \pm 1$ and $N \ge 4$. In this case,
\begin{equation}
S_\mu = 1 + \frac{k^2}{8} \left[N^2 + N \left(k^2 - 8\right)
	+ \frac{k^2 - 1}{3}\right]
\end{equation}
for $\mu_1 = \mu_2 \pm k$, which gives $S_\mu = 1$ for $k = 0$, $S_\mu = 1 +
\frac{N (N - 7)}{8} \ge -\frac{1}{2}$ for $k = 1$, and $S_\mu > 1$ for $k >
1$.

For $\abs{\mu_3 - \mu_4} \ge 2$, $S_\mu \ge 1$ for all values of $\mu_1,
\mu_2$.

Thus, there are only three types of tableaux we must consider to ensure that
$X > 0$: those with $\mu_3 = \mu_4$ (or $N < 4$) and $\mu_1 = \mu_2 \pm 1$ or
$\mu_1 = \mu_2 \pm 2$, and those with $\mu_3 = \mu_4 \pm 1$ and $\mu_1 = \mu_2
\pm 1$.

First, note that $S_\mu \ge 0$ for all $T$ when $N = 2$, so $X \ge 0$ for
$\SU(2)$.

For $N > 2$, it will be useful to consider the collective contribution
$\tilde{S}_\mu := \sum_{\pi \in \gS_N} S_{\pi(\mu)}$ from all diagrams of a
given weight $\mu$ and all of its permutations. To see why, we briefly
introduce some facts about the Kostka numbers. The number of SSYT of shape
$\lambda$ and weight $\mu$ is given by the \emph{Kostka number} $K_{\lambda
\mu}$ \cite{KostkaSymmFunctions}. A useful fact is that the Kostka number
$K_{\lambda \mu}$ is invariant under permutations of the weight $\mu$
\cite{StanleyEnumComb2}, i.e., $K_{\lambda \mu} = K_{\lambda \pi(\mu)}$ for
any permutation $\pi \in \gS_N$. Thus, every term in $\sum_{\pi \in \gS_N}
S_{\pi(\mu)}$ will be proportional to $K_{\lambda \mu}$. Another useful fact
is that if $\mu, \mu'$ are \emph{partitions} (i.e., they are weakly
decreasing) and $\mu \trianglerighteq \mu'$, then $K_{\lambda \mu'} \ge
K_{\lambda \mu}$ \cite{FayersKostka}. Here, $\trianglerighteq$ is the
dominance order, under which $(\mu_1, \dots, \mu_N) \trianglerighteq (\mu'_1,
\dots, \mu'_N)$ if $\sum_{i = 1}^k \mu_i \ge \sum_{i = 1}^k \mu'_i$ for all $1
\le k \le N$.

We now move on to consider $N > 2$. We will carefully elaborate the argument
for $N = 3$, and then appeal to similar reasoning for $N > 3$. For $N = 3$,
the only tableaux that contribute negatively are those with $\mu_1 = \mu_2 \pm
1$, which contribute $S_\mu = -\frac{1}{2}$. Consider the summed contribution
$\tilde{S}_\mu$ for weight $\mu = (j, j + 1, k')$ with fixed $j, k'$. We
know the result will be proportional to $K_{\lambda \mu}$, and the
contributions $S_\mu$ each only depend on the relative differences of the
$\mu_i$, so for notational simplicity we can instead refer to a ``relative
weight'' $\mu - \mu_1 = [0, 1, k]$ for $k = k' - j$, where we are using square
brackets to indicate that the entries are taken relative to $\mu_1$.
Furthermore, we will still account for all possible cases if we assume $k \ge
1$, because $S_\mu$ does not depend on the sign of the differences $\mu_1 -
\mu_2$ and $\mu_3 - \mu_4$. The tableaux of weights $[0, 1, k]$ and $[1, 0,
k]$ each contribute $-K_{\lambda \mu} / 2$. We have the cases $k = 1$, $k =
2$, and $k > 2$. In the case $k = 1$, the contribution from tableaux of weight
$(1, 1, 0)$ is $+K_{\lambda \mu}$, as $S_\mu = 1$ for all tableaux with $\mu_1
= \mu_2$. In the case $k > 2$, the contribution from tableaux of weight $(0,
k, 1)$ is at least $+K_{\lambda \mu}$, as $S_\mu \ge 1$ for all tableaux with
$\abs{\mu_1 - \mu_2} > 2$. The only case in question is then $k = 2$, in which
case the weights $[1, 2, 0]$ and $[2, 1, 0]$ each contribute an additional
$-K_{\lambda \mu} / 2$ and the weights $[0, 2, 1]$ and $[2, 0, 1]$ contribute
nothing. Thus, $\tilde{S}_{[2, 1, 0]} = -2 K_{\lambda \mu}$ and all other
contributions $\tilde{S}_\mu$ are non-negative. We will defer the discussion
of this case to later in this section.

Similar arguments can be considered for $N > 3$. Using such arguments, we can
see that if the relative weight $[0, \mu_2 - \mu_1, \dots, \mu_N - \mu_1]$
contains an entry of $3$ or greater, then $\tilde{S}_\mu$ will be
non-negative, because the negative contributions will be outweighed by large
positive contributions from the differences of at least three in the relevant
permutations. Thus, we restrict to relative weights with only entries $0, 1,
2$. Again, similar arguments lead to the conclusion that the only relative
weight that gives a negative $\tilde{S}_\mu$ is of the form $[2, 1, \dots, 1,
0]$, with a single $2$, a single $0$, and all other entries $1$. Using the
result above for the values of $S_\mu$ for a given $\mu$, we find that in this
case,
\begin{equation}
\tilde{S}_{[2, 1, \dots, 1, 0]} = (1 - N) K_{\lambda \mu}\,, \quad N \ge 3\,.
\end{equation}
This includes the only possible trouble case we found for $N = 3$.

We thus restrict our attention to relative weights of the form $[2, 1, \dots,
1, 0]$ for $N \ge 3$. Our arguments thus far have only dealt with the weight
$\mu$ of the tableaux, but not the shape $\lambda$. Consider a weight $\mu =
(2 + k, 1 + k, \dots, 1 + k, k)$. Note that if the diagram $[\lambda]$ has
fewer than $2 + k$ columns, then $K_{\lambda \mu} = 0$; in fact, $[\lambda]$
must have at least $k + 1$ boxes outside of the first $k + 1$ columns, as
there can be at most $N - 1$ rows in the Young diagram $[\lambda]$
corresponding to an irreducible representation of $\SU(N)$. In the case $k =
0$, $[\lambda]$ can have a single box in the second column and none in the
third. In this case, $[\lambda]$ is the adjoint representation and $\mu = (2,
1, \dots, 1, 0)$. Here, we can see that the negative contribution is $(1 - N)
K_{\lambda \mu} = 1 - N$, which is exactly balanced by the positive
contribution $K_{\lambda \mu'} = N - 1$ of diagrams with weight $\mu' = (1,
\dots, 1)$. These are the only diagrams that contribute, so in this case $X =
0$, as we already knew because the adjoint is one of the representations in
our set of generic matter.

In all remaining cases, there are at least two boxes outside of the first $k +
1$ columns, at least one of which must be in the first row. First, consider
the case that there are two additional boxes in the first row, i.e., the first
row contains at least $k + 3$ boxes, and consider tableaux of weight $\mu =
(k, 2 + k, 1 + k, \dots, 1 + k)$. Because such tableaux have at least $k + 3$
columns and only $k + 2$ instances of the entry $2$, and also have more of
each entry higher than $2$ than of $1$ entries, there must be at least one
column with no entry $2$ but containing higher entries. Thus, from each such
tableau we can produce a new valid tableau of the same shape as follows: in
the leftmost column that contains no $2$ but does contain higher entries, we
replace the lowest entry greater than $2$ with a $2$.
% considering the boxes in order from left to right along the first row and
% then the second row, we find the last box that occurs immediately following
% a $2$ entry, and replace the entry in that box with a $2$. This replacement
% is always valid, as all entries to its left are $2$ or less, and it is
% either in the first row or in the second row below a $1$ entry.
% Equivalently, because such tableaux have at least $k + 3$ columns, and only
% $k + 2$ instances of the entry $2$, there must be at least one column with
% no entry $2$, but containing higher entries; in the leftmost such column,
% there will be exactly one entry that can be validly replaced with a $2$ with
% no modification to other entries in the tableaux, and this is the
% replacement we make.
This maps each tableau of shape $\lambda$ and weight $\mu$ to a new tableau of
shape $\lambda$ and weight $(k, 3 + k, 1 + k, \dots, 1 + k, k, 1 + k, \dots, 1
+ k)$, although which other entry has only $k$ instances depends on the
original tableau. This map is not generally invertible, as it may not be
injective; however, because all $2$ entries occur in the first two rows, the
map is at most two-to-one. Thus, although $K_{\lambda [0, 0, 3, 1, \dots,
1]} \le K_{\lambda \mu}$ (because of the dominance order), we see that
\begin{equation}
K_{\lambda \mu} \le 2 (N - 2) K_{\lambda [0, 0, 3, 1, \dots, 1]}\,,
\end{equation}
where the factor of $N - 2$ accounts for all possible entries higher than $2$
that could have been traded for an additional $2$ in the above map, and the
factor of $2$ accounts for the map possibly being two-to-one. The negative
contribution we must balance is $(1 - N) K_{\lambda \mu}$, and so the more
useful inequality is
\begin{equation}
\label{eq:sunKostkaIneq1}
(N - 1) K_{\lambda \mu} \le
	2 (N - 1) (N - 2) K_{\lambda [0, 0, 3, 1, \dots, 1]}\,.
\end{equation}
Now, if we consider all tableaux of shape $\lambda$ with weight $\mu' = (k, 3
+ k, k, 1 + k, \dots, 1 + k)$ or any permutation thereof, their total
contribution to $X$ is
\begin{equation}
\frac{1}{2} (N - 2) \left(7 N^2 + 9\right) K_{\lambda \mu'}\,.
\end{equation}
Using the inequality \labelcref{eq:sunKostkaIneq1}, we then have
\begin{equation}
\begin{aligned}
\frac{1}{2} (N - 2) \left(7 N^2 + 9\right) K_{\lambda \mu'}
	- (N - 1) K_{\lambda \mu} &\ge
		\frac{1}{2} (N - 2) \left(7 N^2 + 9\right) K_{\lambda \mu'} \\
		&\qquad\quad - 2 (N - 1) (N - 2)  K_{\lambda \mu'} \\
	&= \frac{1}{2} (N - 2) [N (7 N - 4) + 13] K_{\lambda \mu'} \\
	&\ge 0\,.
\end{aligned}
\end{equation}
Thus, the positive contribution from such tableaux is always sufficient to
balance the negative contributions from those of relative weight $[2, 1,
\dots, 1, 0]$ and its permutations, when $[\lambda]$ has at least $k + 3$
boxes in the first row.

In the remaining cases, $[\lambda]$ must have at least two boxes in the column
$k + 2$, and no boxes in column $k + 3$. We can then use a similar approach as
above. In this case, we again consider tableaux of weight $(k, 2 + k, 1 + k,
\dots, 1 + k)$, and note that we can map each to a new tableau by replacing
with a $3$ the least entry greater than $3$ in the leftmost column that
contains no $3$ but does contain higher entries. This is again at most a
two-to-one map. This proves that
\begin{equation}
\label{eq:sunKostkaIneq2}
(N - 1) K_{\lambda \mu} \le
	2 (N - 1) (N - 3) K_{\lambda [0, 0, 2, 2, 1, \dots, 1]}\,.
\end{equation}
Considering all tableaux of weight $\mu' = (k, k, 2 + k, 2 + k, 1 + k \dots, 1
+ k)$ or any permutations thereof, we see that these tableaux collectively
contribute
\begin{equation}
\left[\frac{1}{2} N (N - 1) (N - 2) - 2\right] K_{\lambda \mu'}\,.
\end{equation}
We then have
\begin{equation}
\begin{aligned}
\left[\frac{1}{2} N (N - 1) (N - 2) - 2\right] K_{\lambda \mu'}
	- (N - 1) K_{\lambda \mu} &\ge
		\left[\frac{1}{2} N (N - 1) (N - 2) - 2\right] K_{\lambda \mu'} \\
		&\qquad\quad - 2 (N - 1) (N - 3) K_{\lambda \mu'} \\
	&= \frac{1}{2} (N - 2) [N (N - 5) + 8] K_{\lambda \mu'} \\
	&\ge 0\,.
\end{aligned}
\end{equation}
This shows that the negative contribution from tableaux of relative weight
$[2, 1, \dots, 1, 0]$ and its permutations is compensated in all remaining
cases, completing the proof. Thus, the representations presented in
\cref{tab:generic} are generic for the gauge group $G = \SU(N)$.

%%%%%%%%%%%%%%%%%%%%%%%%%%%%%%%%%%%%%%%%%%%%%%%%%%%%%%%%%%%%%%%%%%%%%%%%%%%%%%
%%%%%%%%%%%%%%%%%%%%%%%%%%%%%%%%%%%%%%%%%%%%%%%%%%%%%%%%%%%%%%%%%%%%%%%%%%%%%%
\subsection{$\U(1)^3$}\label{sec:u13-proof}

Now we consider a theory with gauge group $G = \U(1)^3$. An exchange that
trades a hypermultiplet charged under the non-generic representation $(q_1,
q_2, q_3)$ for some combination of hypermultiplets charged under the canonical
generic matter representations in \cref{tab:generic} increases the number of
uncharged scalars by an amount
\begin{equation}
Y = \frac{1}{16} \left[2 Q_1^2 (Q_2 - 7) + Q_2 (Q_2 - 6)
	+ Q_1^4 + 4 (Q_{2 2} + 3 Q_{1 1}) + 16\right]\,,
\end{equation}
where
\begin{equation}
\begin{aligned}
Q_1 &= q_1 + q_2 + q_3\,, \\
Q_2 &= q_1^2 + q_2^2 + q_3^2\,, \\
Q_{1 1} &= q_1 q_2 + q_1 q_3 + q_2 q_3\,, \\
Q_{2 2} &= q_1^2 q_2^2 + q_1^2 q_3^2 + q_2^2 q_3^2\,.
\end{aligned}
\end{equation}

This polynomial is non-negative for all integer charges $q_i \in \Z$. To see
this, first note that
\begin{equation}
Q_{2 2} + 3 Q_{1 1} = q_1 q_2 (q_1 q_2 + 3) + q_1 q_3 (q_1 q_3 + 3)
	+ q_2 q_3 (q_2 q_3 + 3)
\end{equation}
has a minimum value of $-2$ on the integers, so $4 (Q_{2 2} + 3 Q_{1 1}) + 16
\ge 8$ on the integers. The term $Q_1^4$ is clearly non-negative, as it is a
fourth power. Thus, the only negative contributions can occur whenever $Q_2 <
7$. We can easily enumerate the possible charge combinations for which this is
true: $(0, 0, 0)$, $(1, 0, 0)$, $(1, \pm 1, 0)$, $(1, 1, \pm 1)$, $(\pm 2, 0,
0)$, $(2, \pm 1, 0)$, $(2, 1, \pm 1)$, $(2, -1, -1)$, and their permutations
(and conjugates). However, most of these representations are members of the
set of generic matter in \cref{tab:generic}, and so we already know that their
exchanges do not change the number of uncharged scalars, and we can check the
remaining cases by hand. We find that $Y$ is non-negative for all cases, so
the representations presented in \cref{tab:generic} are generic for the gauge
group $G = \U(1)^3$.

%%%%%%%%%%%%%%%%%%%%%%%%%%%%%%%%%%%%%%%%%%%%%%%%%%%%%%%%%%%%%%%%%%%%%%%%%%%%%%
%%%%%%%%%%%%%%%%%%%%%%%%%%%%%%%%%%%%%%%%%%%%%%%%%%%%%%%%%%%%%%%%%%%%%%%%%%%%%%
\subsection{$\SU(N) \times \U(1)^3$}
\label{sec:sun-u13-proof}

We now consider a theory with gauge group $\SU(N) \times \U(1)^3$. An exchange
that trades a hypermultiplet charged under the non-generic representation
$R_{(q_1, q_2, q_3)}$ for some combination of hypermultiplets charged under
the canonical generic matter representations in \cref{tab:generic} increases
the number of uncharged scalars by an amount
\begin{equation}
P_1 = X + d Y - d + N A_R (Q_2 - Q_{11})\,,
\end{equation}
in terms of the quantities defined in \cref{sec:sun-proof,sec:u13-proof}. We
know already that $X, Y \ge 0$ for any $R$ and $q_1, q_2, q_3 \in \Z$. To deal
with the final two terms, note that $Y - 1 \ge 0$ for all charge combinations
other than $(1, 0, 0)$, $(1, \pm 1, 0)$, $(1, 1, -1)$, $(2, 0, 0)$, $(2, -1,
0)$, and their permutations (and conjugates). For each of these charge
combinations, $Q_2 - Q_{11} \ge 1$, and so $X - d + N A_R (Q_2 - Q_{11}) \ge
0$ because $C_R, g_R \ge 0$. Thus, $P_1 \ge 0$ for all representations
$R_{(q_1, q_2, q_3)}$.

%%%%%%%%%%%%%%%%%%%%%%%%%%%%%%%%%%%%%%%%%%%%%%%%%%%%%%%%%%%%%%%%%%%%%%%%%%%%%%
%%%%%%%%%%%%%%%%%%%%%%%%%%%%%%%%%%%%%%%%%%%%%%%%%%%%%%%%%%%%%%%%%%%%%%%%%%%%%%
\subsection{$\SU(N_1) \times \SU(N_2) \times \U(1)^3$}
\label{sec:sun-sun-u13-proof}

We now consider a theory with gauge group $\SU(N_1) \times \SU(N_2) \times
\U(1)^3$. An exchange that trades a hypermultiplet charged under the
non-generic representation $(R_1, R_2)_{(q_1, q_2, q_3)}$ for some combination
of hypermultiplets charged under the canonical generic matter representations
in \cref{tab:generic} increases the number of uncharged scalars by an amount
\begin{equation}
P_2 = d_1 X_2 + d_2 X_1 + d_1 d_2 Y + N_1 A_1 N_2 A_2 - 2 d_1 d_2
	+ (d_1 N_2 A_2 + d_2 N_1 A_1) (Q_2 - Q_{11})\,,
\end{equation}
in terms of the quantities defined in \cref{sec:sun-proof,sec:u13-proof}, with
$X_i$, $d_i$, and $A_i := A_{R_i}$ for $i = 1, 2$ corresponding to the
respective $\SU(N_i)$ gauge factors. Using the same arguments as in the
previous section, we can account for one factor of $-d_1 d_2$ by absorbing it
into $Y$ for most values of the $q_i$, or into the nonabelian terms in the
remaining cases.

To account for the other factor of $-d_1 d_2$, we appeal to the fact that $N
A_i \ge d_i$ for all representations other than the singlet. This fact can be
demonstrated using the same approach as in \cref{sec:sun-proof}; in this case,
we find that the only ``trouble cases'' are the relative weights of the form
$\mu = [1, \dots, 1]$, which contribute $-K_{\lambda \mu}$. Using the approach
of \cref{sec:sun-proof}, we consider the map between tableaux that replaces
the entry in the box immediately to the right of the final $1$ (which must
exist, as there are $N (1 + k)$ boxes and $1 + k$ instances of entry $1$ in at
most $N - 1$ rows, for some $k$) with a $1$. Unlike the earlier cases, this
map is a bijection, and thus shows that
\begin{equation}
K_{\lambda \mu} \le (N - 1) K_{\lambda [2, 1, \dots, 1, 0]}\,.
\end{equation}
The tableaux of weight $\mu' = (2 + k, 1 + k, \dots, 1 + k, k)$ and its
permutations collectively contribute $N (N + 1) K_{\lambda \mu'}$, and we then
have
\begin{equation}
\begin{aligned}
N (N + 1) K_{\lambda \mu'} - K_{\lambda \mu} &\ge
	[N (N + 1) - (N - 1)] K_{\lambda \mu'} \\
	&= (N^2 + 1) K_{\lambda \mu'} \\
	&\ge 0\,.
\end{aligned}
\end{equation}
Thus, in the case that neither $R_1$ nor $R_2$ is the
singlet, $N_1 A_1 N_2 A_2 \ge d_1 d_2$; in the case that one of them is the
singlet, say $R_1 = \bm{1}$, then $d_2 X_1 = d_1 d_2$, balancing the
remaining $-d_1 d_2$. Thus, $P_2 \ge 0$ for all representations $(R_1,
R_2)_{(q_1, q_2, q_3)}$.

\subsection{$\SU(N_1) \times \dots \times \SU(N_r) \times \U(1)^3$}
\label{sec:suns-u13-proof}

Because the generic matter in \cref{tab:generic} only has matter charged under
at most two of the nonabelian factors, we can see that the result of
\cref{sec:sun-sun-u13-proof} generalizes trivially to an arbitrary number of
$\SU(N)$ factors, completing the proof that the canonical matter in
\cref{tab:generic} is generic for the gauge group $G = \SU(N_1) \times \dots
\times \SU(N_r) \times \U(1)^3$. By taking the all hypermultiplets to be
uncharged under some or all of the $\U(1)$ factors, this also proves the
result for $\SU(N_1) \times \dots \times \SU(N_r) \times \U(1)^s$, $s \le 3$.

%%%%%%%%%%%%%%%%%%%%%%%%%%%%%%%%%%%%%%%%%%%%%%%%%%%%%%%%%%%%%%%%%%%%%%%%%%%%%%
%%%%%%%%%%%%%%%%%%%%%%%%%%%%%%%%%%%%%%%%%%%%%%%%%%%%%%%%%%%%%%%%%%%%%%%%%%%%%%
%%%%%%%%%%%%%%%%%%%%%%%%%%%%%%%%%%%%%%%%%%%%%%%%%%%%%%%%%%%%%%%%%%%%%%%%%%%%%%
\section{Generic $\SU(2)$ matter at small $b$, $T > 0$}
\label{sec:appendix-t}

Here, we will generalize the proofs given in \cref{sec:dimension-issues-T>0}
to higher $T$, using an approach that does not rely on the positivity cone. We
will restrict our attention to $\SU(2)$ theories.
% Consider the odd lattice
% \begin{equation}
% \Gamma = \mat[p]{1 & 0 \\ 0 & -1}\,,
% \end{equation}
% for $T = 1$, which has $-a = (3, -1)$.
Consider the lattice
\begin{equation}
\label{eq:unimodular-lattice}
\Gamma = \diag(1, -1, -1, \dots, -1)\,,
\end{equation}
for $1 \le T < 9$, with corresponding $-a = (3, -1, \dots, -1)$, which is
fixed by the characteristic vector criterion shown in
\cite{MonnierMooreParkQuantization}. (Note that the even lattice
\labelcref{eq:even-lattice}, which was already treated in
\cref{sec:dimension-issues-T>0}, only arises at $T = 1$, so we do not revisit
that case here.)
% This is one of the only two unimodular lattices of signature $(1, 1)$, the
% other being the even lattice
% \begin{equation}
% U = \mat[p]{0 & 1 \\ 1 & 0}\,.
% \end{equation}
% The positivity condition for $(b_0, b_1)$ is $b_0, b_0 + b_1 \ge 0$.
We wish to prove that for sufficiently small values of $b = (b_0, \dots,
b_T)$, there is always a solution of the AC conditions with only fundamentals
and adjoints. We define $n = b \cdot b$ and $\gamma = -a \cdot b$.

The genus is defined as
\begin{equation}
g = \frac{b \cdot (b + a)}{2} + 1 = \frac{n - \gamma}{2} + 1\,.
\end{equation}
This expression is given by the Riemann--Roch formula in the F-theory setting,
but from the low-energy point of view, this can be taken as a definition of
the quantity $g$. By taking the appropriate combination of
\cref{eq:nonabelACA,eq:nonabelACC}, we can write the genus for an $\SU(2)$
model as
\begin{equation}
g = \frac{1}{12} \sum_R x_R (2 C_R - A_R)\,.
\end{equation}
Thus, we can assign to each representation $R$ a genus
\begin{equation}
g_R = \frac{2 C_R - A_R}{12}\,,
\end{equation}
so that $g$ is the sum of the $g_R$ over all hypermultiplets. As mentioned in
\cref{sec:generic-examples}, an arbitrary $\SU(2)$ representation
$\underbrace{\ytableaushort{\none{\none[\dotsm]}\none}*{1}*{2+1}}_k$ has $A_R
= \binom{k + 2}{3}$ and $C_R = \binom{k + 2}{3} \frac{3 k (k + 2) - 4}{10}$.
From these, we see that
\begin{equation}
g_R = \frac{(k + 3) (k + 2) (k + 1) k (k - 1)}{120}\,.
\end{equation}
We see that $g_{\, \tyng{1}} = 0$, $g_{\, \tyng{2}} = 1$, and that the genus
is strictly increasing in $k$. Thus, for a given choice of $b$, the genus must
be non-negative for there to be \emph{any} solutions to the AC equations, so
we must have
\begin{equation}
\label{eq:genusPos}
\frac{n - \gamma}{2} + 1 \ge 0\,.
\end{equation}

Now consider a choice of $b$ for which there is no valid solution (one with
non-negative multiplicities) in terms of generic matter. There is a unique
solution for the multiplicities $x_{\bm{1}}, x_{\, \tyng{1}}, x_{\, \tyng{2}}$
for this $b$, and we know that $x_{\, \tyng{2}} = g$ must be non-negative from
the above argument. We further know that $x_{\bm{1}} \ge 0$, because if this
were not the case, then no exchanges of the form
\labelcref{eq:su2-equivalence} could result in a valid solution, as they all
reduce the number of uncharged scalars in exchange for larger representations.
Thus, we must have $x_{\,
\tyng{1}} < 0$, by the assumption this choice of $b$ does not have a valid
solution with only these three representations. This gives us the further
constraint
\begin{equation}
\label{eq:fundsNeg}
x_{\, \tyng{1}} = 8 \gamma - 2 n < 0\,.
\end{equation}

For there to be a valid solution at all with this choice of $b$, there must be
a sequence of exchanges for larger representations that will result in all
non-negative multiplicities.
%%% THIS IS WRONG %%%
% In order for this to be the case, there must be a
% sufficient number of adjoints to account for the total genus without the
% multiplicity of adjoints becoming negative. As we saw above, for $\SU(2)$ the
% genus of a representation strictly increases with the number of boxes in the
% corresponding Young diagram. Thus, if there is any valid solution, then there
% will be a valid solution with only fundamentals, adjoints, and triple
% symmetrics.
As shown in \cref{eq:su2-equivalence}, exchanges
to higher representations produce fundamentals but consume adjoints, and so
there must be a sufficient number of adjoints in order for exchanges to give a
positive multiplicity $x_{\, \tyng{1}}$ without $x_{\, \tyng{2}}$ becoming
negative. From \cref{eq:su2-equivalence}, an exchange to
$\underbrace{\ytableaushort{\none{\none[\dotsm]}\none}*{1}*{2+1}}_k$ produces
\begin{equation}
\frac{(k +  4) (k + 2) (k + 1) k (k - 2)}{30 \binom{k + 3}{5}} =
	\frac{4 (k +  4) (k - 2)}{(k + 3) (k - 1)}
\end{equation}
fundamentals for every adjoint it consumes. This ratio is strictly increasing
in $k$, and approaches $4$ as $k \to \infty$. Thus, we must have
\begin{equation}
x_{\, \tyng{1}} \ge -4 x_{\, \tyng{2}}
\end{equation}
in order for there to be a sufficient number of adjoints for exchanges to make
all multiplicities positive. This gives us the third constraint
\begin{equation}
\label{eq:enoughAdjs}
8 \gamma - 2 n \ge -4 \left(\frac{n - \gamma}{2} + 1\right)\,.
\end{equation}
This constraint along with \cref{eq:fundsNeg} implies \cref{eq:genusPos}. We
see then that the range of possible $\gamma, n$ where there may be a problem
with generic matter solutions is the set of integer pairs in the range
\begin{equation}
\label{eq:problem-region}
\gamma \geq 0\,, \quad  n > 4 \gamma\,.
\end{equation}

\begin{figure}[h!]
\centering

\pic{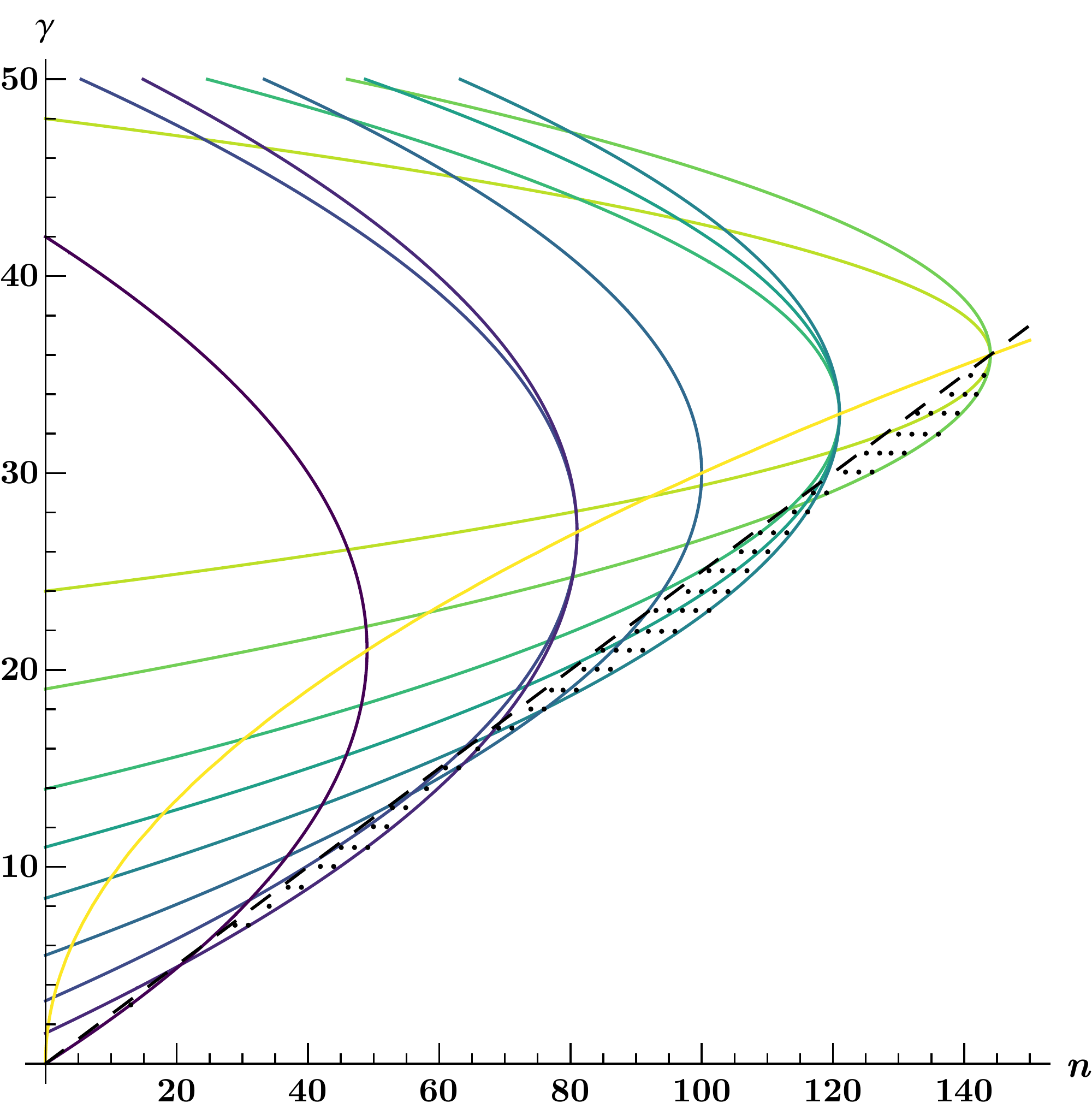}

\caption{A plot of the parabolas $n = b_0^2 - \frac{(\gamma - 3 b_0)^2}{T}$
with the critical value $b_0 = \floor{6 + 2 \sqrt{9 - T} + 1}$, for $T = 1,
\dots, 9$. These values are $b_0 = 12, 12, 11, 11, 11, 10, 9, 9, 7$ for $T =
1, \dots, 9$, respectively. The lightest yellow curve is the parabola $n =
\gamma^2 / 9$, relevant for $T = 0$, and the curves for $T = 1, \dots, 9$
appear in progressively darker colors. The dashed line is $n = 4 \gamma$, the
line below which there exist valid choices of anomaly coefficient that do not
yield generic matter solutions. In black are points within the parabolas for
$T = 1, \dots, 9$ for which there are valid solutions of the AC conditions but
no solutions with generic matter.}
\label{fig:su2-ts}
\end{figure}

% For $1 \leq T < 9$, the only unimodular lattice up to symmetries is
% \begin{equation}
% \label{eq:unimodular-lattice}
% \Gamma = \diag(1, -1, -1, \dots, -1)\,,
% \end{equation}
% with corresponding $-a = (3, -1, \dots, -1)$, which is fixed by the
% characteristic vector criterion shown in \cite{MonnierMooreParkQuantization}.
From $-a = (3, -1. \dots, -1)$, we have
\begin{equation}
n = b_0^2 - \sum_{i = 1}^T b_i^2\,, \quad
	\gamma = 3 b_0 + \sum_{i = 1}^T b_i\,.
\end{equation}
% Let us write $x = b_0, y = \sum_{i = 1}^T b_i / T$.
For each fixed $T, x$, from the inequality
\begin{equation}
\sum_{i = 1}^T b_i^2 \geq \frac{1}{T} \left(\sum_{i = 1}^T b_i\right)^2\,,
\end{equation}
we have
\begin{equation}
n \leq b_0^2 - \frac{1}{T} \left(\sum_{i = 1}^T b_i\right)^2
	= b_0^2 - \frac{(\gamma - 3 b_0)^2}{T}\,.
\end{equation}
We see that for a fixed $T$ the resulting parabola is tangent to the
line $n = 4 \gamma$ when
\begin{equation}
\label{eq:tangentPoint}
b_0 = 6 + 2 \sqrt{9 - T}\,.
\end{equation}
For smaller values of $b_0$, the parabola lies outside the region
\labelcref{eq:problem-region}, so there is always a solution with
non-negative multiplicities of generic matter when $T < 9, b_0 \leq 6$. An
explicit check shows that the first integer value of $b_0$ that exceeds the
value in \cref{eq:tangentPoint} for each $T$ corresponds to a valid choice of
$b$ that does not admit generic matter. For example, at $T = 3$, $b = (11, 2,
2, 2)$ gives $\gamma = 27, n = 109$, which has $n > 4 \gamma$. In
\cref{fig:su2-ts}, we show the parabolas $n = b_0^2 - \frac{(\gamma - 3
b_0)^2}{T}$ with $b_0 = \floor{6 + 2 \sqrt{9 - T} + 1}$ for each $T$, and how
they intersect the line $n = 4 \gamma$.

For $T = 9$ with the unimodular lattice \labelcref{eq:unimodular-lattice}, the
same result holds.  While from the low-energy point of view we cannot rule out
the unimodular lattice $U \oplus \gE_8$ at $T = 9$, the only F-theory
construction with this lattice uses the Enriques surface and does not admit
any gauge group since the canonical class is trivial up to torsion. Similarly,
any F-theory model with $T > 9$ must have a larger gauge group than $\SU(2)$
from non-Higgsable clusters, so this argument shows that there is a
non-negative generic matter spectrum for all $\SU(2)$ models that are not
already in the swampland for other reasons.

% Now, note that
% \begin{equation}
% n = b_0^2 - b_1^2\,, \quad \gamma = 3 b_0 + b_1\,.
% \end{equation}
% We know that $b_0$ and $b_1$ must be integers. Defining $j := b_0$ and $k :=
% b_0 + b_1$, choices of $b$ are parameterized by integer choices of $j, k$. We
% can then write
% \begin{equation}
% n = k (2 j - k)\,, \quad \gamma = 2 j + k\,.
% \end{equation}
% In terms of $j$ and $k$, the constraints \labelcref{eq:fundsNeg,eq:enoughAdjs}
% \wati{would a graph be helpful?}
% \begin{equation}
% \begin{aligned}
% k (2 j - k) - (2 j + k) + 2 &\ge 0\,, \\
% 8 (2 j + k) - 2 k (2 j - k) &< 0\,, \\
% 6 j + 3 k + 2 & \ge 0\,.
% \end{aligned}
% \end{equation}
% These three inequalities have no integer solutions for $j < 12$, which gives
% us our bound on $b_0, b_1$. We see that if we have $b_0 < 12$ or $b_0 + b_1 <
% 5$, then there are always valid solutions with only generic matter.

% For $1 < T < 9$, the only unimodular lattice up to symmetries is
% \begin{equation}
% \Gamma = \diag(1, -1, -1, \dots, -1)\,,
% \end{equation}
% with corresponding $-a = (3, -1, \dots, -1)$, which is fixed by the
% characteristic vector criterion shown in \cite{MonnierMooreParkQuantization}.
% In this case, the proof above follows in the same way, taking $j = b_0$ and $k
% = b_0 + \dots + b_T$, to show that when $b_0 < 12$, there are always valid
% solutions with only generic matter.

Note that this proof does not require a choice of positivity cone, and so is
strictly more powerful than the arguments given in
\cref{sec:dimension-issues-T>0}, even for the case of $T = 1$.

%%%%%%%%%%%%%%%%%%%%%%%%%%%%%%%%%%%%%%%%%%%%%%%%%%%%%%%%%%%%%%%%%%%%%%%%%%%%%%
%%%%%%%%%%%%%%%%%%%%%%%%%%%%%%%%%%%%%%%%%%%%%%%%%%%%%%%%%%%%%%%%%%%%%%%%%%%%%%
%%%%%%%%%%%%%%%%%%%%%%%%%%%%%%%%%%%%%%%%%%%%%%%%%%%%%%%%%%%%%%%%%%%%%%%%%%%%%%
\section{Generic matter spectra for $G = \U(1)^2$}
\label{sec:two-u1-appendix}

In this appendix, we briefly discuss generic matter for the group $G =
\U(1)^2$ with general $T$.

Solving the AC equations \labelcref{eq:abelAC} for the two nine-element
subsets (including $(0, 0)$) of generic $\U(1)^2$ matter that include the $(2,
-1), (-1, 2)$ and $(2, 1), (1, 2)$ pairs of charge combinations, we find that
\begin{equation}
\setlength{\arraycolsep}{4pt}
\renewcommand*{\arraystretch}{1.3}
\mat[p]{
	x_{1, 0} \\
	x_{0, 1} \\
	x_{2, 0} \\
	x_{0, 2} \\
	x_{1, \pm 1} \\
	x_{1, \mp 1} \\
	x_{2, \pm 1} \\
	x_{\pm 1, 2}
}
=
\mat[p]{
	-8 & \pm 7 & 0 & -1 & \pm 2 & -1 & \pm \frac{3}{2} & 0 \\
	0 & \pm 7 & -8 & 0 & \pm \frac{3}{2} & -1 & \pm 2 & -1 \\
	\frac{1}{2} & \mp 1 & 0 & \frac{1}{4} & \mp \frac{1}{2} & 0 & 0 & 0 \\
	0 & \mp 1 & \frac{1}{2} & 0 & 0 & 0 & \mp \frac{1}{2} & \frac{1}{4} \\
	0 & \mp 9 & 0 & 0 & \mp \frac{3}{2} & \frac{1}{2} & \mp \frac{3}{2} & 0 \\
	0 & \pm 1 & 0 & 0 & \mp \frac{1}{2} & \frac{1}{2} & \mp \frac{1}{2} & 0 \\
	0 & \pm 1 & 0 & 0 & \pm \frac{1}{2} & 0 & 0 & 0 \\
	0 & \pm 1 & 0 & 0 & 0 & 0 & \pm \frac{1}{2} & 0
}
\mat[p]{
	a \cdot b_{1 1} \\
	a \cdot b_{1 2} \\
	a \cdot b_{2 2} \\
	b_{1 1}^2 \\
	b_{1 1} \cdot b_{1 2} \\
	b_{1 1} \cdot b_{2 2} + 2 b_{1 2}^2 \\
	b_{2 2} \cdot b_{1 2} \\
	b_{2 2}^2
}\,.
\end{equation}
% By inspecting this matrix, we see that these multiplicities are non-negative
% when
% \begin{equation}
% \begin{aligned}
% \frac{1}{2} b_{1 1} \cdot \left(a + \frac{1}{2} b_{1 1}\right) \ge
% 	\pm b_{1 2} \cdot \left(a + \frac{1}{2} b_{1 1}\right) &\ge 0 \\
% \frac{1}{2} b_{2 2} \cdot \left(a + \frac{1}{2} b_{2 2}\right) \ge
% 	\pm b_{1 2} \cdot \left(a + \frac{1}{2} b_{2 2}\right) &\ge 0 \\
% b_{1 2} \cdot \left[\pm a + b_{1 2} \mp \frac{1}{2} \left(b_{1 1} +
% 	b_{2 2}\right)\right] + \frac{1}{2} b_{1 1} \cdot b_{2 2} &\ge 0 \\
% b_{1 2} \cdot \left[\mp 9 a + b_{1 2} \mp \frac{3}{2} \left(b_{1 1} +
% 	b_{2 2}\right)\right] + \frac{1}{2} b_{1 1} \cdot b_{2 2} &\ge 0 \\
% -8 a \cdot b_{1 1} \pm 7 a \cdot b_{1 2} - b_{1 1}^2 \pm
% 	2 b_{1 1} \cdot b_{1 2} - b_{1 1} \cdot b_{2 2} - 2 b_{1 2}^2 \pm
% 	\frac{3}{2} b_{2 2} \cdot b_{1 2} &\ge 0 \\
% -8 a \cdot b_{2 2} \pm 7 a \cdot b_{1 2} - b_{2 2}^2 \pm
% 	\frac{3}{2} b_{1 1} \cdot b_{1 2} - b_{1 1} \cdot b_{2 2} - 2 b_{1 2}^2
% 	\pm 2 b_{2 2} \cdot b_{1 2} &\ge 0
% \end{aligned}
% \end{equation}
Similarly, solving the AC equations \labelcref{eq:abelAC} for the two
nine-element subsets (including $(0, 0)$) of generic $\U(1)^2$ matter that
include the $(2, -1), (1, 2)$ and $(2, 1), (-1, 2)$ pairs of charge
combinations, we find that
\begin{equation}
\setlength{\arraycolsep}{4pt}
\renewcommand*{\arraystretch}{1.3}
\mat[p]{
	x_{1, 0} \\
	x_{0, 1} \\
	x_{2, 0} \\
	x_{0, 2} \\
	x_{1, \pm 1} \\
	x_{1, \mp 1} \\
	x_{2, \pm 1} \\
	x_{\mp 1, 2}
}
=
\mat[p]{
	-8 & \pm 1 & 0 & -1 & \pm 2 & -1 & \mp \frac{3}{2} & 0 \\
	0 & \mp 1 & -8 & 0 & \pm \frac{3}{2} & -1 & \mp 2 & -1 \\
	\frac{1}{2} & \mp 1 & 0 & \frac{1}{4} & \mp \frac{1}{2} & 0 & 0 & 0 \\
	0 & \pm 1 & \frac{1}{2} & 0 & 0 & 0 & \pm \frac{1}{2} & \frac{1}{4} \\
	0 & \mp 5 & 0 & 0 & \mp \frac{3}{2} & \frac{1}{2} & \pm \frac{1}{2} & 0 \\
	0 & \pm 5 & 0 & 0 & \mp \frac{1}{2} & \frac{1}{2} & \pm \frac{3}{2} & 0 \\
	0 & \pm 1 & 0 & 0 & \pm \frac{1}{2} & 0 & 0 & 0 \\
	0 & \mp 1 & 0 & 0 & 0 & 0 & \mp \frac{1}{2} & 0
}
\mat[p]{
	a \cdot b_{1 1} \\
	a \cdot b_{1 2} \\
	a \cdot b_{2 2} \\
	b_{1 1}^2 \\
	b_{1 1} \cdot b_{1 2} \\
	b_{1 1} \cdot b_{2 2} + 2 b_{1 2}^2 \\
	b_{2 2} \cdot b_{1 2} \\
	b_{2 2}^2
}\,.
\end{equation}

By inspecting these matrices, we see that in all four cases, a good solution
with non-negative multiplicities must have $b_{1 2}$ satisfying
\begin{equation}
\begin{aligned}
\frac{1}{2} b_{1 1} \cdot \left(a + \frac{1}{2} b_{1 1}\right) \ge
	\pm b_{1 2} \cdot \left(a + \frac{1}{2} b_{1 1}\right) &\ge 0 \\
\frac{1}{2} b_{2 2} \cdot \left(a + \frac{1}{2} b_{2 2}\right) \ge
	\pm b_{1 2} \cdot \left(a + \frac{1}{2} b_{2 2}\right) &\ge 0
\end{aligned}
\end{equation}
for the relevant choice of signs. Note that \emph{any} good solution must have
\begin{equation}
-a \cdot b_{1 1}, -a \cdot b_{2 2}, b_{1 1}^2, b_{2 2}^2,
	b_{1 1} \cdot b_{2 2} + 2 b_{1 2}^2 \ge 0\,,
\end{equation}
following directly from the forms of the AC equations
\labelcref{eq:abelAC}.

In \cref{sec:abelian-2-small-b}, we show that when $T = 0$, there is
always a non-negative generic matter spectrum when the anomaly
coefficients $b$ are not too big.

\bibliographystyle{JHEP}
\bibliography{research}

\end{document}